\documentclass[twocolumn,pra,twocolumn,superscriptaddress]{revtex4}
\usepackage{color}
\usepackage{amsmath}
\usepackage{amssymb}
\usepackage{graphicx}
\usepackage{tikz}
\usepackage{multirow}

\usepackage{float}
\usepackage{bbm}
\usepackage{dsfont}

\usepackage[T1]{fontenc}

\usepackage{epsfig}
\usepackage{verbatim}
\usepackage{array}
\usepackage{setspace}
\newcommand*{\red}{\textcolor{black}} 

\usepackage[unicode=true,
bookmarks=true,bookmarksnumbered=false,bookmarksopen=false,
breaklinks=false,pdfborder={0 0 1},backref=false,colorlinks=true]
{hyperref}
\hypersetup{
	linkcolor=magenta, urlcolor=blue, citecolor=blue, pdfstartview={FitH}, hyperfootnotes=false, unicode=true}

\makeatletter
\@ifundefined{textcolor}{}
{%
	\definecolor{BLACK}{gray}{0}
	\definecolor{WHITE}{gray}{1}
	\definecolor{RED}{rgb}{1,0,0}
	\definecolor{GREEN}{rgb}{0,1,0}
	\definecolor{BLUE}{rgb}{0,0,1}
	\definecolor{CYAN}{cmyk}{1,0,0,0}
	\definecolor{MAGENTA}{cmyk}{0,1,0,0}
	\definecolor{YELLOW}{cmyk}{0,0,1,0}
}

\usepackage{amsfonts}\usepackage{tabularx}\usepackage{dcolumn}\usepackage{bm}\usepackage{graphicx}\usepackage{epstopdf}

\hypersetup{urlcolor=blue}

\makeatother

\allowdisplaybreaks
\newcolumntype{C}[1]{>{\centering\arraybackslash$}p{#1}<{$}}

\begin{document}
	
	\title{Microscopic theory on magnetic-field-tuned sweet spot of exchange interactions in multielectron quantum-dot systems}
	
	\author{Guo Xuan Chan}
	\affiliation{Department of Physics, City University of Hong Kong, Tat Chee Avenue, Kowloon, Hong Kong SAR, China, and City University of Hong Kong Shenzhen Research Institute, Shenzhen, Guangdong 518057, China}
	
	\author{Xin Wang}
	\email{x.wang@cityu.edu.hk}
	\affiliation{Department of Physics, City University of Hong Kong, Tat Chee Avenue, Kowloon, Hong Kong SAR, China, and City University of Hong Kong Shenzhen Research Institute, Shenzhen, Guangdong 518057, China}
	\date{\today}
	
	\begin{abstract}		
	The exchange interaction in a singlet-triplet qubit defined by two-electron states in the double-quantum-dot system (``two-electron singlet-triplet qubit'') typically varies monotonically with the exchange interaction and thus carries no sweet spot. Here we study a singlet-triplet qubit defined by four-electron states in the double-quantum-dot system (``four-electron singlet-triplet qubit''). We demonstrate, using configuration-interaction calculations, that in the four-electron singlet-triplet qubit the exchange energy as a function of detuning can be non-monotonic, suggesting existence of sweet spots. We further show that the tuning of the sweet spot and the corresponding exchange energy by perpendicular magnetic field can be related to the variation of orbital splitting. Our results suggest that a singlet-triplet qubit with more than two electrons can have advantages in the realization of quantum computing.
	\end{abstract}

	\maketitle
	
	\section{Introduction}
	Semiconductor quantum-dot (QD) qubits, being one of the most promising candidates for performing quantum information processing, are realized by encoding quantum information into the electrons' degree of freedoms, which are their spins \cite{Koppens.06,Zajac.18,Mills.21,Yoneda.18,Nowack.07,Pioro.08,Jelmer.21,Laird.10,Gaudreau.12,Medford.13,Martins.16,Reed.16,Nichol.17,Takeda.20,Eng.15,Petta.05,Shulman.12,Maune.12,Wu.14,Jock.18,Cerfontaine.20,Harvey.17,Harvey.19,Liu.21,Zhang.18,Li.18,Jock.21,Hendrickx.21,Philips.22}, charge \cite{Hayashi.03,Gorman.05,Shinkai.09,Petersson.10,Shi.13,Dovzhenko.11,Landau.12,Li.15} or a hybrid of both \cite{Shi.14,Kim.15,Cao.16,Thorgrimsson.17,Malinowski.17}. Among different schemes, spin qubit stands out due to its limited access to the charge dipole moment, avoiding strong coupling to the inherent charge noise in QD devices \cite{Cao.13,Shinkai.09,Hayashi.03,Petersson.10,Gorman.05,Shi.13,Dial.13}. The control of single-spin qubit is performed by generating local oscillating magnetic field either by electron spin resonance (ESR) \cite{Koppens.06,Zajac.18,Mills.21} or electron dipole spin resonance technique (EDSR) \cite{Yoneda.18,Nowack.07,Pioro.08,Jelmer.21,Huang.19}. On the other hand, operation of a spin qubit formed by two electrons with anti-parallel spins in a double-quantum-dot (DQD), i.e.~the singlet-triplet ($ST_0$) qubit, can be performed either by ramping the gate voltages \cite{Petta.05,Shulman.12,Eng.15,Barthel.10} or applying oscillating electric field on gate electrodes \cite{Martins.16,Reed.16,Nichol.17,Takeda.20,Malinowski.17}. The advantage of $ST_0$ qubit compared to single-spin qubit is that it allows all electrical control, simplifying the operation complexity.
	
	A starting point to perform fault-tolerant quantum computation \cite{Terhal.15} is to achieve robust single-qubit operations. In the context of $ST_0$ qubit, robust single-qubit gate can be realized if there exists a control protocol which allows tunability of qubit parameters while immune to environmental noises during the time evolution. This has been an enormously arduous task for semiconductor qubit as the control parameter, i.e.~electrical gate voltages, couples directly to the inherent charge noise in QD devices. In general, there are two main techniques to modulate gate voltages, including the resonant pulse and ramping.	The superiority of resonantly driven operations has been demonstrated \cite{Zajac.18,Mills.21,Yoneda.18,Huang.19} as it allows the system to stay at the symmetric point where susceptibility to charge noise is minimized \cite{Medford.13,Takeda.20,Malinowski.17}. However, further improvement for resonant control is challenging as more parameters are required to be carefully calibrated, including the pulse frequencies, Rabi envelope and phase error \cite{Zajac.18,Huang.19}. On the other hand, ramping technique is easier to implement and requires less calibration. Two techniques introduced above can be applied to tune the exchange energy of a two-electron $ST_0$ qubit, either by modulating the relative detuning between neighboring dots \cite{Liu.21,Harvey.19,Petta.05,Shulman.12,Harvey.17,Cerfontaine.20,Wu.14,Maune.12,Jock.18} or the inter-dot barrier height \cite{Reed.16,Martins.16,Yang.17,Yang.18}. 
	The former allows larger tuning range of the exchange energy for a limited amplitude of the voltage pulse while the latter ensures that the system stays in the symmetric point to gain protection against charge noise by sacrificing some tunability \cite{Shim.18}. Even if resonant control on the barrier height shows a relatively promising path \cite{Takeda.20,Malinowski.17}, it is still insightful to explore the potential of detuning ramping if we enter few-electron regime for $ST_0$ qubit architectures.
	
	The conventional two-electron $ST_0$ qubit yields a monotonic increase of exchange energy with respect to detuning \cite{Shulman.12,Petta.05,Barthel.10,Eng.15,Cerfontaine.20,Wu.14,Takeda.20,Maune.12,Jock.18}, which limits the workaround of this system from performing high fidelity operation since the increase of exchange energy by detuning control is accompanied with increased exposure to charge noise dephasing \cite{Dial.13}. However, recent experiments show that coupling a multi-electron QD with a singly-occupied QD leads to a non-monotonic dependence of exchange energy on detuning \cite{Martins.17,Malinowski.18}. This behavior can be attributed to the variation of the total spin of electrons in the fully occupied multi-electron dot, along with the interplay between hybridization of singlet (triplet) with singly-occupied valence orbitals and singlet (triplet) in fully-occupied multi-electron dot. Such new physical phenomenon for multi-electron $ST_0$ qubit gives rise to a sweet spot at larger detuning with larger exchange energy \cite{Martins.17}, which is not observed for a conventional two-electron $ST_0$ qubit, albeit sharing similar electron configurations. \red{Also, this sweet spot is located away from the symmetric operating point (SOP), a point that hosts another sweet spot with smaller exchange energy.} In addition, the sweet spot and the corresponding exchange energy are demonstrated to be tunable by varying the external magnetic field perpendicular to the plane of electron gas \cite{Martins.17}. Microscopically, the origin leading to the interesting behavior of exchange energy in coupled few-electron dots is left unanswered \cite{Martins.17}. Hence, understanding it is important to design high-fidelity quantum information processing in this system. In this paper, we show, using full configuration interaction (full CI) calculations, the tuning of sweet spot and the corresponding exchange energy can be directly attributed to the orbital splitting by magnetic field. In addition, we demonstrate that the tuning of sweet spot and exchange energy can be reproduced accurately using the extended Hubbard model (analytical) by fitting the Hubbard parameters to full CI results (numerical). The correspondence between numerical and analytical results is useful in facilitating the comprehension of the physical mechanism leading to the tuning effect. We also confirm that strong electron correlation and Wigner-molecule physics does not contribute to the tuning of singlet-triplet splitting in the fully-occupied dot. Our results should facilitate the understanding of $ST_0$ qubit realized in few-electron regime, benefiting the realization of quantum information processing beyond the conventional setup.
	
	\section{Model}
	\subsection{Configuration Interaction (CI)}\label{subsec:CI}
	We consider an $n$-electron system \red{$H=\sum h_j + \sum{e^2}/4\pi\epsilon\left|\mathbf{r}_j-\mathbf{r}_k\right|$}
	with the single-particle Hamiltonian \red{$h_j = {(-i\hbar \nabla_j+e \mathbf{A})^2}/{2m^*}+V_\text{DQD}(\mathbf{r})+g^*\mu_B \mathbf{B}\cdot \mathbf{S}$}.
	
	\begin{figure}[t]
		\includegraphics[width=0.9\columnwidth]{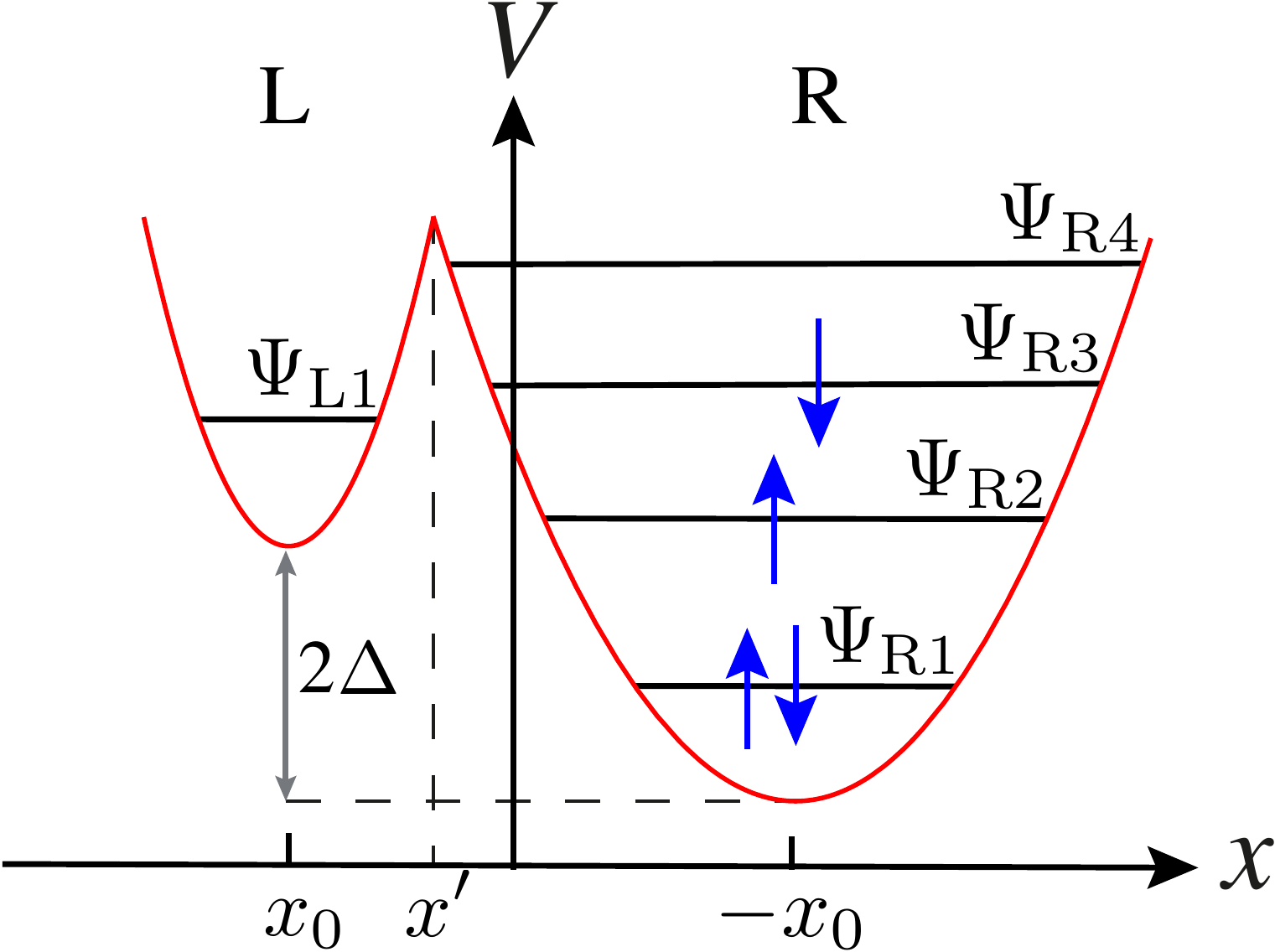}
		\caption{Schematic illustration of the model potential given in Eq.~\eqref{eq:V}}.
		\label{fig:potential}
	\end{figure}

	The confinement potential of a DQD device can be modeled as (cf.~Fig.~\ref{fig:potential}): 
	\begin{equation}\label{eq:V}
		\begin{split}
			V_\text{DQD}(\mathbf{r})&=
			\begin{cases}
				V\left(\mathbf{r} \vert -\mathbf{R},\omega_\text{L}\right) & x < x',\\
				V\left(\mathbf{r}\vert \mathbf{R},\omega_\text{R}\right) & x> x',
			\end{cases}
		\end{split}
	\end{equation}
	where 
	\begin{equation}
		V\left(\mathbf{r} \vert \widetilde{\mathbf{R}},\widetilde{\omega}\right) = \frac{1}{2}m^* \widetilde{\omega}^2\left(\mathbf{r}-\widetilde{\mathbf{R}}\right)^2.
	\end{equation}
	$\mathbf{r}=(x,y)$ is the two dimensional vector in the plane of electron gas while $\mathbf{R} = \left(x_0,0\right)$ is the position of the parabolic well minimum. $x'$ is the potential cut determined by locating the value of $x$ at which the potential values of left and right dot are equal at $y=0$. The effective mass $m^*$ is 0.067 electron mass in GaAs. $\widetilde{\omega}$ is the confinement strength. $\mathbf{B}= B\mathbf{\hat{z}}$ is the perpendicular magnetic field and $\mathbf{S}$ is the total electron spin. The confinement strengths are $\hbar \omega_\text{L} = 7.5$ meV and $\hbar \omega_\text{R} = $ 4 meV for the left (L) and right (R) dots respectively. \red{The inter-dot distance $2x_0$ is 80 nm.}
		
	We focus our study on a four-electron singlet-triplet qubit, of which a single electron occupies the smaller QD, dot L, while three electrons occupy the larger QD, dot R. The eigenvalues of the system are calculated by constructing all possible four-electron Slater determinants for a given number of orbitals. The electron wavefunctions are the orthonormalized Fock-Darwin (F-D) states \cite{Barnes.11}, which are obtained by Cholesky decomposition of the overlap matrix formed by the bare F-D states. As suggested by the convergence of the exchange energy of four electrons occupying a larger quantum-dot (QD) (see Appendix \ref{sec:converg} for details), we retain 10 and 6 lowest orbitals in dot R and dot L respectively, giving a total of 16 orbitals in a DQD device. This setup results in a total of 14,400 Slater determinants.
	
	\subsection{Extended Hubbard Model}\label{sec:HubModel}
	Although CI calculations give accurate descriptions of the eigenvalues, it is hard to interpret the results physically due to the large number of Slater determinants. Hence, it is helpful to fit the CI results into an effective Hamiltonian written in the extended Hubbard Model, giving
	\begin{equation}\label{eq:HubbardModel2ndQuan}
	\begin{split}
		H &= \sum_{j \sigma}\varepsilon_{j \sigma} c^\dagger_{j \sigma} c_{j \sigma}+\sum_{j<k,\sigma} \left(t_{jk\sigma} c^\dagger_{j\sigma} c_{k\sigma}+\mathrm{H. c.}\right)\\
		&\quad+\sum_{j} U_{j} n_{j\downarrow} n_{j\uparrow} +\sum_{\sigma \sigma'}\sum_{j<k} U_{jk} n_{j\sigma} n_{k\sigma'}\\
		&\quad+\sum_{\sigma \sigma'} \sum_{j<k} U^e_{jk} c_{j \sigma}^\dagger c_{k \sigma'}^\dagger c_{j \sigma'} c_{k \sigma},
	\end{split}
	\end{equation}
	where $j$ and $k$ are orbital indices, and $\sigma$ and $\sigma'$ indicate spins. The summations over orbitals $(j,k)$ are from $\text{R}1$ to $\text{R}3$ for the right dot and $\text{L}1$ for the left dot (cf.~Fig.~\ref{fig:potential}), while spins $(\sigma,\sigma')$ can be either up or down. $\varepsilon_{j\sigma}$ denotes the on-site energy at dot $j$ while $t_{jk\sigma}$ denotes the tunneling between the $j$th and $k$th orbital. $U_{j}$ denotes the on-site Coulomb interaction in the $j$th orbital while $U_{jk}$ and $U^e_{jk}$ denotes the direct and exchange Coulomb interaction between the $j$th and $k$th orbital respectively. 
	
	These parameters are calculated from
	\begin{subequations}\label{eq:HubbardPara}
	\begin{align}
		\begin{split}
			U_{j} &= \int{\Psi^*_j (\mathbf{r}_1)\Psi^*_j (\mathbf{r}_2) C(\mathbf{r}_1,\mathbf{r}_2) \Psi_j (\mathbf{r}_1)\Psi_j (\mathbf{r}_2) \text{d}\mathbf{r}^2},
		\end{split}\\
		\begin{split}
			U_{jk} &= \int{\Psi^*_j (\mathbf{r}_1)\Psi^*_k (\mathbf{r}_2) C(\mathbf{r}_1,\mathbf{r}_2) \Psi_j (\mathbf{r}_1)\Psi_k (\mathbf{r}_2) \text{d}\mathbf{r}^2},
		\end{split}\\
		\begin{split}\label{eq:Ue}
			U_{jk}^e &= \int{\Psi^*_j (\mathbf{r}_1)\Psi^*_k (\mathbf{r}_2) C(\mathbf{r}_1,\mathbf{r}_2) \Psi_k (\mathbf{r}_1)\Psi_j (\mathbf{r}_2) \text{d}\mathbf{r}^2},
		\end{split}
		\\
		\begin{split}
			t_{jk}&=  \int{\Psi^*_j (\mathbf{r}) \left[\frac{\hbar^2}{2m^*}\nabla^2+V(\mathbf{r}) \right]\Psi_k (\mathbf{r}) \text{d}\mathbf{r}},
		\end{split}
		\\
		\begin{split}
			\varepsilon_j &=  \int{\Psi^*_j (\mathbf{r}) \left[\frac{\hbar^2}{2m^*}\nabla^2+V(\mathbf{r}) \right]\Psi_j (\mathbf{r}) \text{d}\mathbf{r}},
		\end{split}
		\\
		\begin{split}
			& C(\mathbf{r}_1,\mathbf{r}_2)=\frac{e^2}{\kappa|\mathbf{r}_1-\mathbf{r_2}|}.
		\end{split}
\end{align}
\end{subequations}
In addition, the onsite Coulomb exchange term in the right dot (cf.~Eq.~\eqref{eq:Ue}) can be rewritten as \cite{Malinowski.18,Deng.18}:
	\begin{equation}\label{eq:exchangeTerm}
	\begin{split}
		&\sum_{\sigma \sigma'} \sum_{\text{R}j,\text{R}{k},k\neq j} U^e_{\text{R}j,\text{R}k} c_{\text{R}j \sigma}^\dagger c_{\text{R}k \sigma'}^\dagger c_{\text{R}j \sigma'} c_{\text{R}k \sigma}\\
		&= - 2J^F_{\text{R}j,\text{R}k} \left(\mathbf{S}_{\text{R}j}\cdot \mathbf{S}_{\text{R}k}+\frac{1}{4} n_{\text{R}j} n_{\text{R}k}\right).
	\end{split}
	\end{equation}
	It is observed that the exchange energy as a function of detuning can be reproduced by fitting the lowest part of the energy spectrum to the extended Hubbard Model (Eq.~\eqref{eq:HubbardModel2ndQuan}), which will be discussed in Sec.~\ref{subsec:EvalHubbard}. Note that in Eq.~\eqref{eq:HubbardPara}, $\Psi_j$ is the linear combination of orthonormalized F-D states, whose explicit numeric representation can be inferred from CI results. In this manuscript, only the values of Hubbard parameters are deduced as they give meaningful physical descriptions of the system.
	
	For simpler representations, we denote the four-electron Slater determinants as
	\begin{equation}\label{eq:slaterDet}
		\begin{split}
			\left|S(\uparrow_{j} \downarrow_{j})\right\rangle&= \left|\uparrow_{\Psi_j} \downarrow_{\Psi_j} \right\rangle \left|\uparrow_{\Psi_{\text{R}1}} \downarrow_{\Psi_{\text{R}1}}\right\rangle
			,\\
			\left|S(\uparrow_{j} \downarrow_{k})\right\rangle&= \left(\left|\uparrow_{\Psi_j} \downarrow_{\Psi_k} \right\rangle + \left|\uparrow_{\Psi_k} \downarrow_{\Psi_j} \right\rangle\right)\left|\uparrow_{\Psi_{\text{R}1}} \downarrow_{\Psi_{\text{R}1}}\right\rangle
			,\\
			\left|T(\uparrow_{j} \downarrow_{k})\right\rangle&= \left(\left|\uparrow_{\Psi_j} \downarrow_{\Psi_k} \right\rangle - \left|\uparrow_{\Psi_k} \downarrow_{\Psi_j} \right\rangle\right)\left|\uparrow_{\Psi_{\text{R}1}} \downarrow_{\Psi_{\text{R}1}}\right\rangle,
		\end{split}
	\end{equation}
	where $k\neq j$ and the normalization coefficient is dropped for simplicity. In Eq.~\eqref{eq:slaterDet}, $\left|\uparrow_{\Psi_j}\downarrow_{\Psi_k}\right\rangle\left|\uparrow_{\Psi_{\text{R}1}}\downarrow_{\Psi_{\text{R}1}}\right\rangle$ denotes a four electron Slater determinant $\left|\uparrow_{\Psi_j}\downarrow_{\Psi_k}\uparrow_{\Psi_{\text{R}1}}\downarrow_{\Psi_{\text{R}1}}\right\rangle$. For example, the four-electron state in Fig.~\ref{fig:potential} can be understood as $|T(\uparrow_{\text{R}2} \downarrow_{\text{R}3}) \rangle$. Also, we denote the dot occupation as $(n_\text{L},n_\text{R})$, where $n_\text{L}$ and $n_\text{R}$ are the number of electrons occupying dot L and R respectively.

	\section{Results}
	\subsection{Energy spectrum and Hubbard model}\label{subsec:EvalHubbard}
	Written in the bases: $\{|S\left(\uparrow_{\text{L}1}\downarrow_{\text{R}2}\right)\rangle$, $|S\left(\uparrow_{\text{L}1}\downarrow_{\text{R}3}\right)\rangle$, $|T\left(\uparrow_{\text{L}1}\downarrow_{\text{R}2}\right)\rangle$, $|T\left(\uparrow_{\text{L}1}\downarrow_{\text{R}3}\right)\rangle$, $|S\left(\uparrow_{\text{R}2}\downarrow_{\text{R}2}\right)\rangle$, $|T\left(\uparrow_{\text{R}2}\downarrow_{\text{R}3}\right)\rangle$, $|S\left(\uparrow_{\text{R}2}\downarrow_{\text{R}3}\right)\rangle$, $|S\left(\uparrow_{\text{R}3}\downarrow_{\text{R}3}\right)\rangle$, $|T(\uparrow_{\text{R}3}\downarrow_{\text{R}_4})\rangle\}$, we have the effective Hubbard Hamiltonian: 
	\begin{widetext}
		\begin{equation}\label{eq:Ham}
			\begin{split}
				H = &\left(
				\begin{array}{ccccccccc} 
					U_{ { \text{L} }1,{ \text{R} }2 }+2\Delta  & 0 & 0 & 0 & \sqrt { 2 } t_{ { \text{L} }1,{ \text{R} }2 } \\
					0 & U_{ { \text{L} }1,{ \text{R} }3 }+\Delta E+2\Delta  & 0 & 0 & 0 \\
					0 & 0 & J^{ (1,3) }+U_{ { \text{L} }1,{ \text{R} }2 }+2\Delta  & 0 & 0 \\ 
					0 & 0 & 0 & J^{ (1,3^{ * }) }+U_{ { \text{L} }1,{ \text{R} }3 }+2\Delta +\Delta E & 0 \\
					\sqrt { 2 } t_{ { \text{L} }1,{ \text{R} }2 } & 0 & 0 & 0 & U_{ { \text{R} }2 } \\ 
					0 & 0 & -t_{ { \text{L} }1,{ \text{R} }3 } & t_{ { \text{L} }1,{ \text{R} }2 } & 0 \\ 
					t_{ { \text{L} }1,{ \text{R} }3 } & t_{ { \text{L} }1,{ \text{R} }2 } & 0 & 0 & 0 \\ 
					t'_{\text{\text{L}}1,\text{\text{R}}3} & \sqrt { 2 } t_{ { \text{L} }1,{ \text{R} }3 } & 0 & 0 & 0 \\ 
					0 & 0 & 0 & -t_{ { \text{L} }1,{ \text{R} }4 } & 0 
				\end{array}
				\right.
				\\
				&\quad\left.
				\begin{array}{cccc}
					0 & t_{ { \text{L} }1,{ \text{R} }3 } & t'_{\text{L}1,\text{\text{R}}3} & 0 \\
					0 & t_{ { \text{L} }1,{ \text{R} }2 } & \sqrt { 2 } t_{ { \text{L} }1,{ \text{R} }3 } & 0 \\ 
					-t_{ { \text{L} }1,{ \text{R} }3 } & 0 & 0 & 0 \\ 
					t_{ { \text{L} }1,{ \text{R} }2 } & 0 & 0 & -t_{ { \text{L} }1,{ \text{R} }4 } \\ 
					0 & 0 & 0 & 0 \\ 
					U_{ { \text{R} }2,{ \text{R} }3 }+\Delta E-J^{ F }_{ { \text{R} }2,{ \text{R} }3 } & 0 & 0 & 0 \\ 
					0 & U_{ { \text{R} }2,{ \text{R} }3 }+\Delta E+J^{ F }_{ { R2 },{ R3 } } & 0 & 0 \\ 
					0 & 0 & U_{ { \text{R} }3 }+2\Delta E & 0 \\ 
					0 & 0 & 0 & U_{ { \text{R} }3,{ \text{R} }4 }+\Delta E+\Delta E'-J_{\text{\text{R}}3,\text{\text{R}}4}^F
				\end{array}
				\right),
			\end{split}
		\end{equation}
	\end{widetext}
	
	\begin{figure}[t]
		\includegraphics[width=0.95\columnwidth]{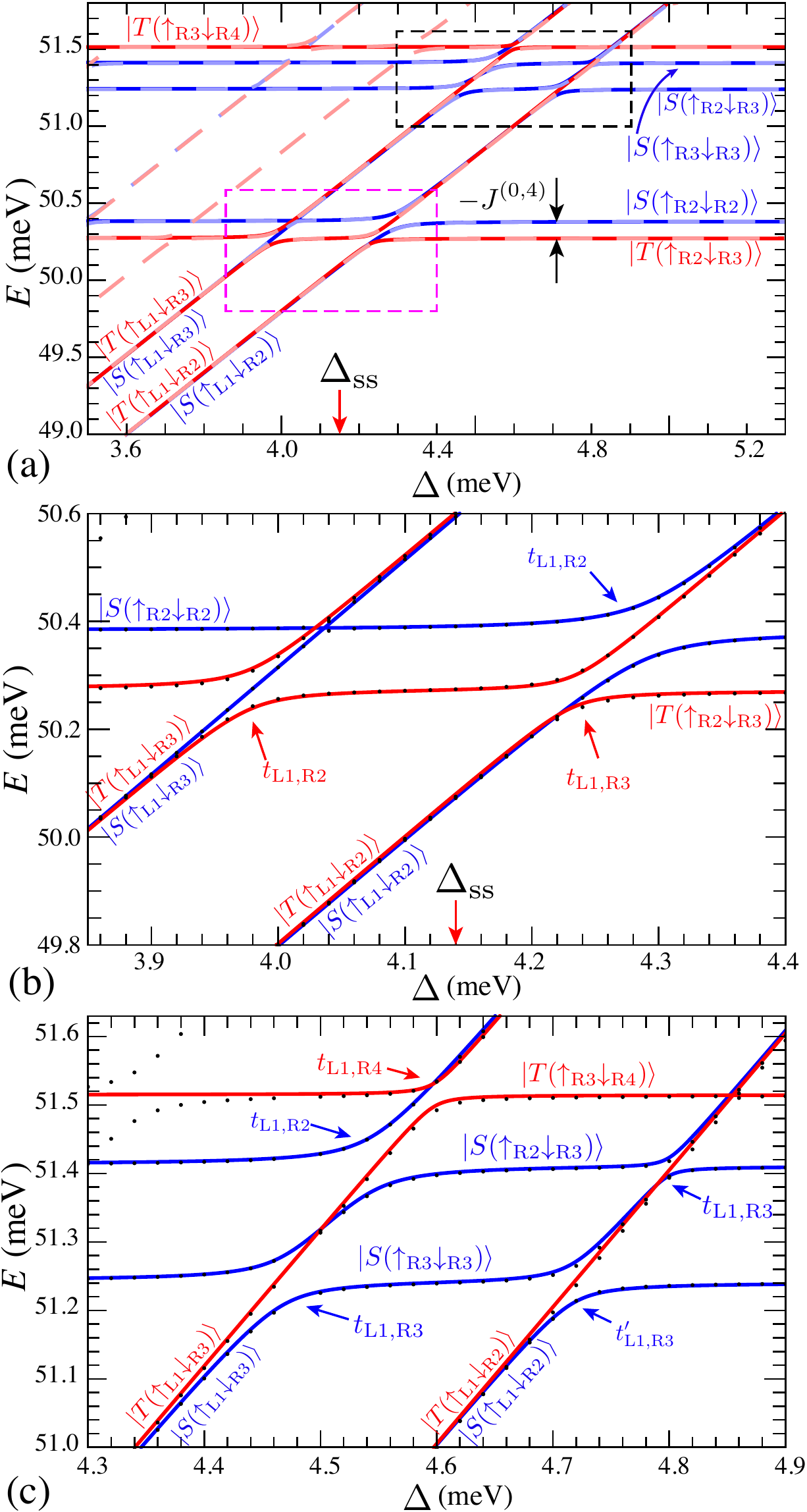}		
		\caption{(a) Lowest eigenvalues of a DQD occupied by four electrons as function of detuning, $\Delta$, at $B = 0.29$ T. The results from CI calculations, interpolated for the detuning range shown in the figure, are plotted as dashed lighter lines while the results from extended Hubbard model are plotted as solid colored lines. The exchange energy in $(n_\text{L},n_\text{R})=(0,4)$ region is labeled as $J^{(0,4)}$. (b) Zoom in of (a) near the anticrossing between $|S(\uparrow_{\text{L}1}\downarrow_{\text{R}2})\rangle$ and $|S(\uparrow_{\text{R}2}\downarrow_{\text{R}2})\rangle$. (c) Zoom in of (a) near the anticrossing between $|S(\uparrow_{\text{L}1}\downarrow_{\text{R}3})\rangle$ and $|S(\uparrow_{\text{R}3}\downarrow_{\text{R}3})\rangle$. The data points obtained using CI calculation are plotted as black circles for $B = 0.29$ T. The sweet spot detuning is marked with the label $\Delta_\text{ss}$.}
		\label{fig:lowestEigenvaluesBfo0p625}
	\end{figure}

	\noindent where $\Delta E = \varepsilon_{\text{R}3} - \varepsilon_{\text{R}2}$, $\Delta E' = \varepsilon_{\text{R}4} - \varepsilon_{\text{R}2}$, $2 \Delta = \varepsilon_{\text{L}1} - \varepsilon_{\text{R}1} $. $t'_{\text{L}1,\text{R}3}$ is the energy of co-tunneling process between $|S\left(\uparrow_{\text{L}1}\downarrow_{\text{R}2}\right)\rangle$ and $|S\left(\uparrow_{\text{R}3}\downarrow_{\text{R}3}\right)\rangle$, which includes the electron tunneling depicted by $\text{L}1\rightarrow\text{R}3$ and $\text{R}2\rightarrow\text{L}1\rightarrow\text{R}3$. $J^{(1,3)}$ and $J^{(1,3^*)}$ are the exchange energies in $(n_\text{L},n_\text{R})=(1,3)$ region with the valence electrons in dot R occupying the lowest and second lowest valence orbital respectively. Note that the diagonal terms in Eq.~\eqref{eq:Ham} are obtained by considering only the valence orbitals in the dot R, i.e.~$\{j,k\} \in \{R2, R3, R4, \cdots\}$. The effect of electrons occupying the core orbitals has been encoded in the Hubbard parameters in Eq.~\eqref{eq:Ham}, which is confirmed by the correspondence between full CI results and the extended Hubbard model, as shown in the following paragraph.
	
	\begin{figure}[b]
		\includegraphics[width=0.95\columnwidth]{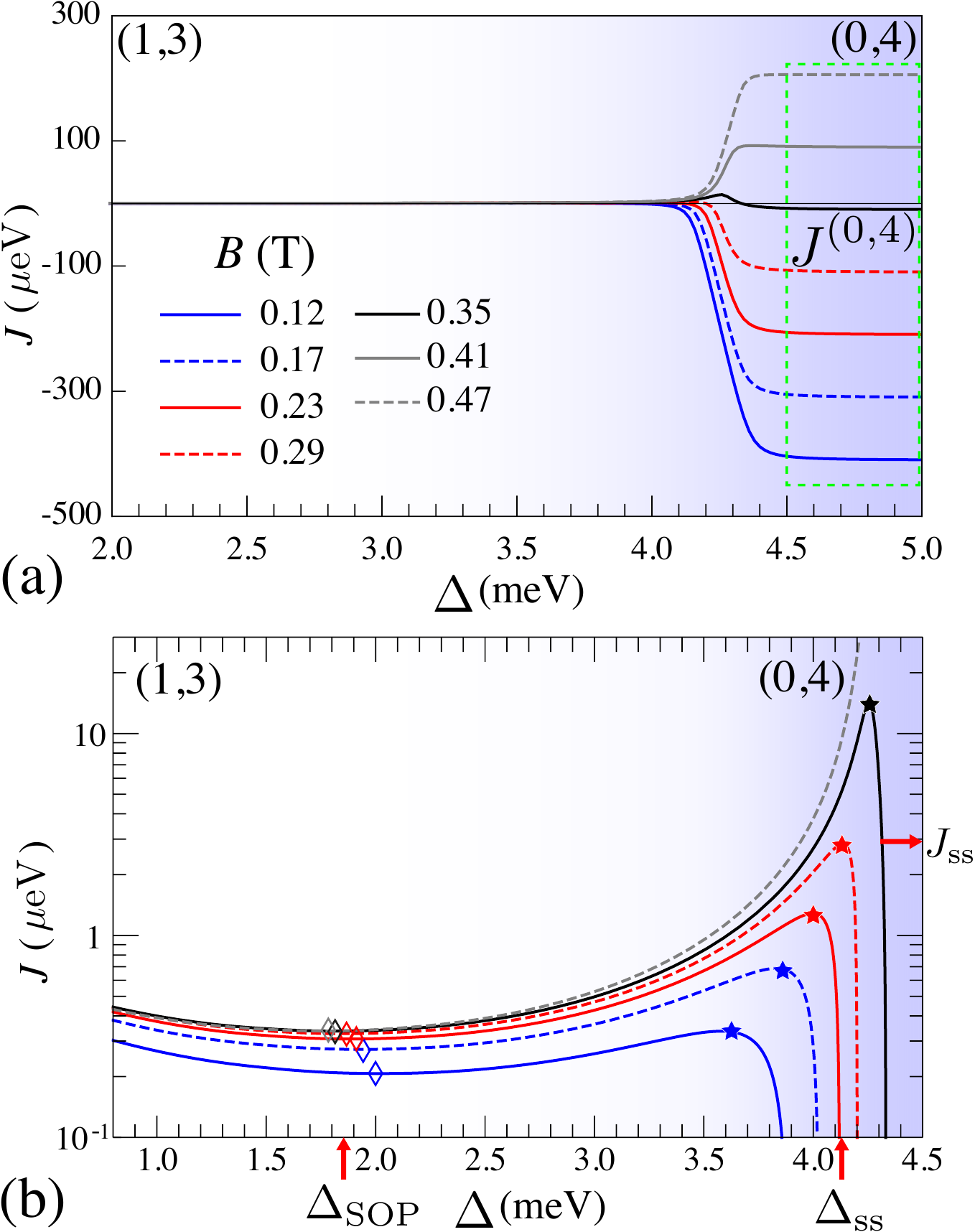}
		\caption{(a) Exchange energy, $J$, as function of detuning, $\Delta$, for different magnetic fields, $B$. (b) A zoom-in of (a) near the sweet spots with y-axis in log scale. The detuning values at which the sweet spots occur and the corresponding exchange energy are denoted as $\Delta_\text{ss}$ and $J_\text{ss}$. Their locations in the exchange energy curves are labeled as filled star symbols. \red{The symmetric operating points (SOP), $\Delta_\text{SOP}$, are labeled as empty diamond symbols} As an example, $\Delta_\text{ss}$, $J_\text{ss}$ and $\Delta_\text{SOP}$ for $B = 0.29$ T are explicitly marked. The labels $(n_\text{L},n_\text{R})$ at top left and right corner indicate the electron occupation of the DQD device. The dashed green box encloses the detuning range which gives the exchange energy in $(n_\text{L},n_\text{R})=(0,4)$ region, denoted as $J^{(0,4)}$.}
		\label{fig:JsmallDelta}
	\end{figure}

	Figure~\ref{fig:lowestEigenvaluesBfo0p625} shows an example of the eigenvalues of a four-electron system at $B=0.29$ T. In Fig.~\ref{fig:lowestEigenvaluesBfo0p625}(a), the solid blue (red) lines show the energies of singlet (triplet) states calculated using Hubbard model, while the lighter dashed lines show the corresponding energies obtained using CI calculations. Fitting to the CI results shows that the tunneling parameters in the extended Hubbard model are $t_{\text{L}1,\text{R}2}=36\mu$eV, $t_{\text{L}1,\text{R}3}=27\mu$eV, $t_{\text{L}1,\text{R}4}=19\mu$eV, $t'_{\text{L}1,\text{R}3}=12\mu$eV. CI results suggest that the tunneling values $t_{j,k}$ are almost constant with respect to the magnetic field strength. The correspondence between the extended Hubbard model and CI results for other magnetic field strengths are shown in Appendix~\ref{sec:CIHub}. It can be observed that the extended Hubbard model reproduces the CI results with high accuracy. \red{The following numerical results are obtained using CI calculations with 16 orbitals (6 and 10 orbitals in the left and right dot respectively, see Sec.~\ref{subsec:CI}) while the analytical results are obtained by fitting the Hubbard parameters in the effective Hamiltonian, Eq.~\eqref{eq:Ham}, based on the correspondence between the extended Hubbard model and CI results}
	
	\subsection{Exchange energy under different magnetic fields}\label{subsec:Ex}
	
	\begin{figure}[b]
		\includegraphics[width=0.95\columnwidth]{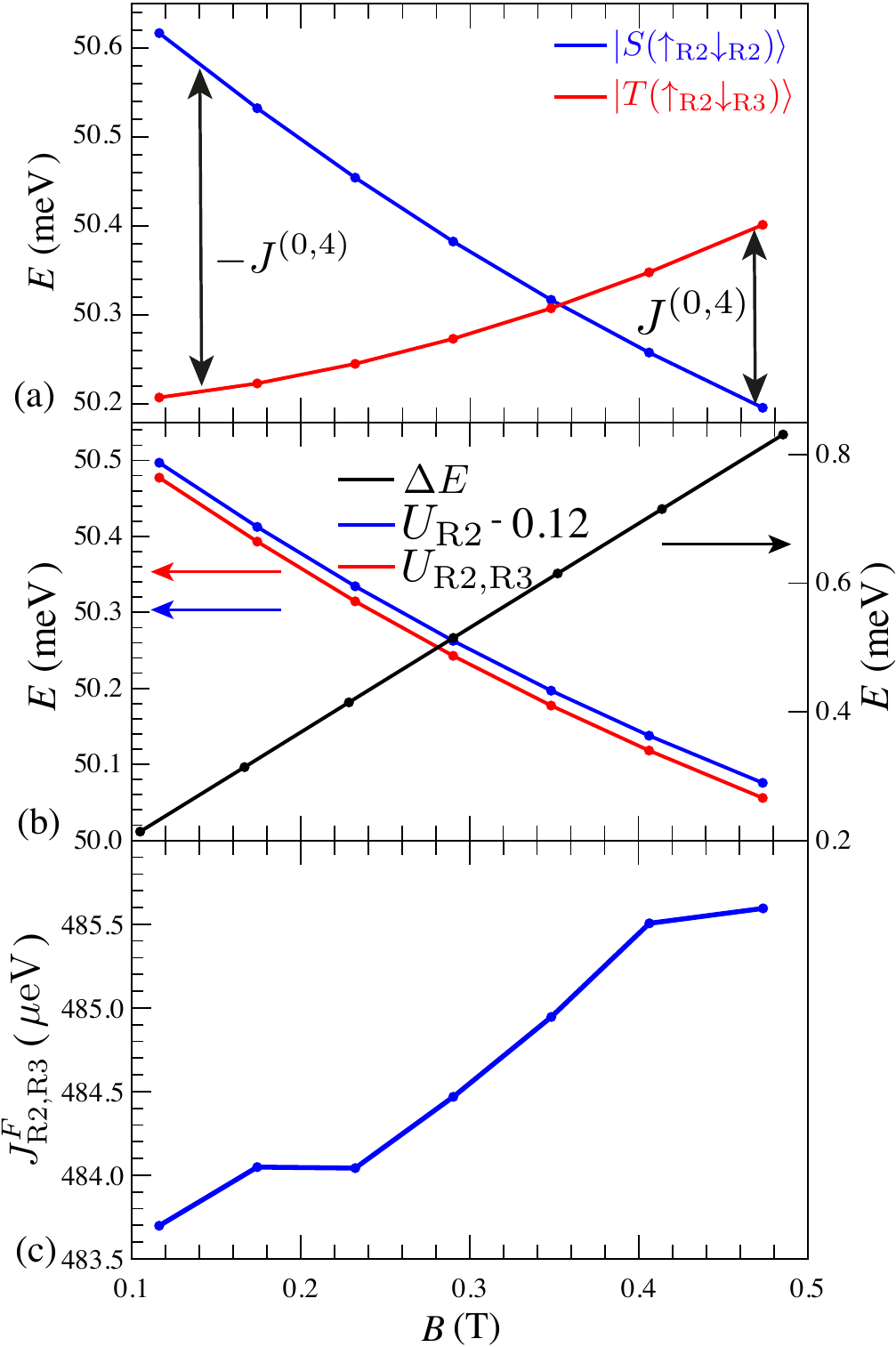}	
		\caption{(a) Eigenvalues of lowest singlet and triplet state in $(n_\text{L},n_\text{R})=(0,4)$ region. (b) Values of parameters in Hubbard model, including the magnetic field induced orbital splitting $\Delta E$ and Coulomb energies, $U_{\text{R}2}$ and $U_{\text{R}2,\text{R}3}$. (c) Ferromagnetic exchange term, $J_{\text{R}2,\text{R}3}^F$, in right dot. The data points are plotted as function of magnetic field.}
		\label{fig:HubbardParaValues}
	\end{figure}

	Figure~\ref{fig:JsmallDelta} shows the exchange energies $J$ as a function of detuning $\Delta$ under several magnetic fields. At weak magnetic fields ($B\lesssim 0.35$ T), $J$ is negative in large $\Delta$ region where $(n_\text{L},n_\text{R})=(0,4)$, i.e.~$J^{(0,4)}<0$. On the other hand, at larger magnetic field ($B\gtrsim 0.41$ T), $J^{(0,4)}>0$. The sign switching of $J^{(0,4)}$ can be understood from the fitted extended Hubbard model (Eq.~\eqref{eq:HubbardModel2ndQuan}). The extracted Hubbard parameters, cf.~Eq.~\eqref{eq:HubbardPara} and Eq.~\eqref{eq:exchangeTerm}, are shown in Fig.~\ref{fig:HubbardParaValues}. It can be observed that the eigenvalues of the lowest singlet and triplet crosses at $B\sim 0.35$ T, resulting in $J^{(0,4)}<0$ for $B< 0.35$ T while $J^{(0,4)}>0$ at $B> 0.35$ T. Fig.~\ref{fig:HubbardParaValues}(b) and Fig.~\ref{fig:HubbardParaValues}(c) show that the sign switching of $J^{(0,4)}$ can be directly attributed to the increase of orbital splitting $\Delta E$ for an increasing magnetic field strength. The increase of $\Delta E$ lifts the energy of the lowest triplet state, $|T(\uparrow_{\text{R}2}\downarrow_{\text{R}3})\rangle$, in relative to the lowest singlet state, $|S(\uparrow_{\text{R}2}\downarrow_{\text{R}2})\rangle$. Although it is observed that the onsite Coulomb energies decreases while the magnetic field increases, the onsite Coulomb energies for $|S(\uparrow_{\text{R}2}\downarrow_{\text{R}2})\rangle$, $U_{\text{R}2}$, and $|T(\uparrow_{\text{R}2}\downarrow_{\text{R}3})\rangle$, $U_{\text{R}2,\text{R}3}$, experience the same degree of reduction, leading to the fact that only $\Delta E$ dominates the behavior of $J^{(0,4)}$ when the magnetic field changes. It is also observed that the onsite ferromagnetic exchange term, $J^F_{\text{R}2,\text{R}3}$, is almost constant for different magnetic fields, as shown in Fig.~\ref{fig:HubbardParaValues}(c), hence it does not contribute to the sign switching of $J^{(0,4)}$ with respect to the magnetic field strength. An in-depth inspection into the electron densities shows that the electron localization effect in terms of Wigner-molecule physics does not play a role in the sign switching of $J^{(0,4)}$ as the variation of the effective confinement length is negligible for the range of magnetic field of interest, rendering only the orbital splitting, $\Delta E$, to be the main factor (see Appendix~\ref{sec:elecDen} for details).
	
	For $B\lesssim 0.35$ T, the negative $J^{(0,4)}$ gives rise to the sweet spots of $J$. Fig.~\ref{fig:JsmallDelta}(b) shows that sweet spots, indicated as colored star symbols, occur at larger detuning values when the magnetic field increases, with the corresponding exchange energies, $J_\text{ss}$, increasing as well. This behavior conforms well qualitatively with experimental results \cite{Malinowski.17}. However, the microscopic mechanism leading to this behavior has not been explained in the literature. In this paper, we show that, with the help of CI results and the fitted extended Hubbard model, we are able to comprehend this phenomenon, as shown in the following section, Sec.~\ref{subsec:ssAna}.
	
	\subsection{Tuning of sweet spot detuning, $\Delta_{\text{ss}}$, and exchange energy, $J_{\text{ss}}$, by magnetic field}\label{subsec:ssAna}
	\begin{figure}[t]
		\includegraphics[width=0.98\columnwidth]{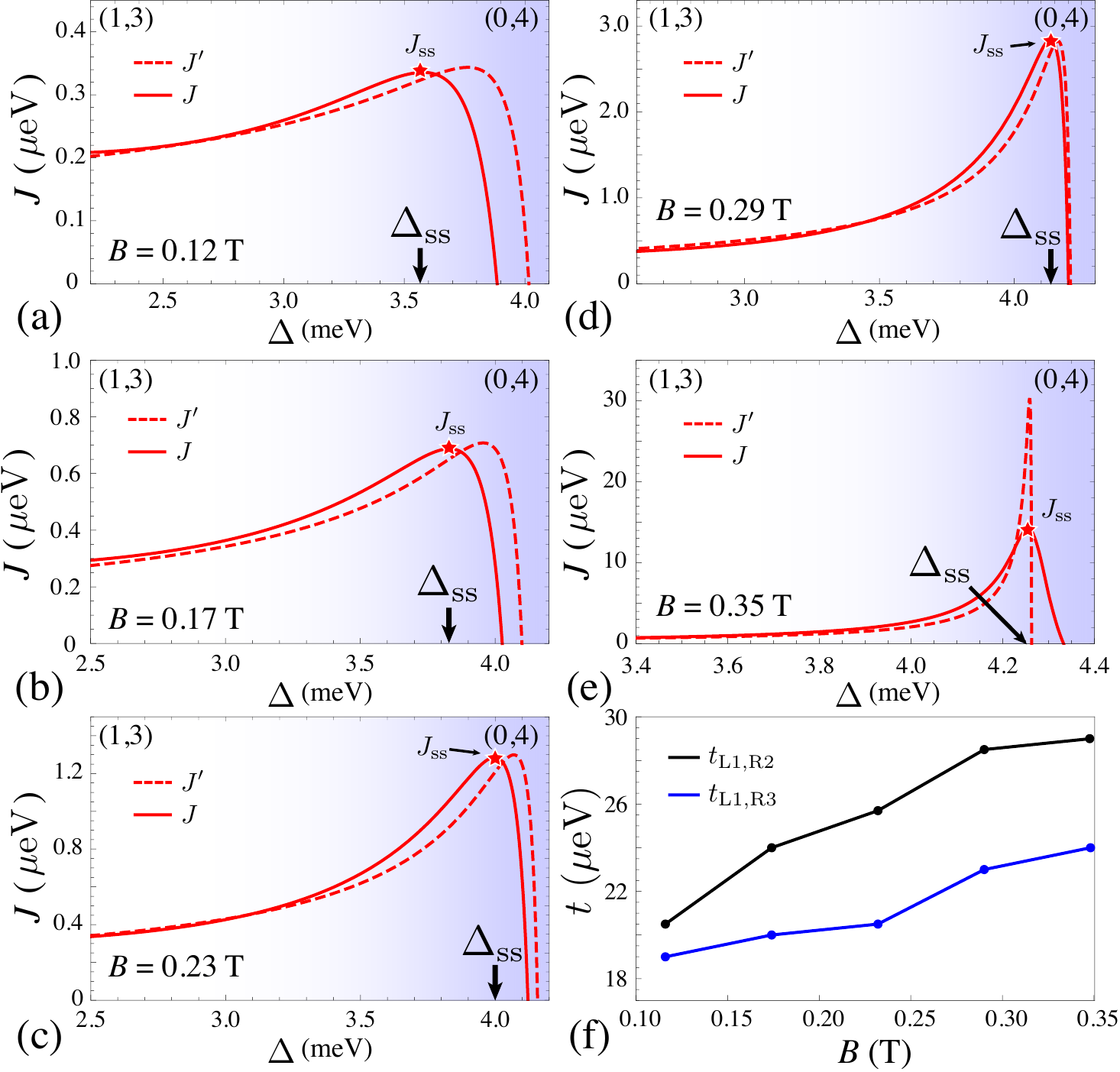}
		\caption{Exchange energy near the sweet spots for $B = $ (a) 0.12 T (b) 0.17 T (c) 0.23 T (d) 0.29 T (e) 0.35 T. The labels $(n_\text{L},n_\text{R})$ at top left and right corner indicate the electron occupation of the DQD device. (f) The presumptive tunneling energies to generate the dashed colored lines in (a)-(e). The sweet spot detunings are labeled as $\Delta_\text{ss}$. The solid color lines are the results obtained from CI calculations, denoted as $J$, while the dashed color lines show the estimation using perturbation theory (Eq.~\eqref{eq:STEvalSW}, denoted as $J'$). (f) The tunneling values employed to calculate $J'$ in (a)-(e).}
		\label{fig:SSDeltaPosition}
	\end{figure}

	Figure \ref{fig:JsmallDelta} shows that at weak magnetic fields, $B\lesssim 0.35$ T, $J>0$ at small detuning (in the $(n_\text{L},n_\text{R})=(1,3)$ region) while $J<0$ at large detuning (in the (0,4) region). The transition of $J$ between those two regions yields a non-monotonic curve, exhibiting a sweet spot at $\Delta_{\text{ss}}$, cf.~Fig.~\ref{fig:JsmallDelta}(b). This fact has appeared in several works in the literature, including experimental results \cite{Martins.17, Malinowski.18}, the extended Hubbard model with presumptive choice for the values of Hubbard parameters \cite{Deng.18} and CI results in this work (cf.~Fig.~\ref{fig:JsmallDelta}). However, its origin is not yet well understood. Here, we present an analytical result based on the perturbation theory to estimate the non-monotonic nature of the transition.
	
	\subsubsection{$J$ at small $\Delta$ at weak magnetic fields}\label{subsubsec:JsmallDeltaAna}
	For $B\lesssim 0.29$ T, $\Delta_\text{ss}$ is well below the first anticrossing between the eigenstates of $(1,3)$ and $(0,4)$ type, i.e.~the $\Delta$ at which the energies of $|T(\uparrow_{\text{L}1}\downarrow_{\text{R}2})\rangle$ and $|T(\uparrow_{\text{R}2}\downarrow_{\text{R}3})\rangle$ are equal, cf.~the anticrossing marked by red arrow and $t_{\text{L}1,\text{R}3}$ in Fig.~\ref{fig:lowestEigenvaluesBfo0p625}(b). For example, at $B=0.29$ T, $\Delta_\text{ss}=4.14$ meV, while the eigenvalue of $|T(\uparrow_{\text{L}1}\downarrow_{\text{R}2})\rangle$ equates with $|T(\uparrow_{\text{R}2}\downarrow_{\text{R}3})\rangle$ at $\Delta=4.24$ meV, see Fig.~\ref{fig:lowestEigenvaluesBfo0p625}(b). $\Delta_\text{ss}$ for other magnetic field strengths can be referred to Fig.~\ref{fig:CIHubOtherB}. The substantial separation between $\Delta_\text{ss}$ and the first anticrossing gives small value of $t_\text{L1,R2}/\left(U_\text{R2}-U_\text{L1,R2}-2\Delta\right)$ and $t_\text{L1,R3}/\left(U_\text{R2,R3}+\Delta E - J^F_\text{R2,R3}-U_\text{L1,R2}-2\Delta \right)$
, allowing us to use the perturbation theory on a truncated effective Hubbard Hamiltonian (Eq.~\eqref{eq:Ham}) to estimate $J$ at small $\Delta$ and the values of $\Delta_\text{ss}$. The truncated Hamiltonian is written in the bases of $|S\left(\uparrow_{\text{L}1}\downarrow_{\text{R}2}\right)\rangle$, $|T\left(\uparrow_{\text{L}1}\downarrow_{\text{R}2}\right)\rangle$, $|S\left(\uparrow_{\text{R}2}\downarrow_{\text{R}2}\right)\rangle$ and $|T\left(\uparrow_{\text{R}2}\downarrow_{\text{R}3}\right)\rangle$, which we will relabel them in a much simpler form, i.e.~$|S\left(1,3\right)\rangle$, $|T\left(1,3\right)\rangle$, $|S\left(0,4\right)\rangle$ and $|T\left(0,4\right)\rangle$, respectively. Physically, $|S\left(n_\text{L},n_\text{R}\right)\rangle$ and $|T\left(n_\text{L},n_\text{R}\right)\rangle$ refers to a singlet and triplet, respectively, formed by $n_\text{L}$ electrons in dot L and $n_\text{R}$ electrons in dot R.

	In small $\Delta$ region, the higher lying states, i.e.~$|S(0,4)\rangle$ and $|T(0,4)\rangle$, shift $|S(1,3)\rangle$ and $T(1,3)\rangle$ down in energy. We first make a global energy shift on the truncated effective Hubbard Hamiltonian such that the energy of the non-hybridized $|S(1,3)\rangle$ state is 0. We then proceed to consider the hybridization of $|S(1,3)\rangle$ and $|T(1,3)\rangle$ with $|S(0,4)\rangle$ and $|T(0,4)\rangle$ respectively. Their eigenvalues, denoted as $E'_{|S(1,3)\rangle}$ and $E'_{|T(1,3)\rangle}$, are
	\begin{equation}
	\begin{split}\label{eq:STEvalSW}
		E'_{|S(1,3)\rangle}&\approx - \frac{2 t_{\text{L}1,\text{R}2}^2}{U_{\text{R}2}-U_{\text{L}1,\text{R}2}-2\Delta} = -\frac{2 t_{\text{L}1,\text{R}2}^2}{\delta S-2\Delta}+\mathcal{O}[t_{j,k}^3],\\
		E'_{|T(1,3)\rangle}&\approx J^{(1,3)}- \frac{t_{\text{L}1,\text{R}3}^2}{U_{\text{R}2,\text{R}3}+\Delta E-J^F_{\text{R}2,\text{R}3}-U_{\text{L}1,\text{R}2}-2\Delta}\\
		&\quad+\mathcal{O}[t_{j,k}^3]\\
		&=J^{(1,3)}- \frac{t_{\text{L}1,\text{R}3}^2}{\delta T-2\Delta}+\mathcal{O}[t_{j,k}^3],
		\\
		J'&=E'_{|T(1,3)\rangle}-E'_{|S(1,3)\rangle},
	\end{split}
	\end{equation}
	where we have neglected $J^{(1,3)}$ in the denominator of $E'_{|T(1,3)\rangle}$ because it is negligible compared to other parameters. $E'_{|S(1,3)\rangle}$ and $E'_{|T(1,3)\rangle}$ are obtained using the time independent perturbation theory up to the second order of tunneling values $t_{jk}$. Also, we have replaced the terms that are not related to $\Delta$ in the denominators of Eq.~\eqref{eq:STEvalSW} as $\delta S$ and $\delta T$. $\delta S$ ($\delta T$) signifies the energy difference, in the absence of detuning, between $|S(1,3)\rangle$ and $|S(0,4)\rangle$ ($|T(1,3)\rangle$ and $|T(0,4)\rangle$).
	
	In the limit that $|\Delta|\rightarrow0$, we can perform Taylor expansion on Eq.~\eqref{eq:STEvalSW} with respect to $\Delta$ and obtain
	\begin{equation}\label{eq:STEvalSWSmallDelta}
	\begin{split}
		J'&\approx J^{(1,1)}+\left(\frac{2 t_{\text{L}1,\text{R}2}^2}{\delta S}-\frac{t_{\text{L}1,\text{R}3}^2}{\delta T}\right)\\
		&\quad+\left(\frac{2 t_{\text{L}1,\text{R}2}^2}{\delta S^2}-\frac{t_{\text{L}1,\text{R}3}^2}{\delta T^2}\right)2\Delta+\mathcal{O}[\Delta^2],
	\end{split}
	\end{equation}
	Although $\delta S > \delta T$, which gives $J^{(0,4)}<0$ at weak magnetic fields, the overall ratio yields the relation: $\sqrt{2}t_{\text{L}1,\text{R}2}/\delta S > t_{\text{L}1,\text{R}3}/\delta T$, 
	as confirmed by the fitted Hubbard parameters, cf.~Sec.~\ref{subsec:EvalHubbard}. This gives a positive coefficient for $2\Delta$ in Eq.~\eqref{eq:STEvalSWSmallDelta}, such that~$\partial J'/\partial \Delta>0$, resulting in the increase of $J$ as a function of $\Delta$ in the small $\Delta$ regime. Fig.~\ref{fig:SSDeltaPosition} shows the results calculated using Eq.~\eqref{eq:STEvalSW}. It can be observed that for $\Delta\ll\Delta_\text{ss}$, $J'$ (dashed red lines) evaluated using Eq.~\eqref{eq:STEvalSW} agrees well with CI results (solid red lines with the legend $J$). Note that the tunneling values used are slightly different than those presented in Fig.~\ref{fig:lowestEigenvaluesBfo0p625} as the energy values are too large in comparison to the exchange energies in the region shown in Fig.~\ref{fig:SSDeltaPosition}. However, the relation of $t_{\text{L}1,\text{R}2} > t_{\text{L}1,\text{R}3}$ remains.
	
	The analysis in this subsection shows that although $J^{(0,4)}<0$, in small $\Delta$ region, $J$ is positive and increases as a function of $\Delta$ due to the larger inter-dot tunneling for singlets as compared to triplets, i.e.~$t_{\text{L}1,\text{R}2} > t_{\text{L}1,\text{R}3}$.
	
	\subsubsection{Detuning value of sweet spot, $\Delta_\text{ss}$}\label{subsec:Deltass}
	\begin{figure}[t]
		\includegraphics[width=0.98\columnwidth]{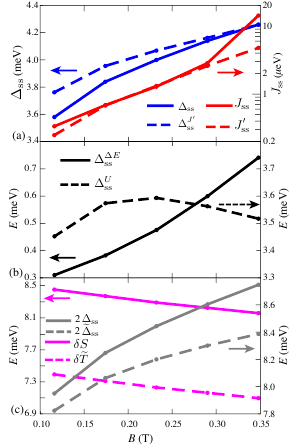}
		\caption{(a) Sweet spot, $\Delta_\text{ss}$, and the corresponding exchange energy, $J_\text{ss}$, as function of the magnetic field. Solid blue and red line, labeled as $\Delta_{\text{ss}}$ and $J_\text{ss}$ respectively, shows the results extracted from CI calculations while dashed blue and red line, labeled as $\Delta_\text{ss}^{J'}$ and $J'_\text{ss}$ respectively, shows the estimation based on perturbation theory. $J'_\text{ss}$ is evaluated using Eq.~\eqref{eq:exchangeSSTaylor} up to $k=40$. (b) $\Delta_{\text{ss}}^{\Delta E}$ (solid black line) and $\Delta_{\text{ss}}^U$ (dashed black line), which is defined in Eq.~\eqref{eq:Deltass}, as function of the magnetic field. (c) Sweet spots,  $\Delta_{\text{ss}}$ and $\widetilde{\Delta}_{\text{ss}}$, and energy differences between two lowest singlets and triplets, denoted as $\delta S$ and $\delta \widetilde{T}$ respectively, as function of the magnetic field, cf. Eq.~\eqref{eq:exchangeSSTaylor}.}
		\label{fig:DeltaSSCIandAna}
	\end{figure}

	In the previous section (Sec.~\ref{subsubsec:JsmallDeltaAna}), we have shown that $J>0$ and $\partial J/\partial \Delta>0$ in the small $\Delta$ region. For $B\lesssim 0.35$ T, there must exist a detuning value where $\partial J/\partial \Delta = 0$ to enable the transition into $J^{(0,4)}<0$ in the large $\Delta$ region, giving rise to a sweet spot. The value of $\Delta_\text{ss}$ can be estimated by equating the derivatives of $E'_{|S(1,1)\rangle}$ and $E'_{|T(1,1)\rangle}$ with respect to detuning, $\Delta$. Taking the minimum root, the approximated $\Delta_\text{ss}$, denoted as $\Delta_\text{ss}^{J'}$, is
	\begin{subequations}\label{eq:Deltass}
	\begin{align}
		\begin{split}
			\Delta_\text{ss}^{J'}&=\Delta_\text{ss}^{\Delta E}+\Delta'_\text{ss},
		\end{split}
		\\
		\begin{split}
			\Delta_\text{ss}^{\Delta E}&=\frac{\sqrt{2}}{2 \left(2 t_{\text{L}1,\text{R}2}^2-t_{\text{L}1,\text{R}3}^2\right)} \\
			&\times t_{\text{L}1,\text{R}2} \left(\sqrt{2} t_{\text{L}1,\text{R}2}+ t_{\text{L}1,\text{R}3}\right) \Delta E,
		\end{split}
		\\
		\begin{split}
			\Delta^U_\text{ss}&=\frac{1}{2 \left(2 t_{\text{L}1,\text{R}2}^2-t_{\text{L}1,\text{R}3}^2\right)}
			\\
			&\times\Big[-t_{\text{L}1,\text{R}3}^2\left(U_{\text{R}2}-U_{\text{L}1,\text{R}2}\right)
			\\
			&- \sqrt{2} t_{\text{L}1,\text{R}2} t_{\text{L}1,\text{R}3}\left|U_{\text{R}2}-U_{\text{R}2,\text{R}3}+J^F_{\text{R}2,\text{R}3}\right|\Big],
		\end{split}
	\end{align}
	\end{subequations}
	where $\Delta_\text{ss}^{\Delta E}$ is the part of $\Delta_\text{ss}^{J'}$ whose variable is $\Delta E$ while $\Delta^U_\text{ss}$ constitutes the remaining part of $\Delta_\text{ss}^{J'}$. 
	
	Figure~\ref{fig:DeltaSSCIandAna}(a) shows the sweet spot detunings as a function of the magnetic field, including both the exact values, $\Delta_\text{ss}$ (solid blue lines), and the approximated ones, $\Delta_\text{ss}^{J'}$ (dashed blue lines). $\Delta_\text{ss}$ is extracted from CI results (cf.~Fig.~\ref{fig:JsmallDelta}) while $\Delta_\text{ss}^{J'}$ is evaluated using Eq.~\eqref{eq:Deltass}.
	In Fig.~\ref{fig:DeltaSSCIandAna}(a), it is observed that the main qualitative result is correctly estimated by Eq.~\eqref{eq:Deltass} with slight deviation at weak magnetic fields. Fig.~\ref{fig:DeltaSSCIandAna}(b) shows that the variation of $\Delta^U_\text{ss}$ with respect to the magnetic field is limited as compared to $\Delta_\text{ss}^{\Delta E}$. According to Eq.~\eqref{eq:Deltass}, the positive prefactor of $\Delta E$, as $t_{j,k} t_{m,n} > 0 $ and $2t_\text{L1,R2}^2-t_\text{L1,R3}^2>0$, leads to the increase of $\Delta_{\text{ss}}^{J'}$ as a function of $\Delta E$. Hence, we can conclude in this subsection that $\Delta_\text{ss}\propto B$ as $\Delta_\text{ss}\propto \Delta E$ (cf.~Eq.~\eqref{eq:Deltass}) while $\Delta E \propto B$ (cf.~Fig.~\ref{fig:HubbardParaValues}(b)).
	
	\subsubsection{Exchange energy at sweet spot, $J_\text{ss}$}
	Exchange energies at sweet spots, denoted as $J_\text{ss}$, can be estimated by performing a Taylor expansion with respect to $2\Delta/\delta S$ and $2\widetilde{\Delta}/\delta \widetilde{T}$ in Eq.~\eqref{eq:STEvalSW}, giving
	\begin{equation}\label{eq:exchangeSSTaylor}
		\begin{split}
			J'_\text{ss} &\approx E'_{|T(1,1)\rangle}-E'_{|S(1,1)\rangle}\\
			&=\sum_{k=0}^{\infty}\left. \left[\frac{2t_{\text{L}1,\text{R}2}^2}{\delta S} \left(\frac{ 2\Delta}{\delta S}\right)^k-\frac{t_{\text{L}1,\text{R}3}^2}{\delta \widetilde{T}}\left(\frac{2\widetilde{\Delta}}{\delta \widetilde{T}}\right)^k\right] \right|_{\Delta=\Delta_{\text{ss}}},
		\end{split}
	\end{equation}
	where $\widetilde{\Delta} = \Delta - \Delta E/2$ and $\delta \widetilde{T} =\delta T - \Delta E$ are modified by the orbital splitting $\Delta E$, the detuning and the energy difference between two lowest triplet states, respectively. Since $2\Delta/\delta S$ and $2\widetilde{\Delta}/\delta \widetilde{T}$ at $\Delta = \Delta_\text{ss}$ are not small, higher-order terms have to be kept in the Taylor expansion. By keeping the Taylor expanded terms up to $k=40$, Fig.~\ref{fig:DeltaSSCIandAna}(a) shows that $J'_\text{ss}$ (dashed red lines) agrees well with the sweet spot exchange energies extracted from CI results, $J_\text{ss}$ (solid red lines). The deviation at $B=0.35$ T is due to the closeness between the sweet spot and the anticrossing between the lowest triplets, rendering the Taylor expansion inapplicable. Fig.~\ref{fig:DeltaSSCIandAna}(c) shows that the variations of $\delta S$ and $\delta \widetilde{T}$ are limited compared to $2\Delta_\text{ss}$ and $2 \widetilde{\Delta}_\text{ss}$, indicating that the increase of $J_\text{ss}$ for an increasing magnetic field can be mainly ascribed to the increase of $\Delta_{\text{ss}}$. As discussed in Sec.~\ref{subsec:Deltass}, $\Delta_\text{ss} \propto \Delta E$ implies that the increase of $J_\text{ss}$ can be directly related to the variation of orbital spitting induced by the magnetic field.
	
	\red{In addition, the values of $J_\text{ss}$ in this work exhibit comparable strengths as those demonstrated in experimental works \cite{Shulman.12,Reed.16,Martins.16,Nichol.17,Takeda.20,Cerfontaine.20}, suggesting the experimental feasibility for performing high-fidelity quantum gates in a four-electron DQD device with an asymmetric electron occupation.}
	
	\section{Conclusion and discussion}
	We have shown, using full CI calculations, that sweet spots can be realized in DQD devices with one electron occupying a small dot while three electrons occupying a larger multi-electron dot, similar to experimental results \cite{Martins.17,Malinowski.18} in which around 50 to 100 electrons occupy the multi-electron dot. Also, we show that sweet spots and the corresponding exchange energies are tunable by the perpendicular magnetic field. By combining CI results and the extended Hubbard model, we identify the main contributing factor of the tuning effect as the magnetic-field-induced orbital splitting, but not the effect of Wigner-molecule physics. 
	
	\red{We have shown that the physical mechanism leading to $\Delta_\text{ss}$ can be attributed to the competition between the hybridizations of the singlet and triplet states. This competition, together with the introduced dipole at $\Delta_\text{ss}$, gives rise to sweets spots of the effective single-qubit exchange energies and capacitive coupling for a pair of capacitively-coupled four-electron singlet-triplet qubits. The possibility for high-fidelity capacitive gates under a realistic noise environment can be attributed to the enhanced capacitive coupling at the effective single-qubit exchange energy sweet spots, which are found to be close to the capacitive coupling sweets spot \cite{ChanGX.22.1}.}
	
	Our results should facilitate the realization of high-fidelity singlet-triplet qubit hosted in few-electron systems.

		\section*{Acknowledgements} We acknowledge support from the Key-Area Research and Development Program of GuangDong Province  (Grant No.~2018B030326001), the National Natural Science Foundation of China (Grant No.~11874312), the Research Grants Council of Hong Kong (Grant No.~11303617), and the Guangdong Innovative and Entrepreneurial Research Team Program (Grant No.~2016ZT06D348). The calculations involved in this work are mostly performed on the Tianhe-2 supercomputer at the National Supercomputer Center in Guangzhou, China.

	\appendix
	
	\section{Convergence}\label{sec:converg}
	\begin{figure}[t]
		\includegraphics[width=\linewidth]{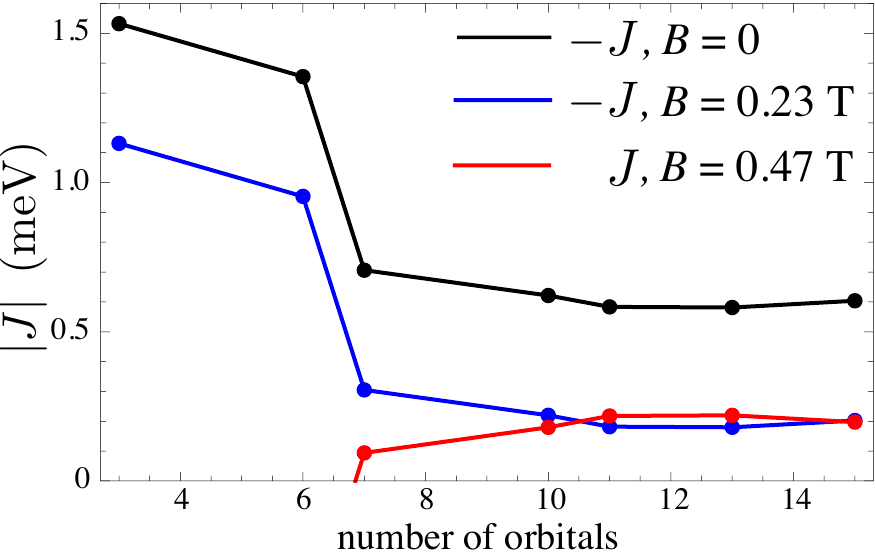}
		\caption{Exchange energy of four electrons occupying a quantum-dot as function of the number of lowest orbitals retained in the Full CI calculation with quantum-dot confinement strength $\hbar \omega_0 = 4$meV.}
		\label{fig:exchangeVSOrbital}
	\end{figure}

	We obtain the eigenvalues of a quantum-dot occupied by four electrons by constructing the Hamiltonian written in all the possible Slater determinants for a given number of orbitals. Fig.~\ref{fig:exchangeVSOrbital} shows the  exchange energies as a function of the number of orbitals retained in the full CI calculations. It is observed that the exchange energy converges when the number of orbitals retained in the calculation $\geq$ 10. Hence, all the results in the main text are obtained by retaining 10 orbitals for the QD hosting more than one electron, i.e.~the right larger QD. We judiciously choose to retain 6 orbitals for the left smaller QD as it is only occupied by one electron in the detuning range giving $(n_\text{L},n_\text{R})=(1,3)$.
	
	\begin{figure}[t]
		\includegraphics[width=0.9\linewidth]{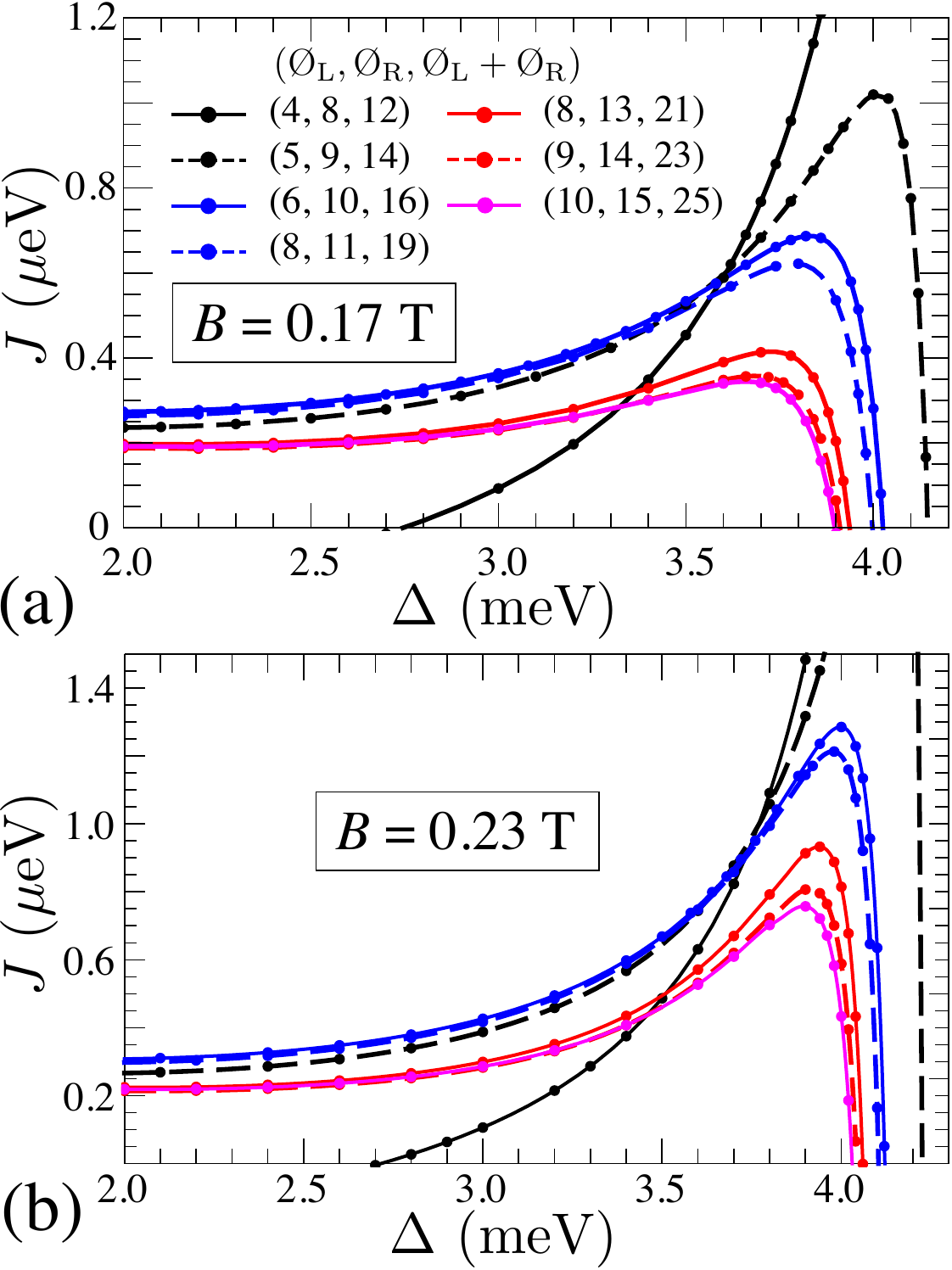}
		\caption{\red{Exchange energy, $J$, as function of $\Delta$ for different numbers of orbitals retained in the CI calculations at (a) $B=$ 0.17 T and (b) $B =$ 0.23 T. The number of orbitals for the left and right dots are denoted as \O${}_\text{L}$ and \O${}_\text{R}$ respectively.}}
		\label{fig:JVSOrbital}
	\end{figure}

	\begin{figure}[t]
		\includegraphics[width=0.9\linewidth]{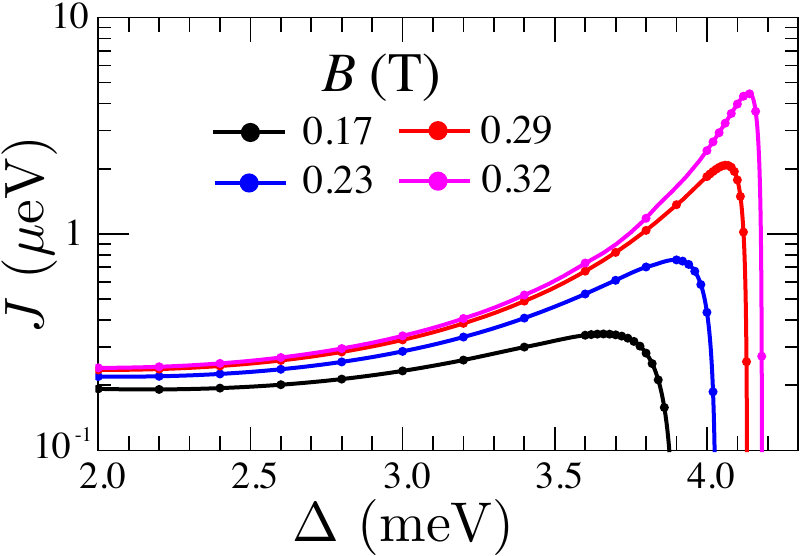}
		\caption{\red{Exchange energy, $J$, as function of $\Delta$ for different magnetic fields $B$. $J$ is obtained for (\O${}_\text{L},{}$\O${}_\text{R})=(10,15)$.}}
		\label{fig:JVSOrbital2}
	\end{figure}

	\red{Figure \ref{fig:JVSOrbital} shows exchange energy $J$ evaluated for different numbers of orbitals retained in the full CI calculations, where the number of orbitals in the left and right dots are denoted as \O${}_\text{L}$ and \O${}_\text{R}$ respectively. Fig.~\ref{fig:JVSOrbital} shows that the main qualitative effect as discussed in the main text is evident for (\O${}_\text{L},{}$\O${}_\text{R})=(5,9)$. Quantitatively, convergence is achieved for $J$ when (\O${}_\text{L},{}$\O${}_\text{R})=(10,15)$. Fig.~\ref{fig:JVSOrbital2} shows $J$ as function of $\Delta$ obtained with (\O${}_\text{L},{}$\O${}_\text{R})=(10,15)$ for different magnetic fields. The key results in the main text, i.e. the magnetic-field-tuned sweet spots and the corresponding exchange energies at sweet spots, remain when higher numerical accuracy of $J$ is obtained using (\O${}_\text{L},{}$\O${}_\text{R})=(10,15)$ in full CI calculations. Therefore, the discussions made in the main text based on the results obtained with (\O${}_\text{L},{}$\O${}_\text{R})=(6,10)$ are valid while retaining higher orbitals will not change our analyses and conclusions in any important way.}
	
	\section{Correspondence between CI and extended Hubbard Model}\label{sec:CIHub}
	\begin{figure*}[t]
		\includegraphics[width=\linewidth]{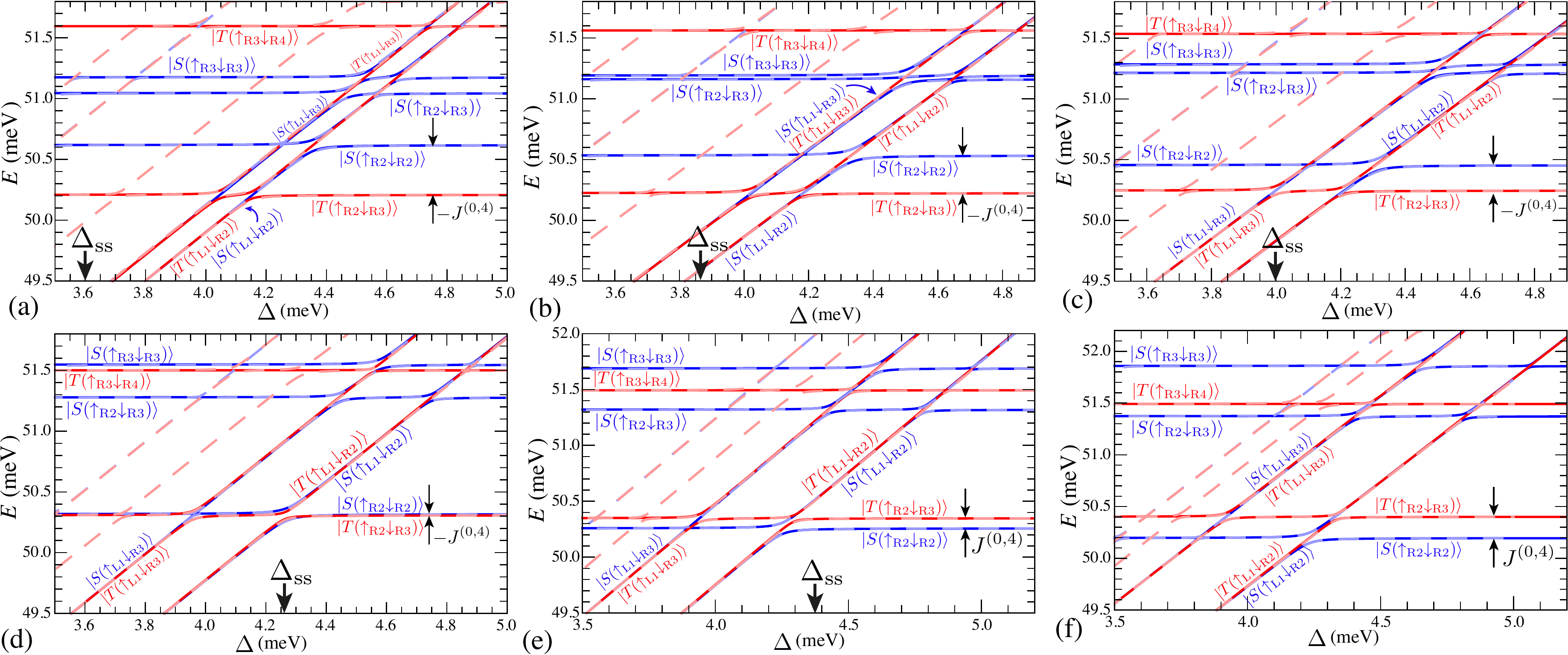}
		\caption{Lowest eigenvalues of a DQD occupied by four electrons as function of detuning, $\Delta$. The energy spectrum obtained using Hubbard model are plotted as colored solid line, while CI calculations are plotted as dashed lighter lines. The magnetic field strength is (a) $B = 0.12$ T, (b) $B = 0.17$ T, (c) $B = 0.23$ T, (d) $B = 0.35$ T, (e) $B = 0.41$ T and (d) $B = 0.47$ T. The exchange energy in $(n_\text{L},n_\text{R})=(0,4)$ region is labeled as $J^{(0,4)}$ while the sweet spot detunings are marked with the label $\Delta_{\text{ss}}$.}
		\label{fig:CIHubOtherB}
	\end{figure*}
	Figure \ref{fig:CIHubOtherB} shows the eigenvalues obtained from CI results (dashed lighter lines) and the extended Hubbard model (solid colored lines). The values of Hubbard parameters can be derived from CI results. Fig.~\ref{fig:CIHubOtherB} shows that the results agree well with each other.
	
	\section{Absence of electron densities redistribution under the tuning of magnetic field}\label{sec:elecDen}
	\begin{figure}
		\includegraphics[width=\linewidth]{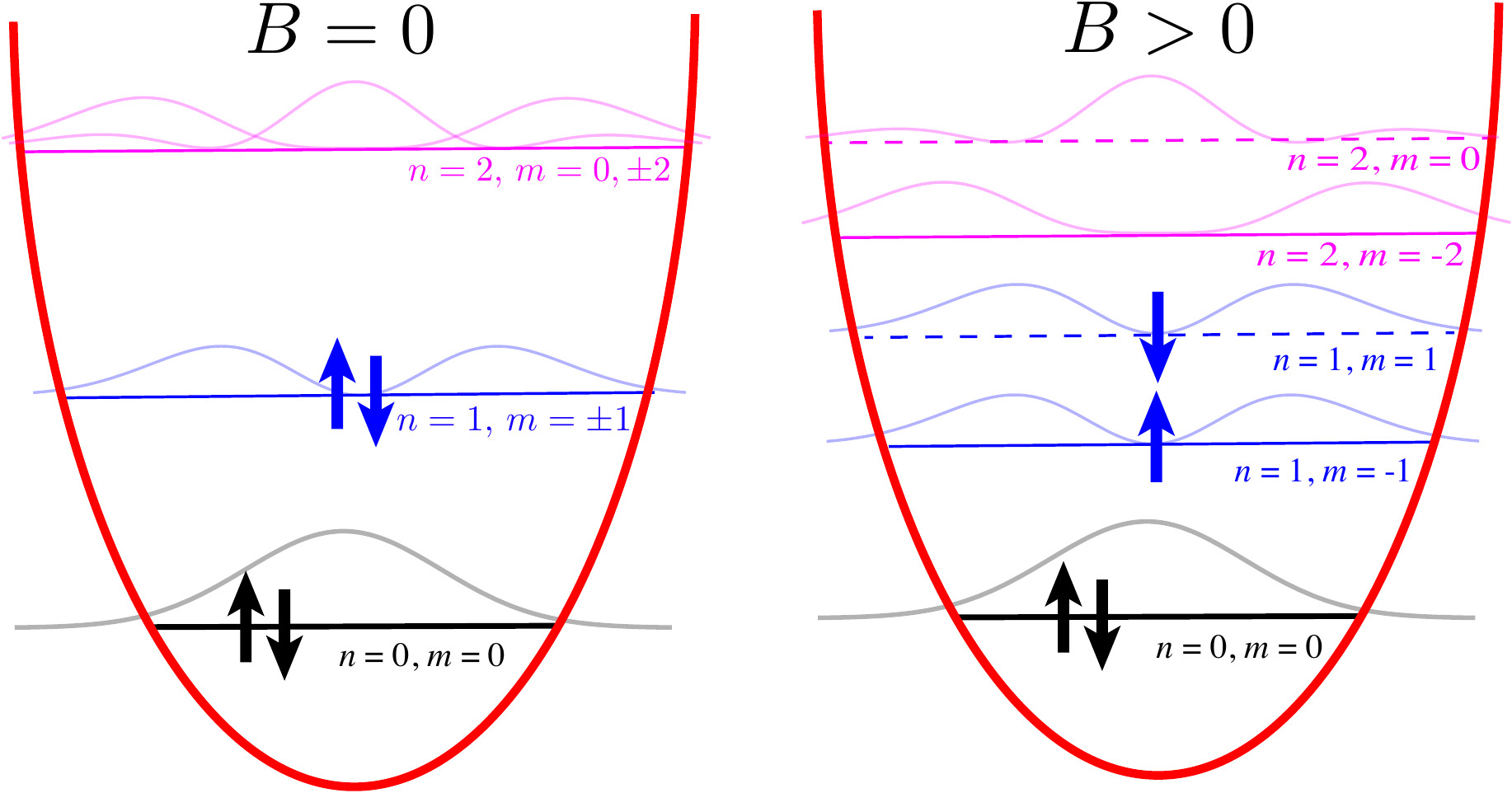}
		\caption{Schematic plot of the single-electron densities for $B = 0$ (left panel) and $B > 0$ (right panel). The single-electron densities (square norm of orthonormalized F-D states) are labeled with the corresponding principle, $n$, and magnetic quantum number, $m$. The black and blue arrows indicates the electron configuration of $\left|T(\uparrow_{\text{R}2}\downarrow_{\text{R}3})\right\rangle$.}
		\label{fig:wfSingPar}
	\end{figure}
	\begin{figure}[t]
		\includegraphics[width=\linewidth]{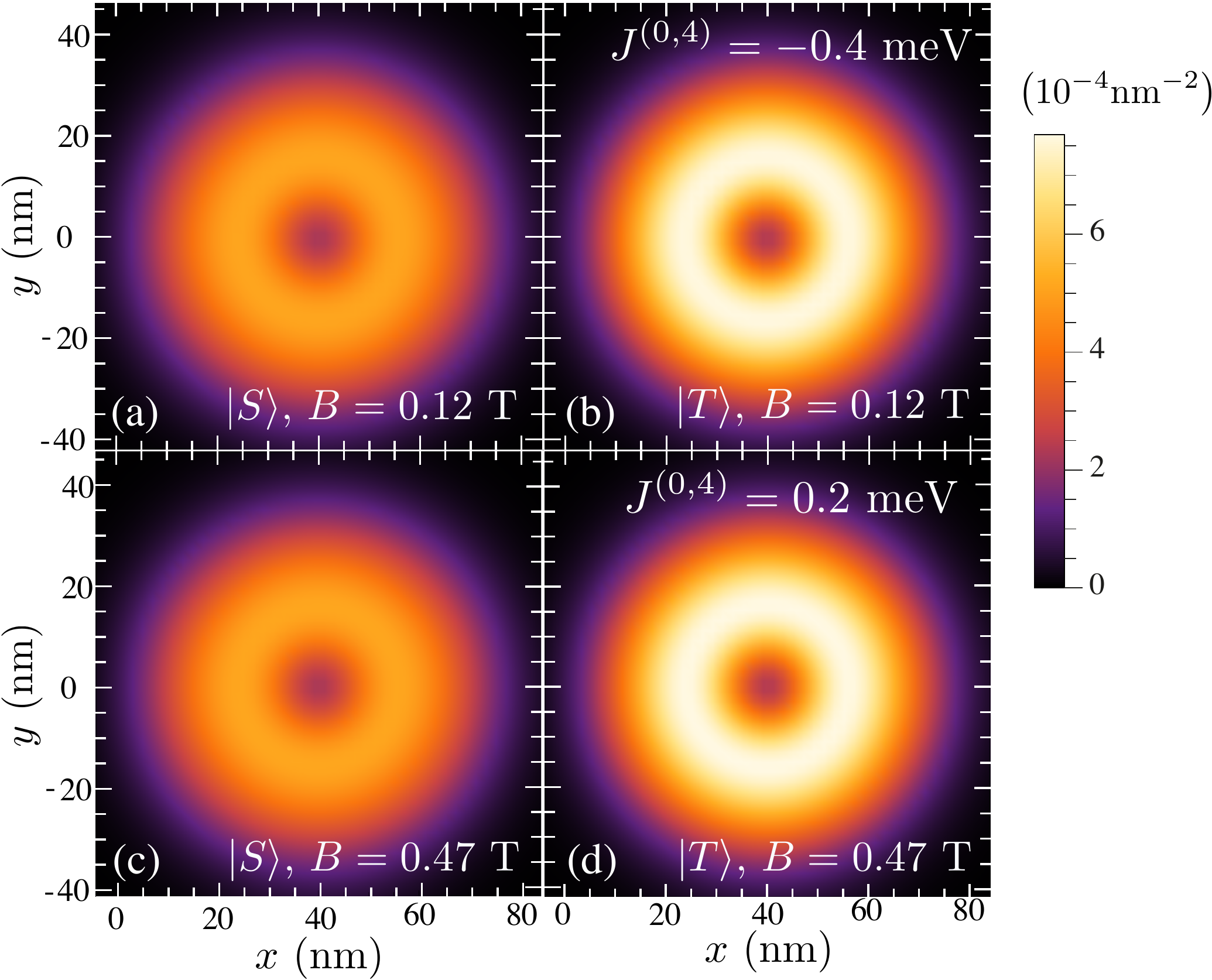}
		\caption{Four-electron densities obtained from full CI calculation. The electron densities are plotted with ground F-D state of the right larger dot excluded to signify the effect of valence orbitals.}
		\label{fig:electonDen}
	\end{figure}
	
	In the main text, it is observed that the exchange energy in $(n_\text{L},n_\text{R})=(0,4)$, denoted as $J^{(0,4)}$, ranges from negative values for $B\lesssim 0.35$ T to positive values for $B\gtrsim 0.41$ T (cf.~Fig.~\ref{fig:JsmallDelta} in the main text). This can be understood by the relative changes of four-electron singlet and triplet eigenvalues, which is found to be tunable by the magnetic field. Conventionally, the variation of singlet-triplet splitting in a QD can be attributed to Wigner-molecule physics \cite{Bryant.87,Yannouleas.99,Cioslowski.00,Reimann.00,Filinov.01,Reusch.01,Szafran.04,Rontani.06,Ghosal.07,Cavaliere.09,Shapir.19,Mintairov.18,Singha.10,Ellenberger.06,Kristinsdottir.11,Pecker.13,Egger.99,Ercan.21,Corrigan.21}. In analogy to the case in which two electrons occupying a QD \cite{Ercan.21,Corrigan.21}, we use the dimensionless measure of the electron-electron interaction strength, defined as $R_\text{w}=E_{\text{e-e}}/\hbar \omega$, where $E_{\text{e-e}}$ is the on-site Coulomb energy while $\hbar \omega$ is the orbital splitting between the lowest and first excited valence orbitals. For the two-electron case mentioned above, $R_\text{w}$ is applicable for the ground singlet state. When $R_\text{w}\gg 1$, the singlet state yields larger composition of excited orbital, hence rendering the singlet and triplet states sharing similar electron densities, resulting in the suppression of singlet-triplet splitting. For the four-electron case in a QD considered in this work, the quantity $R_\text{w}$ is not applicable for the ground singlet state as the ground and first excited valence orbitals share the same electron densities, cf.~single-electron densities for $n=1, m=1$ and $n=1, m=-1$ in Fig.~\ref{fig:wfSingPar}. However, we can still consider the $R_\text{w}$ for triplet state. For a non-zero magnetic field, assuming $\hbar\omega\gg \hbar \omega_c$, where $\omega_c\propto B$ \cite{Barnes.11}, the orbital energy of $n=1,m=1$ increases for $\hbar \omega_c/2$ while $n=2,m=-2$ decreases for $\hbar \omega_c$. Hence for a moderate magnetic field strength, $R_\text{w}\gg 1$ can be achieved for a triplet state as $n=1,m=1$ and $n=2,m=-2$ are brought closer energetically, modifying the singlet-triplet splitting. However, we show in the following paragraph that Wigner-molecule physics does not play a role in the tuning of $J^{(0,4)}$ by magnetic field.

	Figure \ref{fig:electonDen} shows, in the detuning region which gives $(n_\text{L},n_\text{R})=(0,4)$, the electron densities of (left column) the lowest singlet, $|S\rangle$, and (right column) triplet state, $|T\rangle$, with the corresponding exchange energies denoted as $J^{(0,4)}$. The electron densities for $J^{\left(0,4\right)}<0$ and $J^{\left(0,4\right)}>0$ are shown in the top and bottom row of Fig.~\ref{fig:electonDen}, which corresponds to $B=0.21$ T and $B=0.47$ T respectively, cf.~Fig.~\ref{fig:JsmallDelta}. In the range of magnetic field giving rise to the sign switching of $J^{(0,4)}$, the variation of magnetic field is insufficient to enforce localization effect. This is confirmed by the almost identical electron densities, for both the lowest singlet and triplet states, between the cases in which $J^{(0,4)}<0$ ($B=0.12$ T) and $J^{(0,4)}>0$ ($B=0.47$ T), cf.~Fig.~\ref{fig:electonDen}. The discussion in Sec.~\ref{subsec:Ex} shows that the main contributing factor to the sign switching of $J^{(0,4)}$ is the energy splitting between $n=1,m=\pm1$ orbitals, which yields
	\begin{equation}\label{eq:DeltaE}
		\Delta E = \varepsilon_{\text{R}3}-\varepsilon_{\text{R}2} =  \hbar \omega_c,
	\end{equation}
 	where $\omega_c = e B/m^*$ is the cyclotron frequency and $e$ is the electron charge. Note that in Eq.~\eqref{eq:DeltaE}, in contrast to the fitted Hubbard parameters (cf.~Fig.~\ref{fig:HubbardParaValues}), $\Delta E$ is merely a rough estimation with the assumption that $|T(\uparrow_{\text{R}2}\downarrow_{\text{R}3})\rangle$ is formed by the valence electrons occupying orthonormalized F-D states of quantum numbers $n=1,m=\pm1$. Although $\Delta E$ yields a value large enough to induce the sign switching of $J^{(0,4)}$, the variation of effective QD confinement length, $l_B^2=l_0^2/\sqrt{1+e^2B^2l_0^4/4\hbar^2}$, where $l_0=\sqrt{\hbar/m^*\omega_0}$, is negligible for the range of magnetic field of interest, hence resulting in near equivalence of the electron densities for both $J^{(0,4)}<0$ and $J^{(0,4)}>0$ regime, as shown in Fig.~\ref{fig:electonDen}. In particular, for $\hbar \omega_0 = 4$ meV, $l_B = 16.886$ nm when $B=0.12$ T while $l_B = 16.844$ nm when $B=0.47$ T. 
It has been demonstrated experimentally \cite{Kalliakos.08,Kouwenhoven.97}	
that in the case of $J^{(0,4)}$ switching sign, the electron-density  localization effect is absent. Here, we take the discussion further from the point of view of the extended Hubbard model and the plots of electron densities.

	\bibliographystyle{apsrev4-1}

\begin{thebibliography}{76}%
\makeatletter
\providecommand \@ifxundefined [1]{%
 \@ifx{#1\undefined}
}%
\providecommand \@ifnum [1]{%
 \ifnum #1\expandafter \@firstoftwo
 \else \expandafter \@secondoftwo
 \fi
}%
\providecommand \@ifx [1]{%
 \ifx #1\expandafter \@firstoftwo
 \else \expandafter \@secondoftwo
 \fi
}%
\providecommand \natexlab [1]{#1}%
\providecommand \enquote  [1]{``#1''}%
\providecommand \bibnamefont  [1]{#1}%
\providecommand \bibfnamefont [1]{#1}%
\providecommand \citenamefont [1]{#1}%
\providecommand \href@noop [0]{\@secondoftwo}%
\providecommand \href [0]{\begingroup \@sanitize@url \@href}%
\providecommand \@href[1]{\@@startlink{#1}\@@href}%
\providecommand \@@href[1]{\endgroup#1\@@endlink}%
\providecommand \@sanitize@url [0]{\catcode `\\12\catcode `\$12\catcode
  `\&12\catcode `\#12\catcode `\^12\catcode `\_12\catcode `\%12\relax}%
\providecommand \@@startlink[1]{}%
\providecommand \@@endlink[0]{}%
\providecommand \url  [0]{\begingroup\@sanitize@url \@url }%
\providecommand \@url [1]{\endgroup\@href {#1}{\urlprefix }}%
\providecommand \urlprefix  [0]{URL }%
\providecommand \Eprint [0]{\href }%
\providecommand \doibase [0]{http://dx.doi.org/}%
\providecommand \selectlanguage [0]{\@gobble}%
\providecommand \bibinfo  [0]{\@secondoftwo}%
\providecommand \bibfield  [0]{\@secondoftwo}%
\providecommand \translation [1]{[#1]}%
\providecommand \BibitemOpen [0]{}%
\providecommand \bibitemStop [0]{}%
\providecommand \bibitemNoStop [0]{.\EOS\space}%
\providecommand \EOS [0]{\spacefactor3000\relax}%
\providecommand \BibitemShut  [1]{\csname bibitem#1\endcsname}%
\let\auto@bib@innerbib\@empty
\bibitem [{\citenamefont {Koppens}\ \emph {et~al.}(2006)\citenamefont
  {Koppens}, \citenamefont {Buizert}, \citenamefont {Tielrooij}, \citenamefont
  {Vink}, \citenamefont {Nowack}, \citenamefont {Meunier}, \citenamefont
  {Kouwenhoven},\ and\ \citenamefont {Vandersypen}}]{Koppens.06}%
  \BibitemOpen
  \bibfield  {author} {\bibinfo {author} {\bibfnamefont {F.~H.~L.}\
  \bibnamefont {Koppens}}, \bibinfo {author} {\bibfnamefont {C.}~\bibnamefont
  {Buizert}}, \bibinfo {author} {\bibfnamefont {K.~J.}\ \bibnamefont
  {Tielrooij}}, \bibinfo {author} {\bibfnamefont {I.~T.}\ \bibnamefont {Vink}},
  \bibinfo {author} {\bibfnamefont {K.~C.}\ \bibnamefont {Nowack}}, \bibinfo
  {author} {\bibfnamefont {T.}~\bibnamefont {Meunier}}, \bibinfo {author}
  {\bibfnamefont {L.~P.}\ \bibnamefont {Kouwenhoven}}, \ and\ \bibinfo {author}
  {\bibfnamefont {L.~M.~K.}\ \bibnamefont {Vandersypen}},\ }\href
  {https://doi.org/10.1038/nature05065} {\bibfield  {journal} {\bibinfo
  {journal} {Nature}\ }\textbf {\bibinfo {volume} {442}},\ \bibinfo {pages}
  {766} (\bibinfo {year} {2006})}\BibitemShut {NoStop}%
\bibitem [{\citenamefont {Zajac}\ \emph {et~al.}(2018)\citenamefont {Zajac},
  \citenamefont {Sigillito}, \citenamefont {Russ}, \citenamefont {Borjans},
  \citenamefont {Taylor}, \citenamefont {Burkard},\ and\ \citenamefont
  {Petta}}]{Zajac.18}%
  \BibitemOpen
  \bibfield  {author} {\bibinfo {author} {\bibfnamefont {D.~M.}\ \bibnamefont
  {Zajac}}, \bibinfo {author} {\bibfnamefont {A.~J.}\ \bibnamefont
  {Sigillito}}, \bibinfo {author} {\bibfnamefont {M.}~\bibnamefont {Russ}},
  \bibinfo {author} {\bibfnamefont {F.}~\bibnamefont {Borjans}}, \bibinfo
  {author} {\bibfnamefont {J.~M.}\ \bibnamefont {Taylor}}, \bibinfo {author}
  {\bibfnamefont {G.}~\bibnamefont {Burkard}}, \ and\ \bibinfo {author}
  {\bibfnamefont {J.~R.}\ \bibnamefont {Petta}},\ }\href
  {https://www.science.org/doi/abs/10.1126/science.aao5965} {\bibfield
  {journal} {\bibinfo  {journal} {Science}\ }\textbf {\bibinfo {volume}
  {359}},\ \bibinfo {pages} {439} (\bibinfo {year} {2018})}\BibitemShut
  {NoStop}%
\bibitem [{\citenamefont {Mills}\ \emph {et~al.}()\citenamefont {Mills},
  \citenamefont {Guinn}, \citenamefont {Gullans}, \citenamefont {Sigillito},
  \citenamefont {Feldman}, \citenamefont {Nielsen},\ and\ \citenamefont
  {Petta}}]{Mills.21}%
  \BibitemOpen
  \bibfield  {author} {\bibinfo {author} {\bibfnamefont {A.~R.}\ \bibnamefont
  {Mills}}, \bibinfo {author} {\bibfnamefont {C.~R.}\ \bibnamefont {Guinn}},
  \bibinfo {author} {\bibfnamefont {M.~J.}\ \bibnamefont {Gullans}}, \bibinfo
  {author} {\bibfnamefont {A.~J.}\ \bibnamefont {Sigillito}}, \bibinfo {author}
  {\bibfnamefont {M.~M.}\ \bibnamefont {Feldman}}, \bibinfo {author}
  {\bibfnamefont {E.}~\bibnamefont {Nielsen}}, \ and\ \bibinfo {author}
  {\bibfnamefont {J.~R.}\ \bibnamefont {Petta}},\ }\href
  {https://arxiv.org/abs/2111.11937} {\bibinfo  {journal} {arXiv:2111.11937}\
  }\BibitemShut {NoStop}%
\bibitem [{\citenamefont {Yoneda}\ \emph {et~al.}(2018)\citenamefont {Yoneda},
  \citenamefont {Takeda}, \citenamefont {Otsuka}, \citenamefont {Nakajima},
  \citenamefont {Delbecq}, \citenamefont {Allison}, \citenamefont {Honda},
  \citenamefont {Kodera}, \citenamefont {Oda}, \citenamefont {Hoshi},
  \citenamefont {Usami}, \citenamefont {Itoh},\ and\ \citenamefont
  {Tarucha}}]{Yoneda.18}%
  \BibitemOpen
\bibfield  {journal} {  }\bibfield  {author} {\bibinfo {author} {\bibfnamefont
  {J.}~\bibnamefont {Yoneda}}, \bibinfo {author} {\bibfnamefont
  {K.}~\bibnamefont {Takeda}}, \bibinfo {author} {\bibfnamefont
  {T.}~\bibnamefont {Otsuka}}, \bibinfo {author} {\bibfnamefont
  {T.}~\bibnamefont {Nakajima}}, \bibinfo {author} {\bibfnamefont {M.~R.}\
  \bibnamefont {Delbecq}}, \bibinfo {author} {\bibfnamefont {G.}~\bibnamefont
  {Allison}}, \bibinfo {author} {\bibfnamefont {T.}~\bibnamefont {Honda}},
  \bibinfo {author} {\bibfnamefont {T.}~\bibnamefont {Kodera}}, \bibinfo
  {author} {\bibfnamefont {S.}~\bibnamefont {Oda}}, \bibinfo {author}
  {\bibfnamefont {Y.}~\bibnamefont {Hoshi}}, \bibinfo {author} {\bibfnamefont
  {N.}~\bibnamefont {Usami}}, \bibinfo {author} {\bibfnamefont {K.~M.}\
  \bibnamefont {Itoh}}, \ and\ \bibinfo {author} {\bibfnamefont
  {S.}~\bibnamefont {Tarucha}},\ }\href
  {https://doi.org/10.1038/s41565-017-0014-x} {\bibfield  {journal} {\bibinfo
  {journal} {Nat. Nanotechnol.}\ }\textbf {\bibinfo {volume} {13}},\ \bibinfo
  {pages} {102} (\bibinfo {year} {2018})}\BibitemShut {NoStop}%
\bibitem [{\citenamefont {Nowack}\ \emph {et~al.}(2007)\citenamefont {Nowack},
  \citenamefont {Koppens}, \citenamefont {Nazarov},\ and\ \citenamefont
  {Vandersypen}}]{Nowack.07}%
  \BibitemOpen
  \bibfield  {author} {\bibinfo {author} {\bibfnamefont {K.~C.}\ \bibnamefont
  {Nowack}}, \bibinfo {author} {\bibfnamefont {F.~H.~L.}\ \bibnamefont
  {Koppens}}, \bibinfo {author} {\bibfnamefont {Y.~V.}\ \bibnamefont
  {Nazarov}}, \ and\ \bibinfo {author} {\bibfnamefont {L.~M.~K.}\ \bibnamefont
  {Vandersypen}},\ }\href
  {https://www.science.org/doi/abs/10.1126/science.1148092} {\bibfield
  {journal} {\bibinfo  {journal} {Science}\ }\textbf {\bibinfo {volume}
  {318}},\ \bibinfo {pages} {1430} (\bibinfo {year} {2007})}\BibitemShut
  {NoStop}%
\bibitem [{\citenamefont {Pioro-Ladri{\`e}re}\ \emph
  {et~al.}(2008)\citenamefont {Pioro-Ladri{\`e}re}, \citenamefont {Obata},
  \citenamefont {Tokura}, \citenamefont {Shin}, \citenamefont {Kubo},
  \citenamefont {Yoshida}, \citenamefont {Taniyama},\ and\ \citenamefont
  {Tarucha}}]{Pioro.08}%
  \BibitemOpen
  \bibfield  {author} {\bibinfo {author} {\bibfnamefont {M.}~\bibnamefont
  {Pioro-Ladri{\`e}re}}, \bibinfo {author} {\bibfnamefont {T.}~\bibnamefont
  {Obata}}, \bibinfo {author} {\bibfnamefont {Y.}~\bibnamefont {Tokura}},
  \bibinfo {author} {\bibfnamefont {Y.~S.}\ \bibnamefont {Shin}}, \bibinfo
  {author} {\bibfnamefont {T.}~\bibnamefont {Kubo}}, \bibinfo {author}
  {\bibfnamefont {K.}~\bibnamefont {Yoshida}}, \bibinfo {author} {\bibfnamefont
  {T.}~\bibnamefont {Taniyama}}, \ and\ \bibinfo {author} {\bibfnamefont
  {S.}~\bibnamefont {Tarucha}},\ }\href {https://doi.org/10.1038/nphys1053}
  {\bibfield  {journal} {\bibinfo  {journal} {Nat. Phys.}\ }\textbf {\bibinfo
  {volume} {4}},\ \bibinfo {pages} {776} (\bibinfo {year} {2008})}\BibitemShut
  {NoStop}%
\bibitem [{\citenamefont {Boter}\ \emph {et~al.}()\citenamefont {Boter},
  \citenamefont {Dehollain}, \citenamefont {van Dijk}, \citenamefont {Xu},
  \citenamefont {Hensgens}, \citenamefont {Versluis}, \citenamefont {Naus},
  \citenamefont {Clarke}, \citenamefont {Veldhorst}, \citenamefont
  {Sebastiano},\ and\ \citenamefont {Vandersypen}}]{Jelmer.21}%
  \BibitemOpen
  \bibfield  {author} {\bibinfo {author} {\bibfnamefont {J.~M.}\ \bibnamefont
  {Boter}}, \bibinfo {author} {\bibfnamefont {J.~P.}\ \bibnamefont
  {Dehollain}}, \bibinfo {author} {\bibfnamefont {J.~P.~G.}\ \bibnamefont {van
  Dijk}}, \bibinfo {author} {\bibfnamefont {Y.}~\bibnamefont {Xu}}, \bibinfo
  {author} {\bibfnamefont {T.}~\bibnamefont {Hensgens}}, \bibinfo {author}
  {\bibfnamefont {R.}~\bibnamefont {Versluis}}, \bibinfo {author}
  {\bibfnamefont {H.~W.~L.}\ \bibnamefont {Naus}}, \bibinfo {author}
  {\bibfnamefont {J.~S.}\ \bibnamefont {Clarke}}, \bibinfo {author}
  {\bibfnamefont {M.}~\bibnamefont {Veldhorst}}, \bibinfo {author}
  {\bibfnamefont {F.}~\bibnamefont {Sebastiano}}, \ and\ \bibinfo {author}
  {\bibfnamefont {L.~M.~K.}\ \bibnamefont {Vandersypen}},\ }\href
  {https://arxiv.org/abs/2110.00189} {\bibinfo  {journal} {arXiv:2110.00189}\
  }\BibitemShut {NoStop}%
\bibitem [{\citenamefont {Laird}\ \emph {et~al.}(2010)\citenamefont {Laird},
  \citenamefont {Taylor}, \citenamefont {DiVincenzo}, \citenamefont {Marcus},
  \citenamefont {Hanson},\ and\ \citenamefont {Gossard}}]{Laird.10}%
  \BibitemOpen
\bibfield  {journal} {  }\bibfield  {author} {\bibinfo {author} {\bibfnamefont
  {E.~A.}\ \bibnamefont {Laird}}, \bibinfo {author} {\bibfnamefont {J.~M.}\
  \bibnamefont {Taylor}}, \bibinfo {author} {\bibfnamefont {D.~P.}\
  \bibnamefont {DiVincenzo}}, \bibinfo {author} {\bibfnamefont {C.~M.}\
  \bibnamefont {Marcus}}, \bibinfo {author} {\bibfnamefont {M.~P.}\
  \bibnamefont {Hanson}}, \ and\ \bibinfo {author} {\bibfnamefont {A.~C.}\
  \bibnamefont {Gossard}},\ }\href
  {https://link.aps.org/doi/10.1103/PhysRevB.82.075403} {\bibfield  {journal}
  {\bibinfo  {journal} {Phys. Rev. B}\ }\textbf {\bibinfo {volume} {82}},\
  \bibinfo {pages} {075403} (\bibinfo {year} {2010})}\BibitemShut {NoStop}%
\bibitem [{\citenamefont {Gaudreau}\ \emph {et~al.}(2012)\citenamefont
  {Gaudreau}, \citenamefont {Granger}, \citenamefont {Kam}, \citenamefont
  {Aers}, \citenamefont {Studenikin}, \citenamefont {Zawadzki}, \citenamefont
  {Pioro-Ladri{\`e}re}, \citenamefont {Wasilewski},\ and\ \citenamefont
  {Sachrajda}}]{Gaudreau.12}%
  \BibitemOpen
  \bibfield  {author} {\bibinfo {author} {\bibfnamefont {L.}~\bibnamefont
  {Gaudreau}}, \bibinfo {author} {\bibfnamefont {G.}~\bibnamefont {Granger}},
  \bibinfo {author} {\bibfnamefont {A.}~\bibnamefont {Kam}}, \bibinfo {author}
  {\bibfnamefont {G.~C.}\ \bibnamefont {Aers}}, \bibinfo {author}
  {\bibfnamefont {S.~A.}\ \bibnamefont {Studenikin}}, \bibinfo {author}
  {\bibfnamefont {P.}~\bibnamefont {Zawadzki}}, \bibinfo {author}
  {\bibfnamefont {M.}~\bibnamefont {Pioro-Ladri{\`e}re}}, \bibinfo {author}
  {\bibfnamefont {Z.~R.}\ \bibnamefont {Wasilewski}}, \ and\ \bibinfo {author}
  {\bibfnamefont {A.~S.}\ \bibnamefont {Sachrajda}},\ }\href
  {https://doi.org/10.1038/nphys2149} {\bibfield  {journal} {\bibinfo
  {journal} {Nat. Phys.}\ }\textbf {\bibinfo {volume} {8}},\ \bibinfo {pages}
  {54} (\bibinfo {year} {2012})}\BibitemShut {NoStop}%
\bibitem [{\citenamefont {Medford}\ \emph {et~al.}(2013)\citenamefont
  {Medford}, \citenamefont {Beil}, \citenamefont {Taylor}, \citenamefont
  {Rashba}, \citenamefont {Lu}, \citenamefont {Gossard},\ and\ \citenamefont
  {Marcus}}]{Medford.13}%
  \BibitemOpen
  \bibfield  {author} {\bibinfo {author} {\bibfnamefont {J.}~\bibnamefont
  {Medford}}, \bibinfo {author} {\bibfnamefont {J.}~\bibnamefont {Beil}},
  \bibinfo {author} {\bibfnamefont {J.~M.}\ \bibnamefont {Taylor}}, \bibinfo
  {author} {\bibfnamefont {E.~I.}\ \bibnamefont {Rashba}}, \bibinfo {author}
  {\bibfnamefont {H.}~\bibnamefont {Lu}}, \bibinfo {author} {\bibfnamefont
  {A.~C.}\ \bibnamefont {Gossard}}, \ and\ \bibinfo {author} {\bibfnamefont
  {C.~M.}\ \bibnamefont {Marcus}},\ }\href
  {https://link.aps.org/doi/10.1103/PhysRevLett.111.050501} {\bibfield
  {journal} {\bibinfo  {journal} {Phys. Rev. Lett.}\ }\textbf {\bibinfo
  {volume} {111}},\ \bibinfo {pages} {050501} (\bibinfo {year}
  {2013})}\BibitemShut {NoStop}%
\bibitem [{\citenamefont {Martins}\ \emph {et~al.}(2016)\citenamefont
  {Martins}, \citenamefont {Malinowski}, \citenamefont {Nissen}, \citenamefont
  {Barnes}, \citenamefont {Fallahi}, \citenamefont {Gardner}, \citenamefont
  {Manfra}, \citenamefont {Marcus},\ and\ \citenamefont
  {Kuemmeth}}]{Martins.16}%
  \BibitemOpen
  \bibfield  {author} {\bibinfo {author} {\bibfnamefont {F.}~\bibnamefont
  {Martins}}, \bibinfo {author} {\bibfnamefont {F.~K.}\ \bibnamefont
  {Malinowski}}, \bibinfo {author} {\bibfnamefont {P.~D.}\ \bibnamefont
  {Nissen}}, \bibinfo {author} {\bibfnamefont {E.}~\bibnamefont {Barnes}},
  \bibinfo {author} {\bibfnamefont {S.}~\bibnamefont {Fallahi}}, \bibinfo
  {author} {\bibfnamefont {G.~C.}\ \bibnamefont {Gardner}}, \bibinfo {author}
  {\bibfnamefont {M.~J.}\ \bibnamefont {Manfra}}, \bibinfo {author}
  {\bibfnamefont {C.~M.}\ \bibnamefont {Marcus}}, \ and\ \bibinfo {author}
  {\bibfnamefont {F.}~\bibnamefont {Kuemmeth}},\ }\href
  {https://link.aps.org/doi/10.1103/PhysRevLett.116.116801} {\bibfield
  {journal} {\bibinfo  {journal} {Phys. Rev. Lett.}\ }\textbf {\bibinfo
  {volume} {116}},\ \bibinfo {pages} {116801} (\bibinfo {year}
  {2016})}\BibitemShut {NoStop}%
\bibitem [{\citenamefont {Reed}\ \emph {et~al.}(2016)\citenamefont {Reed},
  \citenamefont {Maune}, \citenamefont {Andrews}, \citenamefont {Borselli},
  \citenamefont {Eng}, \citenamefont {Jura}, \citenamefont {Kiselev},
  \citenamefont {Ladd}, \citenamefont {Merkel}, \citenamefont {Milosavljevic},
  \citenamefont {Pritchett}, \citenamefont {Rakher}, \citenamefont {Ross},
  \citenamefont {Schmitz}, \citenamefont {Smith}, \citenamefont {Wright},
  \citenamefont {Gyure},\ and\ \citenamefont {Hunter}}]{Reed.16}%
  \BibitemOpen
  \bibfield  {author} {\bibinfo {author} {\bibfnamefont {M.~D.}\ \bibnamefont
  {Reed}}, \bibinfo {author} {\bibfnamefont {B.~M.}\ \bibnamefont {Maune}},
  \bibinfo {author} {\bibfnamefont {R.~W.}\ \bibnamefont {Andrews}}, \bibinfo
  {author} {\bibfnamefont {M.~G.}\ \bibnamefont {Borselli}}, \bibinfo {author}
  {\bibfnamefont {K.}~\bibnamefont {Eng}}, \bibinfo {author} {\bibfnamefont
  {M.~P.}\ \bibnamefont {Jura}}, \bibinfo {author} {\bibfnamefont {A.~A.}\
  \bibnamefont {Kiselev}}, \bibinfo {author} {\bibfnamefont {T.~D.}\
  \bibnamefont {Ladd}}, \bibinfo {author} {\bibfnamefont {S.~T.}\ \bibnamefont
  {Merkel}}, \bibinfo {author} {\bibfnamefont {I.}~\bibnamefont
  {Milosavljevic}}, \bibinfo {author} {\bibfnamefont {E.~J.}\ \bibnamefont
  {Pritchett}}, \bibinfo {author} {\bibfnamefont {M.~T.}\ \bibnamefont
  {Rakher}}, \bibinfo {author} {\bibfnamefont {R.~S.}\ \bibnamefont {Ross}},
  \bibinfo {author} {\bibfnamefont {A.~E.}\ \bibnamefont {Schmitz}}, \bibinfo
  {author} {\bibfnamefont {A.}~\bibnamefont {Smith}}, \bibinfo {author}
  {\bibfnamefont {J.~A.}\ \bibnamefont {Wright}}, \bibinfo {author}
  {\bibfnamefont {M.~F.}\ \bibnamefont {Gyure}}, \ and\ \bibinfo {author}
  {\bibfnamefont {A.~T.}\ \bibnamefont {Hunter}},\ }\href
  {https://link.aps.org/doi/10.1103/PhysRevLett.116.110402} {\bibfield
  {journal} {\bibinfo  {journal} {Phys. Rev. Lett.}\ }\textbf {\bibinfo
  {volume} {116}},\ \bibinfo {pages} {110402} (\bibinfo {year}
  {2016})}\BibitemShut {NoStop}%
\bibitem [{\citenamefont {Nichol}\ \emph {et~al.}(2017)\citenamefont {Nichol},
  \citenamefont {Orona}, \citenamefont {Harvey}, \citenamefont {Fallahi},
  \citenamefont {Gardner}, \citenamefont {Manfra},\ and\ \citenamefont
  {Yacoby}}]{Nichol.17}%
  \BibitemOpen
  \bibfield  {author} {\bibinfo {author} {\bibfnamefont {J.~M.}\ \bibnamefont
  {Nichol}}, \bibinfo {author} {\bibfnamefont {L.~A.}\ \bibnamefont {Orona}},
  \bibinfo {author} {\bibfnamefont {S.~P.}\ \bibnamefont {Harvey}}, \bibinfo
  {author} {\bibfnamefont {S.}~\bibnamefont {Fallahi}}, \bibinfo {author}
  {\bibfnamefont {G.~C.}\ \bibnamefont {Gardner}}, \bibinfo {author}
  {\bibfnamefont {M.~J.}\ \bibnamefont {Manfra}}, \ and\ \bibinfo {author}
  {\bibfnamefont {A.}~\bibnamefont {Yacoby}},\ }\href
  {https://doi.org/10.1038/s41534-016-0003-1} {\bibfield  {journal} {\bibinfo
  {journal} {npj Quantum Information}\ }\textbf {\bibinfo {volume} {3}},\
  \bibinfo {pages} {3} (\bibinfo {year} {2017})}\BibitemShut {NoStop}%
\bibitem [{\citenamefont {Takeda}\ \emph {et~al.}(2020)\citenamefont {Takeda},
  \citenamefont {Noiri}, \citenamefont {Yoneda}, \citenamefont {Nakajima},\
  and\ \citenamefont {Tarucha}}]{Takeda.20}%
  \BibitemOpen
  \bibfield  {author} {\bibinfo {author} {\bibfnamefont {K.}~\bibnamefont
  {Takeda}}, \bibinfo {author} {\bibfnamefont {A.}~\bibnamefont {Noiri}},
  \bibinfo {author} {\bibfnamefont {J.}~\bibnamefont {Yoneda}}, \bibinfo
  {author} {\bibfnamefont {T.}~\bibnamefont {Nakajima}}, \ and\ \bibinfo
  {author} {\bibfnamefont {S.}~\bibnamefont {Tarucha}},\ }\href
  {https://link.aps.org/doi/10.1103/PhysRevLett.124.117701} {\bibfield
  {journal} {\bibinfo  {journal} {Phys. Rev. Lett.}\ }\textbf {\bibinfo
  {volume} {124}},\ \bibinfo {pages} {117701} (\bibinfo {year}
  {2020})}\BibitemShut {NoStop}%
\bibitem [{\citenamefont {Eng}\ \emph {et~al.}(2015)\citenamefont {Eng},
  \citenamefont {Ladd}, \citenamefont {Smith}, \citenamefont {Borselli},
  \citenamefont {Kiselev}, \citenamefont {Fong}, \citenamefont {Holabird},
  \citenamefont {Hazard}, \citenamefont {Huang}, \citenamefont {Deelman},
  \citenamefont {Milosavljevic}, \citenamefont {Schmitz}, \citenamefont {Ross},
  \citenamefont {Gyure},\ and\ \citenamefont {Hunter}}]{Eng.15}%
  \BibitemOpen
  \bibfield  {author} {\bibinfo {author} {\bibfnamefont {K.}~\bibnamefont
  {Eng}}, \bibinfo {author} {\bibfnamefont {T.~D.}\ \bibnamefont {Ladd}},
  \bibinfo {author} {\bibfnamefont {A.}~\bibnamefont {Smith}}, \bibinfo
  {author} {\bibfnamefont {M.~G.}\ \bibnamefont {Borselli}}, \bibinfo {author}
  {\bibfnamefont {A.~A.}\ \bibnamefont {Kiselev}}, \bibinfo {author}
  {\bibfnamefont {B.~H.}\ \bibnamefont {Fong}}, \bibinfo {author}
  {\bibfnamefont {K.~S.}\ \bibnamefont {Holabird}}, \bibinfo {author}
  {\bibfnamefont {T.~M.}\ \bibnamefont {Hazard}}, \bibinfo {author}
  {\bibfnamefont {B.}~\bibnamefont {Huang}}, \bibinfo {author} {\bibfnamefont
  {P.~W.}\ \bibnamefont {Deelman}}, \bibinfo {author} {\bibfnamefont
  {I.}~\bibnamefont {Milosavljevic}}, \bibinfo {author} {\bibfnamefont {A.~E.}\
  \bibnamefont {Schmitz}}, \bibinfo {author} {\bibfnamefont {R.~S.}\
  \bibnamefont {Ross}}, \bibinfo {author} {\bibfnamefont {M.~F.}\ \bibnamefont
  {Gyure}}, \ and\ \bibinfo {author} {\bibfnamefont {A.~T.}\ \bibnamefont
  {Hunter}},\ }\href {https://advances.sciencemag.org/content/1/4/e1500214}
  {\bibfield  {journal} {\bibinfo  {journal} {Sci. Adv.}\ }\textbf {\bibinfo
  {volume} {1}},\ \bibinfo {pages} {e1500214} (\bibinfo {year}
  {2015})}\BibitemShut {NoStop}%
\bibitem [{\citenamefont {Petta}\ \emph {et~al.}(2005)\citenamefont {Petta},
  \citenamefont {Johnson}, \citenamefont {Taylor}, \citenamefont {Laird},
  \citenamefont {Yacoby}, \citenamefont {Lukin}, \citenamefont {Marcus},
  \citenamefont {Hanson},\ and\ \citenamefont {Gossard}}]{Petta.05}%
  \BibitemOpen
  \bibfield  {author} {\bibinfo {author} {\bibfnamefont {J.~R.}\ \bibnamefont
  {Petta}}, \bibinfo {author} {\bibfnamefont {A.~C.}\ \bibnamefont {Johnson}},
  \bibinfo {author} {\bibfnamefont {J.~M.}\ \bibnamefont {Taylor}}, \bibinfo
  {author} {\bibfnamefont {E.~A.}\ \bibnamefont {Laird}}, \bibinfo {author}
  {\bibfnamefont {A.}~\bibnamefont {Yacoby}}, \bibinfo {author} {\bibfnamefont
  {M.~D.}\ \bibnamefont {Lukin}}, \bibinfo {author} {\bibfnamefont {C.~M.}\
  \bibnamefont {Marcus}}, \bibinfo {author} {\bibfnamefont {M.~P.}\
  \bibnamefont {Hanson}}, \ and\ \bibinfo {author} {\bibfnamefont {A.~C.}\
  \bibnamefont {Gossard}},\ }\href
  {https://science.sciencemag.org/content/309/5744/2180} {\bibfield  {journal}
  {\bibinfo  {journal} {Science}\ }\textbf {\bibinfo {volume} {309}},\ \bibinfo
  {pages} {2180} (\bibinfo {year} {2005})}\BibitemShut {NoStop}%
\bibitem [{\citenamefont {Shulman}\ \emph {et~al.}(2012)\citenamefont
  {Shulman}, \citenamefont {Dial}, \citenamefont {Harvey}, \citenamefont
  {Bluhm}, \citenamefont {Umansky},\ and\ \citenamefont {Yacoby}}]{Shulman.12}%
  \BibitemOpen
  \bibfield  {author} {\bibinfo {author} {\bibfnamefont {M.~D.}\ \bibnamefont
  {Shulman}}, \bibinfo {author} {\bibfnamefont {O.~E.}\ \bibnamefont {Dial}},
  \bibinfo {author} {\bibfnamefont {S.~P.}\ \bibnamefont {Harvey}}, \bibinfo
  {author} {\bibfnamefont {H.}~\bibnamefont {Bluhm}}, \bibinfo {author}
  {\bibfnamefont {V.}~\bibnamefont {Umansky}}, \ and\ \bibinfo {author}
  {\bibfnamefont {A.}~\bibnamefont {Yacoby}},\ }\href
  {https://www.science.org/doi/abs/10.1126/science.1217692} {\bibfield
  {journal} {\bibinfo  {journal} {Science}\ }\textbf {\bibinfo {volume}
  {336}},\ \bibinfo {pages} {202} (\bibinfo {year} {2012})}\BibitemShut
  {NoStop}%
\bibitem [{\citenamefont {Maune}\ \emph {et~al.}(2012)\citenamefont {Maune},
  \citenamefont {Borselli}, \citenamefont {Huang}, \citenamefont {Ladd},
  \citenamefont {Deelman}, \citenamefont {Holabird}, \citenamefont {Kiselev},
  \citenamefont {Alvarado-Rodriguez}, \citenamefont {Ross}, \citenamefont
  {Schmitz}, \citenamefont {Sokolich}, \citenamefont {Watson}, \citenamefont
  {Gyure},\ and\ \citenamefont {Hunter}}]{Maune.12}%
  \BibitemOpen
  \bibfield  {author} {\bibinfo {author} {\bibfnamefont {B.~M.}\ \bibnamefont
  {Maune}}, \bibinfo {author} {\bibfnamefont {M.~G.}\ \bibnamefont {Borselli}},
  \bibinfo {author} {\bibfnamefont {B.}~\bibnamefont {Huang}}, \bibinfo
  {author} {\bibfnamefont {T.~D.}\ \bibnamefont {Ladd}}, \bibinfo {author}
  {\bibfnamefont {P.~W.}\ \bibnamefont {Deelman}}, \bibinfo {author}
  {\bibfnamefont {K.~S.}\ \bibnamefont {Holabird}}, \bibinfo {author}
  {\bibfnamefont {A.~A.}\ \bibnamefont {Kiselev}}, \bibinfo {author}
  {\bibfnamefont {I.}~\bibnamefont {Alvarado-Rodriguez}}, \bibinfo {author}
  {\bibfnamefont {R.~S.}\ \bibnamefont {Ross}}, \bibinfo {author}
  {\bibfnamefont {A.~E.}\ \bibnamefont {Schmitz}}, \bibinfo {author}
  {\bibfnamefont {M.}~\bibnamefont {Sokolich}}, \bibinfo {author}
  {\bibfnamefont {C.~A.}\ \bibnamefont {Watson}}, \bibinfo {author}
  {\bibfnamefont {M.~F.}\ \bibnamefont {Gyure}}, \ and\ \bibinfo {author}
  {\bibfnamefont {A.~T.}\ \bibnamefont {Hunter}},\ }\href
  {https://doi.org/10.1038/nature10707} {\bibfield  {journal} {\bibinfo
  {journal} {Nature}\ }\textbf {\bibinfo {volume} {481}},\ \bibinfo {pages}
  {344} (\bibinfo {year} {2012})}\BibitemShut {NoStop}%
\bibitem [{\citenamefont {Wu}\ \emph {et~al.}(2014)\citenamefont {Wu},
  \citenamefont {Ward}, \citenamefont {Prance}, \citenamefont {Kim},
  \citenamefont {Gamble}, \citenamefont {Mohr}, \citenamefont {Shi},
  \citenamefont {Savage}, \citenamefont {Lagally}, \citenamefont {Friesen},
  \citenamefont {Coppersmith},\ and\ \citenamefont {Eriksson}}]{Wu.14}%
  \BibitemOpen
  \bibfield  {author} {\bibinfo {author} {\bibfnamefont {X.}~\bibnamefont
  {Wu}}, \bibinfo {author} {\bibfnamefont {D.~R.}\ \bibnamefont {Ward}},
  \bibinfo {author} {\bibfnamefont {J.~R.}\ \bibnamefont {Prance}}, \bibinfo
  {author} {\bibfnamefont {D.}~\bibnamefont {Kim}}, \bibinfo {author}
  {\bibfnamefont {J.~K.}\ \bibnamefont {Gamble}}, \bibinfo {author}
  {\bibfnamefont {R.~T.}\ \bibnamefont {Mohr}}, \bibinfo {author}
  {\bibfnamefont {Z.}~\bibnamefont {Shi}}, \bibinfo {author} {\bibfnamefont
  {D.~E.}\ \bibnamefont {Savage}}, \bibinfo {author} {\bibfnamefont {M.~G.}\
  \bibnamefont {Lagally}}, \bibinfo {author} {\bibfnamefont {M.}~\bibnamefont
  {Friesen}}, \bibinfo {author} {\bibfnamefont {S.~N.}\ \bibnamefont
  {Coppersmith}}, \ and\ \bibinfo {author} {\bibfnamefont {M.~A.}\ \bibnamefont
  {Eriksson}},\ }\href {https://www.pnas.org/content/111/33/11938} {\bibfield
  {journal} {\bibinfo  {journal} {Proc. Natl. Acad. Sci. U.S.A.}\ }\textbf
  {\bibinfo {volume} {111}},\ \bibinfo {pages} {11938} (\bibinfo {year}
  {2014})}\BibitemShut {NoStop}%
\bibitem [{\citenamefont {Jock}\ \emph {et~al.}(2018)\citenamefont {Jock},
  \citenamefont {Jacobson}, \citenamefont {Harvey-Collard}, \citenamefont
  {Mounce}, \citenamefont {Srinivasa}, \citenamefont {Ward}, \citenamefont
  {Anderson}, \citenamefont {Manginell}, \citenamefont {Wendt}, \citenamefont
  {Rudolph}, \citenamefont {Pluym}, \citenamefont {Gamble}, \citenamefont
  {Baczewski}, \citenamefont {Witzel},\ and\ \citenamefont
  {Carroll}}]{Jock.18}%
  \BibitemOpen
  \bibfield  {author} {\bibinfo {author} {\bibfnamefont {R.~M.}\ \bibnamefont
  {Jock}}, \bibinfo {author} {\bibfnamefont {N.~T.}\ \bibnamefont {Jacobson}},
  \bibinfo {author} {\bibfnamefont {P.}~\bibnamefont {Harvey-Collard}},
  \bibinfo {author} {\bibfnamefont {A.~M.}\ \bibnamefont {Mounce}}, \bibinfo
  {author} {\bibfnamefont {V.}~\bibnamefont {Srinivasa}}, \bibinfo {author}
  {\bibfnamefont {D.~R.}\ \bibnamefont {Ward}}, \bibinfo {author}
  {\bibfnamefont {J.}~\bibnamefont {Anderson}}, \bibinfo {author}
  {\bibfnamefont {R.}~\bibnamefont {Manginell}}, \bibinfo {author}
  {\bibfnamefont {J.~R.}\ \bibnamefont {Wendt}}, \bibinfo {author}
  {\bibfnamefont {M.}~\bibnamefont {Rudolph}}, \bibinfo {author} {\bibfnamefont
  {T.}~\bibnamefont {Pluym}}, \bibinfo {author} {\bibfnamefont {J.~K.}\
  \bibnamefont {Gamble}}, \bibinfo {author} {\bibfnamefont {A.~D.}\
  \bibnamefont {Baczewski}}, \bibinfo {author} {\bibfnamefont {W.~M.}\
  \bibnamefont {Witzel}}, \ and\ \bibinfo {author} {\bibfnamefont {M.~S.}\
  \bibnamefont {Carroll}},\ }\href {https://doi.org/10.1038/s41467-018-04200-0}
  {\bibfield  {journal} {\bibinfo  {journal} {Nat. Commun.}\ }\textbf {\bibinfo
  {volume} {9}},\ \bibinfo {pages} {1768} (\bibinfo {year} {2018})}\BibitemShut
  {NoStop}%
\bibitem [{\citenamefont {Cerfontaine}\ \emph {et~al.}(2020)\citenamefont
  {Cerfontaine}, \citenamefont {Botzem}, \citenamefont {Ritzmann},
  \citenamefont {Humpohl}, \citenamefont {Ludwig}, \citenamefont {Schuh},
  \citenamefont {Bougeard}, \citenamefont {Wieck},\ and\ \citenamefont
  {Bluhm}}]{Cerfontaine.20}%
  \BibitemOpen
  \bibfield  {author} {\bibinfo {author} {\bibfnamefont {P.}~\bibnamefont
  {Cerfontaine}}, \bibinfo {author} {\bibfnamefont {T.}~\bibnamefont {Botzem}},
  \bibinfo {author} {\bibfnamefont {J.}~\bibnamefont {Ritzmann}}, \bibinfo
  {author} {\bibfnamefont {S.~S.}\ \bibnamefont {Humpohl}}, \bibinfo {author}
  {\bibfnamefont {A.}~\bibnamefont {Ludwig}}, \bibinfo {author} {\bibfnamefont
  {D.}~\bibnamefont {Schuh}}, \bibinfo {author} {\bibfnamefont
  {D.}~\bibnamefont {Bougeard}}, \bibinfo {author} {\bibfnamefont {A.~D.}\
  \bibnamefont {Wieck}}, \ and\ \bibinfo {author} {\bibfnamefont
  {H.}~\bibnamefont {Bluhm}},\ }\href
  {https://doi.org/10.1038/s41467-020-17865-3} {\bibfield  {journal} {\bibinfo
  {journal} {Nat. Commun.}\ }\textbf {\bibinfo {volume} {11}},\ \bibinfo
  {pages} {4144} (\bibinfo {year} {2020})}\BibitemShut {NoStop}%
\bibitem [{\citenamefont {{Harvey-Collard}}\ \emph {et~al.}(2017)\citenamefont
  {{Harvey-Collard}}, \citenamefont {{Jock}}, \citenamefont {{Jacobson}},
  \citenamefont {{Baczewski}}, \citenamefont {{Mounce}}, \citenamefont
  {{Curry}}, \citenamefont {{Ward}}, \citenamefont {{Anderson}}, \citenamefont
  {{Manginell}}, \citenamefont {{Wendt}}, \citenamefont {{Rudolph}},
  \citenamefont {{Pluym}}, \citenamefont {{Lilly}}, \citenamefont
  {{Pioro-Ladrière}},\ and\ \citenamefont {{Carroll}}}]{Harvey.17}%
  \BibitemOpen
  \bibfield  {author} {\bibinfo {author} {\bibfnamefont {P.}~\bibnamefont
  {{Harvey-Collard}}}, \bibinfo {author} {\bibfnamefont {R.~M.}\ \bibnamefont
  {{Jock}}}, \bibinfo {author} {\bibfnamefont {N.~T.}\ \bibnamefont
  {{Jacobson}}}, \bibinfo {author} {\bibfnamefont {A.~D.}\ \bibnamefont
  {{Baczewski}}}, \bibinfo {author} {\bibfnamefont {A.~M.}\ \bibnamefont
  {{Mounce}}}, \bibinfo {author} {\bibfnamefont {M.~J.}\ \bibnamefont
  {{Curry}}}, \bibinfo {author} {\bibfnamefont {D.~R.}\ \bibnamefont {{Ward}}},
  \bibinfo {author} {\bibfnamefont {J.~M.}\ \bibnamefont {{Anderson}}},
  \bibinfo {author} {\bibfnamefont {R.~P.}\ \bibnamefont {{Manginell}}},
  \bibinfo {author} {\bibfnamefont {J.~R.}\ \bibnamefont {{Wendt}}}, \bibinfo
  {author} {\bibfnamefont {M.}~\bibnamefont {{Rudolph}}}, \bibinfo {author}
  {\bibfnamefont {T.}~\bibnamefont {{Pluym}}}, \bibinfo {author} {\bibfnamefont
  {M.~P.}\ \bibnamefont {{Lilly}}}, \bibinfo {author} {\bibfnamefont
  {M.}~\bibnamefont {{Pioro-Ladrière}}}, \ and\ \bibinfo {author}
  {\bibfnamefont {M.~S.}\ \bibnamefont {{Carroll}}},\ }in\ \href
  {https://ieeexplore.ieee.org/document/8268507} {\emph {\bibinfo {booktitle}
  {2017 IEEE International Electron Devices Meeting (IEDM)}}}\ (\bibinfo {year}
  {IEEE, New York, 2017})\ pp.\ \bibinfo {pages} {36.5.1--36.5.4}\BibitemShut
  {NoStop}%
\bibitem [{\citenamefont {Harvey-Collard}\ \emph {et~al.}(2019)\citenamefont
  {Harvey-Collard}, \citenamefont {Jacobson}, \citenamefont {Bureau-Oxton},
  \citenamefont {Jock}, \citenamefont {Srinivasa}, \citenamefont {Mounce},
  \citenamefont {Ward}, \citenamefont {Anderson}, \citenamefont {Manginell},
  \citenamefont {Wendt}, \citenamefont {Pluym}, \citenamefont {Lilly},
  \citenamefont {Luhman}, \citenamefont {Pioro-Ladri\`ere},\ and\ \citenamefont
  {Carroll}}]{Harvey.19}%
  \BibitemOpen
  \bibfield  {author} {\bibinfo {author} {\bibfnamefont {P.}~\bibnamefont
  {Harvey-Collard}}, \bibinfo {author} {\bibfnamefont {N.~T.}\ \bibnamefont
  {Jacobson}}, \bibinfo {author} {\bibfnamefont {C.}~\bibnamefont
  {Bureau-Oxton}}, \bibinfo {author} {\bibfnamefont {R.~M.}\ \bibnamefont
  {Jock}}, \bibinfo {author} {\bibfnamefont {V.}~\bibnamefont {Srinivasa}},
  \bibinfo {author} {\bibfnamefont {A.~M.}\ \bibnamefont {Mounce}}, \bibinfo
  {author} {\bibfnamefont {D.~R.}\ \bibnamefont {Ward}}, \bibinfo {author}
  {\bibfnamefont {J.~M.}\ \bibnamefont {Anderson}}, \bibinfo {author}
  {\bibfnamefont {R.~P.}\ \bibnamefont {Manginell}}, \bibinfo {author}
  {\bibfnamefont {J.~R.}\ \bibnamefont {Wendt}}, \bibinfo {author}
  {\bibfnamefont {T.}~\bibnamefont {Pluym}}, \bibinfo {author} {\bibfnamefont
  {M.~P.}\ \bibnamefont {Lilly}}, \bibinfo {author} {\bibfnamefont {D.~R.}\
  \bibnamefont {Luhman}}, \bibinfo {author} {\bibfnamefont {M.}~\bibnamefont
  {Pioro-Ladri\`ere}}, \ and\ \bibinfo {author} {\bibfnamefont {M.~S.}\
  \bibnamefont {Carroll}},\ }\href
  {https://link.aps.org/doi/10.1103/PhysRevLett.122.217702} {\bibfield
  {journal} {\bibinfo  {journal} {Phys. Rev. Lett.}\ }\textbf {\bibinfo
  {volume} {122}},\ \bibinfo {pages} {217702} (\bibinfo {year}
  {2019})}\BibitemShut {NoStop}%
\bibitem [{\citenamefont {Liu}\ \emph {et~al.}(2021)\citenamefont {Liu},
  \citenamefont {Orona}, \citenamefont {Neyens}, \citenamefont {MacQuarrie},
  \citenamefont {Eriksson},\ and\ \citenamefont {Yacoby}}]{Liu.21}%
  \BibitemOpen
  \bibfield  {author} {\bibinfo {author} {\bibfnamefont {Y.-Y.}\ \bibnamefont
  {Liu}}, \bibinfo {author} {\bibfnamefont {L.}~\bibnamefont {Orona}}, \bibinfo
  {author} {\bibfnamefont {S.~F.}\ \bibnamefont {Neyens}}, \bibinfo {author}
  {\bibfnamefont {E.}~\bibnamefont {MacQuarrie}}, \bibinfo {author}
  {\bibfnamefont {M.}~\bibnamefont {Eriksson}}, \ and\ \bibinfo {author}
  {\bibfnamefont {A.}~\bibnamefont {Yacoby}},\ }\href
  {https://link.aps.org/doi/10.1103/PhysRevApplied.16.024029} {\bibfield
  {journal} {\bibinfo  {journal} {Phys. Rev. Applied}\ }\textbf {\bibinfo
  {volume} {16}},\ \bibinfo {pages} {024029} (\bibinfo {year}
  {2021})}\BibitemShut {NoStop}%
\bibitem [{\citenamefont {Zhang}\ \emph {et~al.}(2018)\citenamefont {Zhang},
  \citenamefont {Yang},\ and\ \citenamefont {Wang}}]{Zhang.18}%
  \BibitemOpen
  \bibfield  {author} {\bibinfo {author} {\bibfnamefont {C.}~\bibnamefont
  {Zhang}}, \bibinfo {author} {\bibfnamefont {X.-C.}\ \bibnamefont {Yang}}, \
  and\ \bibinfo {author} {\bibfnamefont {X.}~\bibnamefont {Wang}},\ }\href
  {https://link.aps.org/doi/10.1103/PhysRevA.97.042326} {\bibfield  {journal}
  {\bibinfo  {journal} {Phys. Rev. A}\ }\textbf {\bibinfo {volume} {97}},\
  \bibinfo {pages} {042326} (\bibinfo {year} {2018})}\BibitemShut {NoStop}%
\bibitem [{\citenamefont {Li}\ \emph {et~al.}(2018)\citenamefont {Li},
  \citenamefont {Petit}, \citenamefont {Franke}, \citenamefont {Dehollain},
  \citenamefont {Helsen}, \citenamefont {Steudtner}, \citenamefont {Thomas},
  \citenamefont {Yoscovits}, \citenamefont {Singh}, \citenamefont {Wehner},
  \citenamefont {Vandersypen}, \citenamefont {Clarke},\ and\ \citenamefont
  {Veldhorst}}]{Li.18}%
  \BibitemOpen
  \bibfield  {author} {\bibinfo {author} {\bibfnamefont {R.}~\bibnamefont
  {Li}}, \bibinfo {author} {\bibfnamefont {L.}~\bibnamefont {Petit}}, \bibinfo
  {author} {\bibfnamefont {D.~P.}\ \bibnamefont {Franke}}, \bibinfo {author}
  {\bibfnamefont {J.~P.}\ \bibnamefont {Dehollain}}, \bibinfo {author}
  {\bibfnamefont {J.}~\bibnamefont {Helsen}}, \bibinfo {author} {\bibfnamefont
  {M.}~\bibnamefont {Steudtner}}, \bibinfo {author} {\bibfnamefont {N.~K.}\
  \bibnamefont {Thomas}}, \bibinfo {author} {\bibfnamefont {Z.~R.}\
  \bibnamefont {Yoscovits}}, \bibinfo {author} {\bibfnamefont {K.~J.}\
  \bibnamefont {Singh}}, \bibinfo {author} {\bibfnamefont {S.}~\bibnamefont
  {Wehner}}, \bibinfo {author} {\bibfnamefont {L.~M.~K.}\ \bibnamefont
  {Vandersypen}}, \bibinfo {author} {\bibfnamefont {J.~S.}\ \bibnamefont
  {Clarke}}, \ and\ \bibinfo {author} {\bibfnamefont {M.}~\bibnamefont
  {Veldhorst}},\ }\href
  {https://www.science.org/doi/abs/10.1126/sciadv.aar3960} {\bibfield
  {journal} {\bibinfo  {journal} {Sci. Adv.}\ }\textbf {\bibinfo {volume}
  {4}},\ \bibinfo {pages} {eaar3960} (\bibinfo {year} {2018})}\BibitemShut
  {NoStop}%
\bibitem [{\citenamefont {Jock}\ \emph {et~al.}(2022)\citenamefont {Jock},
  \citenamefont {Jacobson}, \citenamefont {Rudolph}, \citenamefont {Ward},
  \citenamefont {Carroll},\ and\ \citenamefont {Luhman}}]{Jock.21}%
  \BibitemOpen
  \bibfield  {author} {\bibinfo {author} {\bibfnamefont {R.~M.}\ \bibnamefont
  {Jock}}, \bibinfo {author} {\bibfnamefont {N.~T.}\ \bibnamefont {Jacobson}},
  \bibinfo {author} {\bibfnamefont {M.}~\bibnamefont {Rudolph}}, \bibinfo
  {author} {\bibfnamefont {D.~R.}\ \bibnamefont {Ward}}, \bibinfo {author}
  {\bibfnamefont {M.~S.}\ \bibnamefont {Carroll}}, \ and\ \bibinfo {author}
  {\bibfnamefont {D.~R.}\ \bibnamefont {Luhman}},\ }\href
  {https://doi.org/10.1038/s41467-022-28302-y} {\bibfield  {journal} {\bibinfo
  {journal} {Nat. Commun.}\ }\textbf {\bibinfo {volume} {13}},\ \bibinfo
  {pages} {641} (\bibinfo {year} {2022})}\BibitemShut {NoStop}%
\bibitem [{\citenamefont {Hendrickx}\ \emph {et~al.}(2021)\citenamefont
  {Hendrickx}, \citenamefont {Lawrie}, \citenamefont {Russ}, \citenamefont {van
  Riggelen}, \citenamefont {de~Snoo}, \citenamefont {Schouten}, \citenamefont
  {Sammak}, \citenamefont {Scappucci},\ and\ \citenamefont
  {Veldhorst}}]{Hendrickx.21}%
  \BibitemOpen
  \bibfield  {author} {\bibinfo {author} {\bibfnamefont {N.~W.}\ \bibnamefont
  {Hendrickx}}, \bibinfo {author} {\bibfnamefont {W.~I.~L.}\ \bibnamefont
  {Lawrie}}, \bibinfo {author} {\bibfnamefont {M.}~\bibnamefont {Russ}},
  \bibinfo {author} {\bibfnamefont {F.}~\bibnamefont {van Riggelen}}, \bibinfo
  {author} {\bibfnamefont {S.~L.}\ \bibnamefont {de~Snoo}}, \bibinfo {author}
  {\bibfnamefont {R.~N.}\ \bibnamefont {Schouten}}, \bibinfo {author}
  {\bibfnamefont {A.}~\bibnamefont {Sammak}}, \bibinfo {author} {\bibfnamefont
  {G.}~\bibnamefont {Scappucci}}, \ and\ \bibinfo {author} {\bibfnamefont
  {M.}~\bibnamefont {Veldhorst}},\ }\href
  {https://doi.org/10.1038/s41586-021-03332-6} {\bibfield  {journal} {\bibinfo
  {journal} {Nature}\ }\textbf {\bibinfo {volume} {591}},\ \bibinfo {pages}
  {580} (\bibinfo {year} {2021})}\BibitemShut {NoStop}%
\bibitem [{\citenamefont {Philips}\ \emph {et~al.}()\citenamefont {Philips},
  \citenamefont {M\k{a}dzik}, \citenamefont {Amitonov}, \citenamefont
  {de~Snoo}, \citenamefont {Russ}, \citenamefont {Kalhor}, \citenamefont
  {Volk}, \citenamefont {Lawrie}, \citenamefont {Brousse}, \citenamefont
  {Tryputen}, \citenamefont {Wuetz}, \citenamefont {Sammak}, \citenamefont
  {Veldhorst}, \citenamefont {Scappucci},\ and\ \citenamefont
  {Vandersypen}}]{Philips.22}%
  \BibitemOpen
  \bibfield  {author} {\bibinfo {author} {\bibfnamefont {S.~G.~J.}\
  \bibnamefont {Philips}}, \bibinfo {author} {\bibfnamefont {M.~T.}\
  \bibnamefont {M\k{a}dzik}}, \bibinfo {author} {\bibfnamefont {S.~V.}\
  \bibnamefont {Amitonov}}, \bibinfo {author} {\bibfnamefont {S.~L.}\
  \bibnamefont {de~Snoo}}, \bibinfo {author} {\bibfnamefont {M.}~\bibnamefont
  {Russ}}, \bibinfo {author} {\bibfnamefont {N.}~\bibnamefont {Kalhor}},
  \bibinfo {author} {\bibfnamefont {C.}~\bibnamefont {Volk}}, \bibinfo {author}
  {\bibfnamefont {W.~I.~L.}\ \bibnamefont {Lawrie}}, \bibinfo {author}
  {\bibfnamefont {D.}~\bibnamefont {Brousse}}, \bibinfo {author} {\bibfnamefont
  {L.}~\bibnamefont {Tryputen}}, \bibinfo {author} {\bibfnamefont {B.~P.}\
  \bibnamefont {Wuetz}}, \bibinfo {author} {\bibfnamefont {A.}~\bibnamefont
  {Sammak}}, \bibinfo {author} {\bibfnamefont {M.}~\bibnamefont {Veldhorst}},
  \bibinfo {author} {\bibfnamefont {G.}~\bibnamefont {Scappucci}}, \ and\
  \bibinfo {author} {\bibfnamefont {L.~M.~K.}\ \bibnamefont {Vandersypen}},\
  }\href {https://doi.org/10.48550/arXiv.2202.09252} {\bibinfo  {journal}
  {arXiv:2202.09252}\ }\BibitemShut {NoStop}%
\bibitem [{\citenamefont {Hayashi}\ \emph {et~al.}(2003)\citenamefont
  {Hayashi}, \citenamefont {Fujisawa}, \citenamefont {Cheong}, \citenamefont
  {Jeong},\ and\ \citenamefont {Hirayama}}]{Hayashi.03}%
  \BibitemOpen
\bibfield  {journal} {  }\bibfield  {author} {\bibinfo {author} {\bibfnamefont
  {T.}~\bibnamefont {Hayashi}}, \bibinfo {author} {\bibfnamefont
  {T.}~\bibnamefont {Fujisawa}}, \bibinfo {author} {\bibfnamefont {H.~D.}\
  \bibnamefont {Cheong}}, \bibinfo {author} {\bibfnamefont {Y.~H.}\
  \bibnamefont {Jeong}}, \ and\ \bibinfo {author} {\bibfnamefont
  {Y.}~\bibnamefont {Hirayama}},\ }\href
  {https://link.aps.org/doi/10.1103/PhysRevLett.91.226804} {\bibfield
  {journal} {\bibinfo  {journal} {Phys. Rev. Lett.}\ }\textbf {\bibinfo
  {volume} {91}},\ \bibinfo {pages} {226804} (\bibinfo {year}
  {2003})}\BibitemShut {NoStop}%
\bibitem [{\citenamefont {Gorman}\ \emph {et~al.}(2005)\citenamefont {Gorman},
  \citenamefont {Hasko},\ and\ \citenamefont {Williams}}]{Gorman.05}%
  \BibitemOpen
  \bibfield  {author} {\bibinfo {author} {\bibfnamefont {J.}~\bibnamefont
  {Gorman}}, \bibinfo {author} {\bibfnamefont {D.~G.}\ \bibnamefont {Hasko}}, \
  and\ \bibinfo {author} {\bibfnamefont {D.~A.}\ \bibnamefont {Williams}},\
  }\href {https://link.aps.org/doi/10.1103/PhysRevLett.95.090502} {\bibfield
  {journal} {\bibinfo  {journal} {Phys. Rev. Lett.}\ }\textbf {\bibinfo
  {volume} {95}},\ \bibinfo {pages} {090502} (\bibinfo {year}
  {2005})}\BibitemShut {NoStop}%
\bibitem [{\citenamefont {Shinkai}\ \emph {et~al.}(2009)\citenamefont
  {Shinkai}, \citenamefont {Hayashi}, \citenamefont {Ota},\ and\ \citenamefont
  {Fujisawa}}]{Shinkai.09}%
  \BibitemOpen
  \bibfield  {author} {\bibinfo {author} {\bibfnamefont {G.}~\bibnamefont
  {Shinkai}}, \bibinfo {author} {\bibfnamefont {T.}~\bibnamefont {Hayashi}},
  \bibinfo {author} {\bibfnamefont {T.}~\bibnamefont {Ota}}, \ and\ \bibinfo
  {author} {\bibfnamefont {T.}~\bibnamefont {Fujisawa}},\ }\href
  {https://link.aps.org/doi/10.1103/PhysRevLett.103.056802} {\bibfield
  {journal} {\bibinfo  {journal} {Phys. Rev. Lett.}\ }\textbf {\bibinfo
  {volume} {103}},\ \bibinfo {pages} {056802} (\bibinfo {year}
  {2009})}\BibitemShut {NoStop}%
\bibitem [{\citenamefont {Petersson}\ \emph {et~al.}(2010)\citenamefont
  {Petersson}, \citenamefont {Petta}, \citenamefont {Lu},\ and\ \citenamefont
  {Gossard}}]{Petersson.10}%
  \BibitemOpen
  \bibfield  {author} {\bibinfo {author} {\bibfnamefont {K.~D.}\ \bibnamefont
  {Petersson}}, \bibinfo {author} {\bibfnamefont {J.~R.}\ \bibnamefont
  {Petta}}, \bibinfo {author} {\bibfnamefont {H.}~\bibnamefont {Lu}}, \ and\
  \bibinfo {author} {\bibfnamefont {A.~C.}\ \bibnamefont {Gossard}},\ }\href
  {https://link.aps.org/doi/10.1103/PhysRevLett.105.246804} {\bibfield
  {journal} {\bibinfo  {journal} {Phys. Rev. Lett.}\ }\textbf {\bibinfo
  {volume} {105}},\ \bibinfo {pages} {246804} (\bibinfo {year}
  {2010})}\BibitemShut {NoStop}%
\bibitem [{\citenamefont {Shi}\ \emph {et~al.}(2013)\citenamefont {Shi},
  \citenamefont {Simmons}, \citenamefont {Ward}, \citenamefont {Prance},
  \citenamefont {Mohr}, \citenamefont {Koh}, \citenamefont {Gamble},
  \citenamefont {Wu}, \citenamefont {Savage}, \citenamefont {Lagally},
  \citenamefont {Friesen}, \citenamefont {Coppersmith},\ and\ \citenamefont
  {Eriksson}}]{Shi.13}%
  \BibitemOpen
  \bibfield  {author} {\bibinfo {author} {\bibfnamefont {Z.}~\bibnamefont
  {Shi}}, \bibinfo {author} {\bibfnamefont {C.~B.}\ \bibnamefont {Simmons}},
  \bibinfo {author} {\bibfnamefont {D.~R.}\ \bibnamefont {Ward}}, \bibinfo
  {author} {\bibfnamefont {J.~R.}\ \bibnamefont {Prance}}, \bibinfo {author}
  {\bibfnamefont {R.~T.}\ \bibnamefont {Mohr}}, \bibinfo {author}
  {\bibfnamefont {T.~S.}\ \bibnamefont {Koh}}, \bibinfo {author} {\bibfnamefont
  {J.~K.}\ \bibnamefont {Gamble}}, \bibinfo {author} {\bibfnamefont
  {X.}~\bibnamefont {Wu}}, \bibinfo {author} {\bibfnamefont {D.~E.}\
  \bibnamefont {Savage}}, \bibinfo {author} {\bibfnamefont {M.~G.}\
  \bibnamefont {Lagally}}, \bibinfo {author} {\bibfnamefont {M.}~\bibnamefont
  {Friesen}}, \bibinfo {author} {\bibfnamefont {S.~N.}\ \bibnamefont
  {Coppersmith}}, \ and\ \bibinfo {author} {\bibfnamefont {M.~A.}\ \bibnamefont
  {Eriksson}},\ }\href {https://link.aps.org/doi/10.1103/PhysRevB.88.075416}
  {\bibfield  {journal} {\bibinfo  {journal} {Phys. Rev. B}\ }\textbf {\bibinfo
  {volume} {88}},\ \bibinfo {pages} {075416} (\bibinfo {year}
  {2013})}\BibitemShut {NoStop}%
\bibitem [{\citenamefont {Dovzhenko}\ \emph {et~al.}(2011)\citenamefont
  {Dovzhenko}, \citenamefont {Stehlik}, \citenamefont {Petersson},
  \citenamefont {Petta}, \citenamefont {Lu},\ and\ \citenamefont
  {Gossard}}]{Dovzhenko.11}%
  \BibitemOpen
  \bibfield  {author} {\bibinfo {author} {\bibfnamefont {Y.}~\bibnamefont
  {Dovzhenko}}, \bibinfo {author} {\bibfnamefont {J.}~\bibnamefont {Stehlik}},
  \bibinfo {author} {\bibfnamefont {K.~D.}\ \bibnamefont {Petersson}}, \bibinfo
  {author} {\bibfnamefont {J.~R.}\ \bibnamefont {Petta}}, \bibinfo {author}
  {\bibfnamefont {H.}~\bibnamefont {Lu}}, \ and\ \bibinfo {author}
  {\bibfnamefont {A.~C.}\ \bibnamefont {Gossard}},\ }\href
  {https://link.aps.org/doi/10.1103/PhysRevB.84.161302} {\bibfield  {journal}
  {\bibinfo  {journal} {Phys. Rev. B}\ }\textbf {\bibinfo {volume} {84}},\
  \bibinfo {pages} {161302} (\bibinfo {year} {2011})}\BibitemShut {NoStop}%
\bibitem [{\citenamefont {Stehlik}\ \emph {et~al.}(2012)\citenamefont
  {Stehlik}, \citenamefont {Dovzhenko}, \citenamefont {Petta}, \citenamefont
  {Johansson}, \citenamefont {Nori}, \citenamefont {Lu},\ and\ \citenamefont
  {Gossard}}]{Landau.12}%
  \BibitemOpen
  \bibfield  {author} {\bibinfo {author} {\bibfnamefont {J.}~\bibnamefont
  {Stehlik}}, \bibinfo {author} {\bibfnamefont {Y.}~\bibnamefont {Dovzhenko}},
  \bibinfo {author} {\bibfnamefont {J.~R.}\ \bibnamefont {Petta}}, \bibinfo
  {author} {\bibfnamefont {J.~R.}\ \bibnamefont {Johansson}}, \bibinfo {author}
  {\bibfnamefont {F.}~\bibnamefont {Nori}}, \bibinfo {author} {\bibfnamefont
  {H.}~\bibnamefont {Lu}}, \ and\ \bibinfo {author} {\bibfnamefont {A.~C.}\
  \bibnamefont {Gossard}},\ }\href
  {https://link.aps.org/doi/10.1103/PhysRevB.86.121303} {\bibfield  {journal}
  {\bibinfo  {journal} {Phys. Rev. B}\ }\textbf {\bibinfo {volume} {86}},\
  \bibinfo {pages} {121303} (\bibinfo {year} {2012})}\BibitemShut {NoStop}%
\bibitem [{\citenamefont {Li}\ \emph {et~al.}(2015)\citenamefont {Li},
  \citenamefont {Cao}, \citenamefont {Yu}, \citenamefont {Xiao}, \citenamefont
  {Guo}, \citenamefont {Jiang},\ and\ \citenamefont {Guo}}]{Li.15}%
  \BibitemOpen
  \bibfield  {author} {\bibinfo {author} {\bibfnamefont {H.-O.}\ \bibnamefont
  {Li}}, \bibinfo {author} {\bibfnamefont {G.}~\bibnamefont {Cao}}, \bibinfo
  {author} {\bibfnamefont {G.-D.}\ \bibnamefont {Yu}}, \bibinfo {author}
  {\bibfnamefont {M.}~\bibnamefont {Xiao}}, \bibinfo {author} {\bibfnamefont
  {G.-C.}\ \bibnamefont {Guo}}, \bibinfo {author} {\bibfnamefont {H.-W.}\
  \bibnamefont {Jiang}}, \ and\ \bibinfo {author} {\bibfnamefont {G.-P.}\
  \bibnamefont {Guo}},\ }\href {https://doi.org/10.1038/ncomms8681} {\bibfield
  {journal} {\bibinfo  {journal} {Nat. Commun.}\ }\textbf {\bibinfo {volume}
  {6}},\ \bibinfo {pages} {7681} (\bibinfo {year} {2015})}\BibitemShut
  {NoStop}%
\bibitem [{\citenamefont {Shi}\ \emph {et~al.}(2014)\citenamefont {Shi},
  \citenamefont {Simmons}, \citenamefont {Ward}, \citenamefont {Prance},
  \citenamefont {Wu}, \citenamefont {Koh}, \citenamefont {Gamble},
  \citenamefont {Savage}, \citenamefont {Lagally}, \citenamefont {Friesen},
  \citenamefont {Coppersmith},\ and\ \citenamefont {Eriksson}}]{Shi.14}%
  \BibitemOpen
  \bibfield  {author} {\bibinfo {author} {\bibfnamefont {Z.}~\bibnamefont
  {Shi}}, \bibinfo {author} {\bibfnamefont {C.~B.}\ \bibnamefont {Simmons}},
  \bibinfo {author} {\bibfnamefont {D.~R.}\ \bibnamefont {Ward}}, \bibinfo
  {author} {\bibfnamefont {J.~R.}\ \bibnamefont {Prance}}, \bibinfo {author}
  {\bibfnamefont {X.}~\bibnamefont {Wu}}, \bibinfo {author} {\bibfnamefont
  {T.~S.}\ \bibnamefont {Koh}}, \bibinfo {author} {\bibfnamefont {J.~K.}\
  \bibnamefont {Gamble}}, \bibinfo {author} {\bibfnamefont {D.~E.}\
  \bibnamefont {Savage}}, \bibinfo {author} {\bibfnamefont {M.~G.}\
  \bibnamefont {Lagally}}, \bibinfo {author} {\bibfnamefont {M.}~\bibnamefont
  {Friesen}}, \bibinfo {author} {\bibfnamefont {S.~N.}\ \bibnamefont
  {Coppersmith}}, \ and\ \bibinfo {author} {\bibfnamefont {M.~A.}\ \bibnamefont
  {Eriksson}},\ }\href {https://doi.org/10.1038/ncomms4020} {\bibfield
  {journal} {\bibinfo  {journal} {Nat. Commun.}\ }\textbf {\bibinfo {volume}
  {5}},\ \bibinfo {pages} {3020} (\bibinfo {year} {2014})}\BibitemShut
  {NoStop}%
\bibitem [{\citenamefont {Kim}\ \emph {et~al.}(2015)\citenamefont {Kim},
  \citenamefont {Ward}, \citenamefont {Simmons}, \citenamefont {Gamble},
  \citenamefont {Blume-Kohout}, \citenamefont {Nielsen}, \citenamefont
  {Savage}, \citenamefont {Lagally}, \citenamefont {Friesen}, \citenamefont
  {Coppersmith},\ and\ \citenamefont {Eriksson}}]{Kim.15}%
  \BibitemOpen
  \bibfield  {author} {\bibinfo {author} {\bibfnamefont {D.}~\bibnamefont
  {Kim}}, \bibinfo {author} {\bibfnamefont {D.~R.}\ \bibnamefont {Ward}},
  \bibinfo {author} {\bibfnamefont {C.~B.}\ \bibnamefont {Simmons}}, \bibinfo
  {author} {\bibfnamefont {J.~K.}\ \bibnamefont {Gamble}}, \bibinfo {author}
  {\bibfnamefont {R.}~\bibnamefont {Blume-Kohout}}, \bibinfo {author}
  {\bibfnamefont {E.}~\bibnamefont {Nielsen}}, \bibinfo {author} {\bibfnamefont
  {D.~E.}\ \bibnamefont {Savage}}, \bibinfo {author} {\bibfnamefont {M.~G.}\
  \bibnamefont {Lagally}}, \bibinfo {author} {\bibfnamefont {M.}~\bibnamefont
  {Friesen}}, \bibinfo {author} {\bibfnamefont {S.~N.}\ \bibnamefont
  {Coppersmith}}, \ and\ \bibinfo {author} {\bibfnamefont {M.~A.}\ \bibnamefont
  {Eriksson}},\ }\href {https://doi.org/10.1038/nnano.2014.336} {\bibfield
  {journal} {\bibinfo  {journal} {Nat. Nanotechnol.}\ }\textbf {\bibinfo
  {volume} {10}},\ \bibinfo {pages} {243} (\bibinfo {year} {2015})}\BibitemShut
  {NoStop}%
\bibitem [{\citenamefont {Cao}\ \emph {et~al.}(2016)\citenamefont {Cao},
  \citenamefont {Li}, \citenamefont {Yu}, \citenamefont {Wang}, \citenamefont
  {Chen}, \citenamefont {Song}, \citenamefont {Xiao}, \citenamefont {Guo},
  \citenamefont {Jiang}, \citenamefont {Hu},\ and\ \citenamefont
  {Guo}}]{Cao.16}%
  \BibitemOpen
  \bibfield  {author} {\bibinfo {author} {\bibfnamefont {G.}~\bibnamefont
  {Cao}}, \bibinfo {author} {\bibfnamefont {H.-O.}\ \bibnamefont {Li}},
  \bibinfo {author} {\bibfnamefont {G.-D.}\ \bibnamefont {Yu}}, \bibinfo
  {author} {\bibfnamefont {B.-C.}\ \bibnamefont {Wang}}, \bibinfo {author}
  {\bibfnamefont {B.-B.}\ \bibnamefont {Chen}}, \bibinfo {author}
  {\bibfnamefont {X.-X.}\ \bibnamefont {Song}}, \bibinfo {author}
  {\bibfnamefont {M.}~\bibnamefont {Xiao}}, \bibinfo {author} {\bibfnamefont
  {G.-C.}\ \bibnamefont {Guo}}, \bibinfo {author} {\bibfnamefont {H.-W.}\
  \bibnamefont {Jiang}}, \bibinfo {author} {\bibfnamefont {X.}~\bibnamefont
  {Hu}}, \ and\ \bibinfo {author} {\bibfnamefont {G.-P.}\ \bibnamefont {Guo}},\
  }\href {https://link.aps.org/doi/10.1103/PhysRevLett.116.086801} {\bibfield
  {journal} {\bibinfo  {journal} {Phys. Rev. Lett.}\ }\textbf {\bibinfo
  {volume} {116}},\ \bibinfo {pages} {086801} (\bibinfo {year}
  {2016})}\BibitemShut {NoStop}%
\bibitem [{\citenamefont {Thorgrimsson}\ \emph {et~al.}(2017)\citenamefont
  {Thorgrimsson}, \citenamefont {Kim}, \citenamefont {Yang}, \citenamefont
  {Smith}, \citenamefont {Simmons}, \citenamefont {Ward}, \citenamefont
  {Foote}, \citenamefont {Corrigan}, \citenamefont {Savage}, \citenamefont
  {Lagally}, \citenamefont {Friesen}, \citenamefont {Coppersmith},\ and\
  \citenamefont {Eriksson}}]{Thorgrimsson.17}%
  \BibitemOpen
  \bibfield  {author} {\bibinfo {author} {\bibfnamefont {B.}~\bibnamefont
  {Thorgrimsson}}, \bibinfo {author} {\bibfnamefont {D.}~\bibnamefont {Kim}},
  \bibinfo {author} {\bibfnamefont {Y.-C.}\ \bibnamefont {Yang}}, \bibinfo
  {author} {\bibfnamefont {L.~W.}\ \bibnamefont {Smith}}, \bibinfo {author}
  {\bibfnamefont {C.~B.}\ \bibnamefont {Simmons}}, \bibinfo {author}
  {\bibfnamefont {D.~R.}\ \bibnamefont {Ward}}, \bibinfo {author}
  {\bibfnamefont {R.~H.}\ \bibnamefont {Foote}}, \bibinfo {author}
  {\bibfnamefont {J.}~\bibnamefont {Corrigan}}, \bibinfo {author}
  {\bibfnamefont {D.~E.}\ \bibnamefont {Savage}}, \bibinfo {author}
  {\bibfnamefont {M.~G.}\ \bibnamefont {Lagally}}, \bibinfo {author}
  {\bibfnamefont {M.}~\bibnamefont {Friesen}}, \bibinfo {author} {\bibfnamefont
  {S.~N.}\ \bibnamefont {Coppersmith}}, \ and\ \bibinfo {author} {\bibfnamefont
  {M.~A.}\ \bibnamefont {Eriksson}},\ }\href
  {https://doi.org/10.1038/s41534-017-0034-2} {\bibfield  {journal} {\bibinfo
  {journal} {npj Quantum Inf.}\ }\textbf {\bibinfo {volume} {3}},\ \bibinfo
  {pages} {32} (\bibinfo {year} {2017})}\BibitemShut {NoStop}%
\bibitem [{\citenamefont {Malinowski}\ \emph {et~al.}(2017)\citenamefont
  {Malinowski}, \citenamefont {Martins}, \citenamefont {Nissen}, \citenamefont
  {Fallahi}, \citenamefont {Gardner}, \citenamefont {Manfra}, \citenamefont
  {Marcus},\ and\ \citenamefont {Kuemmeth}}]{Malinowski.17}%
  \BibitemOpen
  \bibfield  {author} {\bibinfo {author} {\bibfnamefont {F.~K.}\ \bibnamefont
  {Malinowski}}, \bibinfo {author} {\bibfnamefont {F.}~\bibnamefont {Martins}},
  \bibinfo {author} {\bibfnamefont {P.~D.}\ \bibnamefont {Nissen}}, \bibinfo
  {author} {\bibfnamefont {S.}~\bibnamefont {Fallahi}}, \bibinfo {author}
  {\bibfnamefont {G.~C.}\ \bibnamefont {Gardner}}, \bibinfo {author}
  {\bibfnamefont {M.~J.}\ \bibnamefont {Manfra}}, \bibinfo {author}
  {\bibfnamefont {C.~M.}\ \bibnamefont {Marcus}}, \ and\ \bibinfo {author}
  {\bibfnamefont {F.}~\bibnamefont {Kuemmeth}},\ }\href
  {https://link.aps.org/doi/10.1103/PhysRevB.96.045443} {\bibfield  {journal}
  {\bibinfo  {journal} {Phys. Rev. B}\ }\textbf {\bibinfo {volume} {96}},\
  \bibinfo {pages} {045443} (\bibinfo {year} {2017})}\BibitemShut {NoStop}%
\bibitem [{\citenamefont {Cao}\ \emph {et~al.}(2013)\citenamefont {Cao},
  \citenamefont {Li}, \citenamefont {Tu}, \citenamefont {Wang}, \citenamefont
  {Zhou}, \citenamefont {Xiao}, \citenamefont {Guo}, \citenamefont {Jiang},\
  and\ \citenamefont {Guo}}]{Cao.13}%
  \BibitemOpen
  \bibfield  {author} {\bibinfo {author} {\bibfnamefont {G.}~\bibnamefont
  {Cao}}, \bibinfo {author} {\bibfnamefont {H.-O.}\ \bibnamefont {Li}},
  \bibinfo {author} {\bibfnamefont {T.}~\bibnamefont {Tu}}, \bibinfo {author}
  {\bibfnamefont {L.}~\bibnamefont {Wang}}, \bibinfo {author} {\bibfnamefont
  {C.}~\bibnamefont {Zhou}}, \bibinfo {author} {\bibfnamefont {M.}~\bibnamefont
  {Xiao}}, \bibinfo {author} {\bibfnamefont {G.-C.}\ \bibnamefont {Guo}},
  \bibinfo {author} {\bibfnamefont {H.-W.}\ \bibnamefont {Jiang}}, \ and\
  \bibinfo {author} {\bibfnamefont {G.-P.}\ \bibnamefont {Guo}},\ }\href
  {https://doi.org/10.1038/ncomms2412} {\bibfield  {journal} {\bibinfo
  {journal} {Nat. Commun.}\ }\textbf {\bibinfo {volume} {4}},\ \bibinfo {pages}
  {1401} (\bibinfo {year} {2013})}\BibitemShut {NoStop}%
\bibitem [{\citenamefont {Dial}\ \emph {et~al.}(2013)\citenamefont {Dial},
  \citenamefont {Shulman}, \citenamefont {Harvey}, \citenamefont {Bluhm},
  \citenamefont {Umansky},\ and\ \citenamefont {Yacoby}}]{Dial.13}%
  \BibitemOpen
  \bibfield  {author} {\bibinfo {author} {\bibfnamefont {O.~E.}\ \bibnamefont
  {Dial}}, \bibinfo {author} {\bibfnamefont {M.~D.}\ \bibnamefont {Shulman}},
  \bibinfo {author} {\bibfnamefont {S.~P.}\ \bibnamefont {Harvey}}, \bibinfo
  {author} {\bibfnamefont {H.}~\bibnamefont {Bluhm}}, \bibinfo {author}
  {\bibfnamefont {V.}~\bibnamefont {Umansky}}, \ and\ \bibinfo {author}
  {\bibfnamefont {A.}~\bibnamefont {Yacoby}},\ }\href
  {https://link.aps.org/doi/10.1103/PhysRevLett.110.146804} {\bibfield
  {journal} {\bibinfo  {journal} {Phys. Rev. Lett.}\ }\textbf {\bibinfo
  {volume} {110}},\ \bibinfo {pages} {146804} (\bibinfo {year}
  {2013})}\BibitemShut {NoStop}%
\bibitem [{\citenamefont {Huang}\ \emph {et~al.}(2019)\citenamefont {Huang},
  \citenamefont {Yang}, \citenamefont {Chan}, \citenamefont {Tanttu},
  \citenamefont {Hensen}, \citenamefont {Leon}, \citenamefont {Fogarty},
  \citenamefont {Hwang}, \citenamefont {Hudson}, \citenamefont {Itoh},
  \citenamefont {Morello}, \citenamefont {Laucht},\ and\ \citenamefont
  {Dzurak}}]{Huang.19}%
  \BibitemOpen
  \bibfield  {author} {\bibinfo {author} {\bibfnamefont {W.}~\bibnamefont
  {Huang}}, \bibinfo {author} {\bibfnamefont {C.~H.}\ \bibnamefont {Yang}},
  \bibinfo {author} {\bibfnamefont {K.~W.}\ \bibnamefont {Chan}}, \bibinfo
  {author} {\bibfnamefont {T.}~\bibnamefont {Tanttu}}, \bibinfo {author}
  {\bibfnamefont {B.}~\bibnamefont {Hensen}}, \bibinfo {author} {\bibfnamefont
  {R.~C.~C.}\ \bibnamefont {Leon}}, \bibinfo {author} {\bibfnamefont {M.~A.}\
  \bibnamefont {Fogarty}}, \bibinfo {author} {\bibfnamefont {J.~C.~C.}\
  \bibnamefont {Hwang}}, \bibinfo {author} {\bibfnamefont {F.~E.}\ \bibnamefont
  {Hudson}}, \bibinfo {author} {\bibfnamefont {K.~M.}\ \bibnamefont {Itoh}},
  \bibinfo {author} {\bibfnamefont {A.}~\bibnamefont {Morello}}, \bibinfo
  {author} {\bibfnamefont {A.}~\bibnamefont {Laucht}}, \ and\ \bibinfo {author}
  {\bibfnamefont {A.~S.}\ \bibnamefont {Dzurak}},\ }\href
  {https://doi.org/10.1038/s41586-019-1197-0} {\bibfield  {journal} {\bibinfo
  {journal} {Nature}\ }\textbf {\bibinfo {volume} {569}},\ \bibinfo {pages}
  {532} (\bibinfo {year} {2019})}\BibitemShut {NoStop}%
\bibitem [{\citenamefont {Barthel}\ \emph {et~al.}(2010)\citenamefont
  {Barthel}, \citenamefont {Medford}, \citenamefont {Marcus}, \citenamefont
  {Hanson},\ and\ \citenamefont {Gossard}}]{Barthel.10}%
  \BibitemOpen
  \bibfield  {author} {\bibinfo {author} {\bibfnamefont {C.}~\bibnamefont
  {Barthel}}, \bibinfo {author} {\bibfnamefont {J.}~\bibnamefont {Medford}},
  \bibinfo {author} {\bibfnamefont {C.~M.}\ \bibnamefont {Marcus}}, \bibinfo
  {author} {\bibfnamefont {M.~P.}\ \bibnamefont {Hanson}}, \ and\ \bibinfo
  {author} {\bibfnamefont {A.~C.}\ \bibnamefont {Gossard}},\ }\href
  {https://link.aps.org/doi/10.1103/PhysRevLett.105.266808} {\bibfield
  {journal} {\bibinfo  {journal} {Phys. Rev. Lett.}\ }\textbf {\bibinfo
  {volume} {105}},\ \bibinfo {pages} {266808} (\bibinfo {year}
  {2010})}\BibitemShut {NoStop}%
\bibitem [{\citenamefont {Terhal}(2015)}]{Terhal.15}%
  \BibitemOpen
  \bibfield  {author} {\bibinfo {author} {\bibfnamefont {B.~M.}\ \bibnamefont
  {Terhal}},\ }\href {https://link.aps.org/doi/10.1103/RevModPhys.87.307}
  {\bibfield  {journal} {\bibinfo  {journal} {Rev. Mod. Phys.}\ }\textbf
  {\bibinfo {volume} {87}},\ \bibinfo {pages} {307} (\bibinfo {year}
  {2015})}\BibitemShut {NoStop}%
\bibitem [{\citenamefont {Yang}\ and\ \citenamefont {Wang}(2017)}]{Yang.17}%
  \BibitemOpen
  \bibfield  {author} {\bibinfo {author} {\bibfnamefont {X.-C.}\ \bibnamefont
  {Yang}}\ and\ \bibinfo {author} {\bibfnamefont {X.}~\bibnamefont {Wang}},\
  }\href {https://link.aps.org/doi/10.1103/PhysRevA.96.012318} {\bibfield
  {journal} {\bibinfo  {journal} {Phys. Rev. A}\ }\textbf {\bibinfo {volume}
  {96}},\ \bibinfo {pages} {012318} (\bibinfo {year} {2017})}\BibitemShut
  {NoStop}%
\bibitem [{\citenamefont {Yang}\ and\ \citenamefont {Wang}(2018)}]{Yang.18}%
  \BibitemOpen
  \bibfield  {author} {\bibinfo {author} {\bibfnamefont {X.-C.}\ \bibnamefont
  {Yang}}\ and\ \bibinfo {author} {\bibfnamefont {X.}~\bibnamefont {Wang}},\
  }\href {https://link.aps.org/doi/10.1103/PhysRevA.97.012304} {\bibfield
  {journal} {\bibinfo  {journal} {Phys. Rev. A}\ }\textbf {\bibinfo {volume}
  {97}},\ \bibinfo {pages} {012304} (\bibinfo {year} {2018})}\BibitemShut
  {NoStop}%
\bibitem [{\citenamefont {Shim}\ and\ \citenamefont {Tahan}(2018)}]{Shim.18}%
  \BibitemOpen
  \bibfield  {author} {\bibinfo {author} {\bibfnamefont {Y.-P.}\ \bibnamefont
  {Shim}}\ and\ \bibinfo {author} {\bibfnamefont {C.}~\bibnamefont {Tahan}},\
  }\href {https://link.aps.org/doi/10.1103/PhysRevB.97.155402} {\bibfield
  {journal} {\bibinfo  {journal} {Phys. Rev. B}\ }\textbf {\bibinfo {volume}
  {97}},\ \bibinfo {pages} {155402} (\bibinfo {year} {2018})}\BibitemShut
  {NoStop}%
\bibitem [{\citenamefont {Martins}\ \emph {et~al.}(2017)\citenamefont
  {Martins}, \citenamefont {Malinowski}, \citenamefont {Nissen}, \citenamefont
  {Fallahi}, \citenamefont {Gardner}, \citenamefont {Manfra}, \citenamefont
  {Marcus},\ and\ \citenamefont {Kuemmeth}}]{Martins.17}%
  \BibitemOpen
  \bibfield  {author} {\bibinfo {author} {\bibfnamefont {F.}~\bibnamefont
  {Martins}}, \bibinfo {author} {\bibfnamefont {F.~K.}\ \bibnamefont
  {Malinowski}}, \bibinfo {author} {\bibfnamefont {P.~D.}\ \bibnamefont
  {Nissen}}, \bibinfo {author} {\bibfnamefont {S.}~\bibnamefont {Fallahi}},
  \bibinfo {author} {\bibfnamefont {G.~C.}\ \bibnamefont {Gardner}}, \bibinfo
  {author} {\bibfnamefont {M.~J.}\ \bibnamefont {Manfra}}, \bibinfo {author}
  {\bibfnamefont {C.~M.}\ \bibnamefont {Marcus}}, \ and\ \bibinfo {author}
  {\bibfnamefont {F.}~\bibnamefont {Kuemmeth}},\ }\href
  {https://link.aps.org/doi/10.1103/PhysRevLett.119.227701} {\bibfield
  {journal} {\bibinfo  {journal} {Phys. Rev. Lett.}\ }\textbf {\bibinfo
  {volume} {119}},\ \bibinfo {pages} {227701} (\bibinfo {year}
  {2017})}\BibitemShut {NoStop}%
\bibitem [{\citenamefont {Malinowski}\ \emph {et~al.}(2018)\citenamefont
  {Malinowski}, \citenamefont {Martins}, \citenamefont {Smith}, \citenamefont
  {Bartlett}, \citenamefont {Doherty}, \citenamefont {Nissen}, \citenamefont
  {Fallahi}, \citenamefont {Gardner}, \citenamefont {Manfra}, \citenamefont
  {Marcus},\ and\ \citenamefont {Kuemmeth}}]{Malinowski.18}%
  \BibitemOpen
  \bibfield  {author} {\bibinfo {author} {\bibfnamefont {F.~K.}\ \bibnamefont
  {Malinowski}}, \bibinfo {author} {\bibfnamefont {F.}~\bibnamefont {Martins}},
  \bibinfo {author} {\bibfnamefont {T.~B.}\ \bibnamefont {Smith}}, \bibinfo
  {author} {\bibfnamefont {S.~D.}\ \bibnamefont {Bartlett}}, \bibinfo {author}
  {\bibfnamefont {A.~C.}\ \bibnamefont {Doherty}}, \bibinfo {author}
  {\bibfnamefont {P.~D.}\ \bibnamefont {Nissen}}, \bibinfo {author}
  {\bibfnamefont {S.}~\bibnamefont {Fallahi}}, \bibinfo {author} {\bibfnamefont
  {G.~C.}\ \bibnamefont {Gardner}}, \bibinfo {author} {\bibfnamefont {M.~J.}\
  \bibnamefont {Manfra}}, \bibinfo {author} {\bibfnamefont {C.~M.}\
  \bibnamefont {Marcus}}, \ and\ \bibinfo {author} {\bibfnamefont
  {F.}~\bibnamefont {Kuemmeth}},\ }\href
  {https://link.aps.org/doi/10.1103/PhysRevX.8.011045} {\bibfield  {journal}
  {\bibinfo  {journal} {Phys. Rev. X}\ }\textbf {\bibinfo {volume} {8}},\
  \bibinfo {pages} {011045} (\bibinfo {year} {2018})}\BibitemShut {NoStop}%
\bibitem [{\citenamefont {Barnes}\ \emph {et~al.}(2011)\citenamefont {Barnes},
  \citenamefont {Kestner}, \citenamefont {Nguyen},\ and\ \citenamefont
  {Das~Sarma}}]{Barnes.11}%
  \BibitemOpen
  \bibfield  {author} {\bibinfo {author} {\bibfnamefont {E.}~\bibnamefont
  {Barnes}}, \bibinfo {author} {\bibfnamefont {J.~P.}\ \bibnamefont {Kestner}},
  \bibinfo {author} {\bibfnamefont {N.~T.~T.}\ \bibnamefont {Nguyen}}, \ and\
  \bibinfo {author} {\bibfnamefont {S.}~\bibnamefont {Das~Sarma}},\ }\href
  {https://link.aps.org/doi/10.1103/PhysRevB.84.235309} {\bibfield  {journal}
  {\bibinfo  {journal} {Phys. Rev. B}\ }\textbf {\bibinfo {volume} {84}},\
  \bibinfo {pages} {235309} (\bibinfo {year} {2011})}\BibitemShut {NoStop}%
\bibitem [{\citenamefont {Deng}\ \emph {et~al.}(2018)\citenamefont {Deng},
  \citenamefont {Calderon-Vargas}, \citenamefont {Mayhall},\ and\ \citenamefont
  {Barnes}}]{Deng.18}%
  \BibitemOpen
  \bibfield  {author} {\bibinfo {author} {\bibfnamefont {K.}~\bibnamefont
  {Deng}}, \bibinfo {author} {\bibfnamefont {F.~A.}\ \bibnamefont
  {Calderon-Vargas}}, \bibinfo {author} {\bibfnamefont {N.~J.}\ \bibnamefont
  {Mayhall}}, \ and\ \bibinfo {author} {\bibfnamefont {E.}~\bibnamefont
  {Barnes}},\ }\href {https://link.aps.org/doi/10.1103/PhysRevB.97.245301}
  {\bibfield  {journal} {\bibinfo  {journal} {Phys. Rev. B}\ }\textbf {\bibinfo
  {volume} {97}},\ \bibinfo {pages} {245301} (\bibinfo {year}
  {2018})}\BibitemShut {NoStop}%
\bibitem [{\citenamefont {Chan}\ and\ \citenamefont {Wang}()}]{ChanGX.22.1}%
  \BibitemOpen
  \bibfield  {author} {\bibinfo {author} {\bibfnamefont {G.~X.}\ \bibnamefont
  {Chan}}\ and\ \bibinfo {author} {\bibfnamefont {X.}~\bibnamefont {Wang}},\
  }\href {https://arxiv.org/abs/2201.01583} {\bibinfo  {journal}
  {arXiv:2201.01583}\ }\BibitemShut {NoStop}%
\bibitem [{\citenamefont {Bryant}(1987)}]{Bryant.87}%
  \BibitemOpen
\bibfield  {journal} {  }\bibfield  {author} {\bibinfo {author} {\bibfnamefont
  {G.~W.}\ \bibnamefont {Bryant}},\ }\href
  {https://link.aps.org/doi/10.1103/PhysRevLett.59.1140} {\bibfield  {journal}
  {\bibinfo  {journal} {Phys. Rev. Lett.}\ }\textbf {\bibinfo {volume} {59}},\
  \bibinfo {pages} {1140} (\bibinfo {year} {1987})}\BibitemShut {NoStop}%
\bibitem [{\citenamefont {Yannouleas}\ and\ \citenamefont
  {Landman}(1999)}]{Yannouleas.99}%
  \BibitemOpen
  \bibfield  {author} {\bibinfo {author} {\bibfnamefont {C.}~\bibnamefont
  {Yannouleas}}\ and\ \bibinfo {author} {\bibfnamefont {U.}~\bibnamefont
  {Landman}},\ }\href {https://link.aps.org/doi/10.1103/PhysRevLett.82.5325}
  {\bibfield  {journal} {\bibinfo  {journal} {Phys. Rev. Lett.}\ }\textbf
  {\bibinfo {volume} {82}},\ \bibinfo {pages} {5325} (\bibinfo {year}
  {1999})}\BibitemShut {NoStop}%
\bibitem [{\citenamefont {Cioslowski}\ and\ \citenamefont
  {Pernal}(2000)}]{Cioslowski.00}%
  \BibitemOpen
  \bibfield  {author} {\bibinfo {author} {\bibfnamefont {J.}~\bibnamefont
  {Cioslowski}}\ and\ \bibinfo {author} {\bibfnamefont {K.}~\bibnamefont
  {Pernal}},\ }\href {https://doi.org/10.1063/1.1318767} {\bibfield  {journal}
  {\bibinfo  {journal} {The Journal of Chemical Physics}\ }\textbf {\bibinfo
  {volume} {113}},\ \bibinfo {pages} {8434} (\bibinfo {year}
  {2000})}\BibitemShut {NoStop}%
\bibitem [{\citenamefont {Reimann}\ \emph {et~al.}(2000)\citenamefont
  {Reimann}, \citenamefont {Koskinen},\ and\ \citenamefont
  {Manninen}}]{Reimann.00}%
  \BibitemOpen
  \bibfield  {author} {\bibinfo {author} {\bibfnamefont {S.~M.}\ \bibnamefont
  {Reimann}}, \bibinfo {author} {\bibfnamefont {M.}~\bibnamefont {Koskinen}}, \
  and\ \bibinfo {author} {\bibfnamefont {M.}~\bibnamefont {Manninen}},\ }\href
  {https://link.aps.org/doi/10.1103/PhysRevB.62.8108} {\bibfield  {journal}
  {\bibinfo  {journal} {Phys. Rev. B}\ }\textbf {\bibinfo {volume} {62}},\
  \bibinfo {pages} {8108} (\bibinfo {year} {2000})}\BibitemShut {NoStop}%
\bibitem [{\citenamefont {Filinov}\ \emph {et~al.}(2001)\citenamefont
  {Filinov}, \citenamefont {Bonitz},\ and\ \citenamefont
  {Lozovik}}]{Filinov.01}%
  \BibitemOpen
  \bibfield  {author} {\bibinfo {author} {\bibfnamefont {A.~V.}\ \bibnamefont
  {Filinov}}, \bibinfo {author} {\bibfnamefont {M.}~\bibnamefont {Bonitz}}, \
  and\ \bibinfo {author} {\bibfnamefont {Y.~E.}\ \bibnamefont {Lozovik}},\
  }\href {https://link.aps.org/doi/10.1103/PhysRevLett.86.3851} {\bibfield
  {journal} {\bibinfo  {journal} {Phys. Rev. Lett.}\ }\textbf {\bibinfo
  {volume} {86}},\ \bibinfo {pages} {3851} (\bibinfo {year}
  {2001})}\BibitemShut {NoStop}%
\bibitem [{\citenamefont {Reusch}\ \emph {et~al.}(2001)\citenamefont {Reusch},
  \citenamefont {H\"ausler},\ and\ \citenamefont {Grabert}}]{Reusch.01}%
  \BibitemOpen
  \bibfield  {author} {\bibinfo {author} {\bibfnamefont {B.}~\bibnamefont
  {Reusch}}, \bibinfo {author} {\bibfnamefont {W.}~\bibnamefont {H\"ausler}}, \
  and\ \bibinfo {author} {\bibfnamefont {H.}~\bibnamefont {Grabert}},\ }\href
  {https://link.aps.org/doi/10.1103/PhysRevB.63.113313} {\bibfield  {journal}
  {\bibinfo  {journal} {Phys. Rev. B}\ }\textbf {\bibinfo {volume} {63}},\
  \bibinfo {pages} {113313} (\bibinfo {year} {2001})}\BibitemShut {NoStop}%
\bibitem [{\citenamefont {Szafran}\ \emph {et~al.}(2004)\citenamefont
  {Szafran}, \citenamefont {Peeters}, \citenamefont {Bednarek}, \citenamefont
  {Chwiej},\ and\ \citenamefont {Adamowski}}]{Szafran.04}%
  \BibitemOpen
  \bibfield  {author} {\bibinfo {author} {\bibfnamefont {B.}~\bibnamefont
  {Szafran}}, \bibinfo {author} {\bibfnamefont {F.~M.}\ \bibnamefont
  {Peeters}}, \bibinfo {author} {\bibfnamefont {S.}~\bibnamefont {Bednarek}},
  \bibinfo {author} {\bibfnamefont {T.}~\bibnamefont {Chwiej}}, \ and\ \bibinfo
  {author} {\bibfnamefont {J.}~\bibnamefont {Adamowski}},\ }\href
  {https://link.aps.org/doi/10.1103/PhysRevB.70.035401} {\bibfield  {journal}
  {\bibinfo  {journal} {Phys. Rev. B}\ }\textbf {\bibinfo {volume} {70}},\
  \bibinfo {pages} {035401} (\bibinfo {year} {2004})}\BibitemShut {NoStop}%
\bibitem [{\citenamefont {Rontani}\ \emph {et~al.}(2006)\citenamefont
  {Rontani}, \citenamefont {Cavazzoni}, \citenamefont {Bellucci},\ and\
  \citenamefont {Goldoni}}]{Rontani.06}%
  \BibitemOpen
  \bibfield  {author} {\bibinfo {author} {\bibfnamefont {M.}~\bibnamefont
  {Rontani}}, \bibinfo {author} {\bibfnamefont {C.}~\bibnamefont {Cavazzoni}},
  \bibinfo {author} {\bibfnamefont {D.}~\bibnamefont {Bellucci}}, \ and\
  \bibinfo {author} {\bibfnamefont {G.}~\bibnamefont {Goldoni}},\ }\href
  {https://doi.org/10.1063/1.2179418} {\bibfield  {journal} {\bibinfo
  {journal} {The Journal of Chemical Physics}\ }\textbf {\bibinfo {volume}
  {124}},\ \bibinfo {pages} {124102} (\bibinfo {year} {2006})}\BibitemShut
  {NoStop}%
\bibitem [{\citenamefont {Ghosal}\ \emph {et~al.}(2007)\citenamefont {Ghosal},
  \citenamefont {G\"u\ifmmode~\mbox{\c{c}}\else \c{c}\fi{}l\"u}, \citenamefont
  {Umrigar}, \citenamefont {Ullmo},\ and\ \citenamefont
  {Baranger}}]{Ghosal.07}%
  \BibitemOpen
  \bibfield  {author} {\bibinfo {author} {\bibfnamefont {A.}~\bibnamefont
  {Ghosal}}, \bibinfo {author} {\bibfnamefont {A.~D.}\ \bibnamefont
  {G\"u\ifmmode~\mbox{\c{c}}\else \c{c}\fi{}l\"u}}, \bibinfo {author}
  {\bibfnamefont {C.~J.}\ \bibnamefont {Umrigar}}, \bibinfo {author}
  {\bibfnamefont {D.}~\bibnamefont {Ullmo}}, \ and\ \bibinfo {author}
  {\bibfnamefont {H.~U.}\ \bibnamefont {Baranger}},\ }\href
  {https://link.aps.org/doi/10.1103/PhysRevB.76.085341} {\bibfield  {journal}
  {\bibinfo  {journal} {Phys. Rev. B}\ }\textbf {\bibinfo {volume} {76}},\
  \bibinfo {pages} {085341} (\bibinfo {year} {2007})}\BibitemShut {NoStop}%
\bibitem [{\citenamefont {Cavaliere}\ \emph {et~al.}(2009)\citenamefont
  {Cavaliere}, \citenamefont {Giovannini}, \citenamefont {Sassetti},\ and\
  \citenamefont {Kramer}}]{Cavaliere.09}%
  \BibitemOpen
  \bibfield  {author} {\bibinfo {author} {\bibfnamefont {F.}~\bibnamefont
  {Cavaliere}}, \bibinfo {author} {\bibfnamefont {U.~D.}\ \bibnamefont
  {Giovannini}}, \bibinfo {author} {\bibfnamefont {M.}~\bibnamefont
  {Sassetti}}, \ and\ \bibinfo {author} {\bibfnamefont {B.}~\bibnamefont
  {Kramer}},\ }\href {https://doi.org/10.1088/1367-2630/11/12/123004}
  {\bibfield  {journal} {\bibinfo  {journal} {New J. Phys.}\ }\textbf {\bibinfo
  {volume} {11}},\ \bibinfo {pages} {123004} (\bibinfo {year}
  {2009})}\BibitemShut {NoStop}%
\bibitem [{\citenamefont {Shapir}\ \emph {et~al.}(2019)\citenamefont {Shapir},
  \citenamefont {Hamo}, \citenamefont {Pecker}, \citenamefont {Moca},
  \citenamefont {Legeza}, \citenamefont {Zarand},\ and\ \citenamefont
  {Ilani}}]{Shapir.19}%
  \BibitemOpen
  \bibfield  {author} {\bibinfo {author} {\bibfnamefont {I.}~\bibnamefont
  {Shapir}}, \bibinfo {author} {\bibfnamefont {A.}~\bibnamefont {Hamo}},
  \bibinfo {author} {\bibfnamefont {S.}~\bibnamefont {Pecker}}, \bibinfo
  {author} {\bibfnamefont {C.~P.}\ \bibnamefont {Moca}}, \bibinfo {author}
  {\bibfnamefont {{\"O}.}~\bibnamefont {Legeza}}, \bibinfo {author}
  {\bibfnamefont {G.}~\bibnamefont {Zarand}}, \ and\ \bibinfo {author}
  {\bibfnamefont {S.}~\bibnamefont {Ilani}},\ }\href
  {https://www.science.org/doi/abs/10.1126/science.aat0905} {\bibfield
  {journal} {\bibinfo  {journal} {Science}\ }\textbf {\bibinfo {volume}
  {364}},\ \bibinfo {pages} {870} (\bibinfo {year} {2019})}\BibitemShut
  {NoStop}%
\bibitem [{\citenamefont {Mintairov}\ \emph {et~al.}(2018)\citenamefont
  {Mintairov}, \citenamefont {Kapaldo}, \citenamefont {Merz}, \citenamefont
  {Rouvimov}, \citenamefont {Lebedev}, \citenamefont {Kalyuzhnyy},
  \citenamefont {Mintairov}, \citenamefont {Belyaev}, \citenamefont {Rakhlin},
  \citenamefont {Toropov}, \citenamefont {Brunkov}, \citenamefont {Vlasov},
  \citenamefont {Zadiranov}, \citenamefont {Blundell}, \citenamefont
  {Mozharov}, \citenamefont {Mukhin}, \citenamefont {Yakimov}, \citenamefont
  {Oktyabrsky}, \citenamefont {Shelaev},\ and\ \citenamefont
  {Bykov}}]{Mintairov.18}%
  \BibitemOpen
  \bibfield  {author} {\bibinfo {author} {\bibfnamefont {A.~M.}\ \bibnamefont
  {Mintairov}}, \bibinfo {author} {\bibfnamefont {J.}~\bibnamefont {Kapaldo}},
  \bibinfo {author} {\bibfnamefont {J.~L.}\ \bibnamefont {Merz}}, \bibinfo
  {author} {\bibfnamefont {S.}~\bibnamefont {Rouvimov}}, \bibinfo {author}
  {\bibfnamefont {D.~V.}\ \bibnamefont {Lebedev}}, \bibinfo {author}
  {\bibfnamefont {N.~A.}\ \bibnamefont {Kalyuzhnyy}}, \bibinfo {author}
  {\bibfnamefont {S.~A.}\ \bibnamefont {Mintairov}}, \bibinfo {author}
  {\bibfnamefont {K.~G.}\ \bibnamefont {Belyaev}}, \bibinfo {author}
  {\bibfnamefont {M.~V.}\ \bibnamefont {Rakhlin}}, \bibinfo {author}
  {\bibfnamefont {A.~A.}\ \bibnamefont {Toropov}}, \bibinfo {author}
  {\bibfnamefont {P.~N.}\ \bibnamefont {Brunkov}}, \bibinfo {author}
  {\bibfnamefont {A.~S.}\ \bibnamefont {Vlasov}}, \bibinfo {author}
  {\bibfnamefont {Y.~M.}\ \bibnamefont {Zadiranov}}, \bibinfo {author}
  {\bibfnamefont {S.~A.}\ \bibnamefont {Blundell}}, \bibinfo {author}
  {\bibfnamefont {A.~M.}\ \bibnamefont {Mozharov}}, \bibinfo {author}
  {\bibfnamefont {I.}~\bibnamefont {Mukhin}}, \bibinfo {author} {\bibfnamefont
  {M.}~\bibnamefont {Yakimov}}, \bibinfo {author} {\bibfnamefont
  {S.}~\bibnamefont {Oktyabrsky}}, \bibinfo {author} {\bibfnamefont {A.~V.}\
  \bibnamefont {Shelaev}}, \ and\ \bibinfo {author} {\bibfnamefont {V.~A.}\
  \bibnamefont {Bykov}},\ }\href
  {https://link.aps.org/doi/10.1103/PhysRevB.97.195443} {\bibfield  {journal}
  {\bibinfo  {journal} {Phys. Rev. B}\ }\textbf {\bibinfo {volume} {97}},\
  \bibinfo {pages} {195443} (\bibinfo {year} {2018})}\BibitemShut {NoStop}%
\bibitem [{\citenamefont {Singha}\ \emph {et~al.}(2010)\citenamefont {Singha},
  \citenamefont {Pellegrini}, \citenamefont {Pinczuk}, \citenamefont
  {Pfeiffer}, \citenamefont {West},\ and\ \citenamefont {Rontani}}]{Singha.10}%
  \BibitemOpen
  \bibfield  {author} {\bibinfo {author} {\bibfnamefont {A.}~\bibnamefont
  {Singha}}, \bibinfo {author} {\bibfnamefont {V.}~\bibnamefont {Pellegrini}},
  \bibinfo {author} {\bibfnamefont {A.}~\bibnamefont {Pinczuk}}, \bibinfo
  {author} {\bibfnamefont {L.~N.}\ \bibnamefont {Pfeiffer}}, \bibinfo {author}
  {\bibfnamefont {K.~W.}\ \bibnamefont {West}}, \ and\ \bibinfo {author}
  {\bibfnamefont {M.}~\bibnamefont {Rontani}},\ }\href
  {https://link.aps.org/doi/10.1103/PhysRevLett.104.246802} {\bibfield
  {journal} {\bibinfo  {journal} {Phys. Rev. Lett.}\ }\textbf {\bibinfo
  {volume} {104}},\ \bibinfo {pages} {246802} (\bibinfo {year}
  {2010})}\BibitemShut {NoStop}%
\bibitem [{\citenamefont {Ellenberger}\ \emph {et~al.}(2006)\citenamefont
  {Ellenberger}, \citenamefont {Ihn}, \citenamefont {Yannouleas}, \citenamefont
  {Landman}, \citenamefont {Ensslin}, \citenamefont {Driscoll},\ and\
  \citenamefont {Gossard}}]{Ellenberger.06}%
  \BibitemOpen
  \bibfield  {author} {\bibinfo {author} {\bibfnamefont {C.}~\bibnamefont
  {Ellenberger}}, \bibinfo {author} {\bibfnamefont {T.}~\bibnamefont {Ihn}},
  \bibinfo {author} {\bibfnamefont {C.}~\bibnamefont {Yannouleas}}, \bibinfo
  {author} {\bibfnamefont {U.}~\bibnamefont {Landman}}, \bibinfo {author}
  {\bibfnamefont {K.}~\bibnamefont {Ensslin}}, \bibinfo {author} {\bibfnamefont
  {D.}~\bibnamefont {Driscoll}}, \ and\ \bibinfo {author} {\bibfnamefont
  {A.~C.}\ \bibnamefont {Gossard}},\ }\href
  {https://link.aps.org/doi/10.1103/PhysRevLett.96.126806} {\bibfield
  {journal} {\bibinfo  {journal} {Phys. Rev. Lett.}\ }\textbf {\bibinfo
  {volume} {96}},\ \bibinfo {pages} {126806} (\bibinfo {year}
  {2006})}\BibitemShut {NoStop}%
\bibitem [{\citenamefont {Kristinsd\'ottir}\ \emph {et~al.}(2011)\citenamefont
  {Kristinsd\'ottir}, \citenamefont {Cremon}, \citenamefont {Nilsson},
  \citenamefont {Xu}, \citenamefont {Samuelson}, \citenamefont {Linke},
  \citenamefont {Wacker},\ and\ \citenamefont {Reimann}}]{Kristinsdottir.11}%
  \BibitemOpen
  \bibfield  {author} {\bibinfo {author} {\bibfnamefont {L.~H.}\ \bibnamefont
  {Kristinsd\'ottir}}, \bibinfo {author} {\bibfnamefont {J.~C.}\ \bibnamefont
  {Cremon}}, \bibinfo {author} {\bibfnamefont {H.~A.}\ \bibnamefont {Nilsson}},
  \bibinfo {author} {\bibfnamefont {H.~Q.}\ \bibnamefont {Xu}}, \bibinfo
  {author} {\bibfnamefont {L.}~\bibnamefont {Samuelson}}, \bibinfo {author}
  {\bibfnamefont {H.}~\bibnamefont {Linke}}, \bibinfo {author} {\bibfnamefont
  {A.}~\bibnamefont {Wacker}}, \ and\ \bibinfo {author} {\bibfnamefont {S.~M.}\
  \bibnamefont {Reimann}} (\bibinfo {collaboration} {Nanometer Structure
  Consortium, nmC@LU}),\ }\href
  {https://link.aps.org/doi/10.1103/PhysRevB.83.041101} {\bibfield  {journal}
  {\bibinfo  {journal} {Phys. Rev. B}\ }\textbf {\bibinfo {volume} {83}},\
  \bibinfo {pages} {041101} (\bibinfo {year} {2011})}\BibitemShut {NoStop}%
\bibitem [{\citenamefont {Pecker}\ \emph {et~al.}(2013)\citenamefont {Pecker},
  \citenamefont {Kuemmeth}, \citenamefont {Secchi}, \citenamefont {Rontani},
  \citenamefont {Ralph}, \citenamefont {McEuen},\ and\ \citenamefont
  {Ilani}}]{Pecker.13}%
  \BibitemOpen
  \bibfield  {author} {\bibinfo {author} {\bibfnamefont {S.}~\bibnamefont
  {Pecker}}, \bibinfo {author} {\bibfnamefont {F.}~\bibnamefont {Kuemmeth}},
  \bibinfo {author} {\bibfnamefont {A.}~\bibnamefont {Secchi}}, \bibinfo
  {author} {\bibfnamefont {M.}~\bibnamefont {Rontani}}, \bibinfo {author}
  {\bibfnamefont {D.~C.}\ \bibnamefont {Ralph}}, \bibinfo {author}
  {\bibfnamefont {P.~L.}\ \bibnamefont {McEuen}}, \ and\ \bibinfo {author}
  {\bibfnamefont {S.}~\bibnamefont {Ilani}},\ }\href
  {https://doi.org/10.1038/nphys2692} {\bibfield  {journal} {\bibinfo
  {journal} {Nat. Phys.}\ }\textbf {\bibinfo {volume} {9}},\ \bibinfo {pages}
  {576} (\bibinfo {year} {2013})}\BibitemShut {NoStop}%
\bibitem [{\citenamefont {Egger}\ \emph {et~al.}(1999)\citenamefont {Egger},
  \citenamefont {H\"ausler}, \citenamefont {Mak},\ and\ \citenamefont
  {Grabert}}]{Egger.99}%
  \BibitemOpen
  \bibfield  {author} {\bibinfo {author} {\bibfnamefont {R.}~\bibnamefont
  {Egger}}, \bibinfo {author} {\bibfnamefont {W.}~\bibnamefont {H\"ausler}},
  \bibinfo {author} {\bibfnamefont {C.~H.}\ \bibnamefont {Mak}}, \ and\
  \bibinfo {author} {\bibfnamefont {H.}~\bibnamefont {Grabert}},\ }\href
  {https://link.aps.org/doi/10.1103/PhysRevLett.82.3320} {\bibfield  {journal}
  {\bibinfo  {journal} {Phys. Rev. Lett.}\ }\textbf {\bibinfo {volume} {82}},\
  \bibinfo {pages} {3320} (\bibinfo {year} {1999})}\BibitemShut {NoStop}%
\bibitem [{\citenamefont {Ercan}\ \emph {et~al.}(2021)\citenamefont {Ercan},
  \citenamefont {Coppersmith},\ and\ \citenamefont {Friesen}}]{Ercan.21}%
  \BibitemOpen
  \bibfield  {author} {\bibinfo {author} {\bibfnamefont {H.~E.}\ \bibnamefont
  {Ercan}}, \bibinfo {author} {\bibfnamefont {S.~N.}\ \bibnamefont
  {Coppersmith}}, \ and\ \bibinfo {author} {\bibfnamefont {M.}~\bibnamefont
  {Friesen}},\ }\href {https://link.aps.org/doi/10.1103/PhysRevB.104.235302}
  {\bibfield  {journal} {\bibinfo  {journal} {Phys. Rev. B}\ }\textbf {\bibinfo
  {volume} {104}},\ \bibinfo {pages} {235302} (\bibinfo {year}
  {2021})}\BibitemShut {NoStop}%
\bibitem [{\citenamefont {Corrigan}\ \emph {et~al.}(2021)\citenamefont
  {Corrigan}, \citenamefont {Dodson}, \citenamefont {Ercan}, \citenamefont
  {Abadillo-Uriel}, \citenamefont {Thorgrimsson}, \citenamefont {Knapp},
  \citenamefont {Holman}, \citenamefont {McJunkin}, \citenamefont {Neyens},
  \citenamefont {MacQuarrie}, \citenamefont {Foote}, \citenamefont {Edge},
  \citenamefont {Friesen}, \citenamefont {Coppersmith},\ and\ \citenamefont
  {Eriksson}}]{Corrigan.21}%
  \BibitemOpen
  \bibfield  {author} {\bibinfo {author} {\bibfnamefont {J.}~\bibnamefont
  {Corrigan}}, \bibinfo {author} {\bibfnamefont {J.~P.}\ \bibnamefont
  {Dodson}}, \bibinfo {author} {\bibfnamefont {H.~E.}\ \bibnamefont {Ercan}},
  \bibinfo {author} {\bibfnamefont {J.~C.}\ \bibnamefont {Abadillo-Uriel}},
  \bibinfo {author} {\bibfnamefont {B.}~\bibnamefont {Thorgrimsson}}, \bibinfo
  {author} {\bibfnamefont {T.~J.}\ \bibnamefont {Knapp}}, \bibinfo {author}
  {\bibfnamefont {N.}~\bibnamefont {Holman}}, \bibinfo {author} {\bibfnamefont
  {T.}~\bibnamefont {McJunkin}}, \bibinfo {author} {\bibfnamefont {S.~F.}\
  \bibnamefont {Neyens}}, \bibinfo {author} {\bibfnamefont {E.~R.}\
  \bibnamefont {MacQuarrie}}, \bibinfo {author} {\bibfnamefont {R.~H.}\
  \bibnamefont {Foote}}, \bibinfo {author} {\bibfnamefont {L.~F.}\ \bibnamefont
  {Edge}}, \bibinfo {author} {\bibfnamefont {M.}~\bibnamefont {Friesen}},
  \bibinfo {author} {\bibfnamefont {S.~N.}\ \bibnamefont {Coppersmith}}, \ and\
  \bibinfo {author} {\bibfnamefont {M.~A.}\ \bibnamefont {Eriksson}},\ }\href
  {https://link.aps.org/doi/10.1103/PhysRevLett.127.127701} {\bibfield
  {journal} {\bibinfo  {journal} {Phys. Rev. Lett.}\ }\textbf {\bibinfo
  {volume} {127}},\ \bibinfo {pages} {127701} (\bibinfo {year}
  {2021})}\BibitemShut {NoStop}%
\bibitem [{\citenamefont {Kalliakos}\ \emph {et~al.}(2008)\citenamefont
  {Kalliakos}, \citenamefont {Rontani}, \citenamefont {Pellegrini},
  \citenamefont {Garc{\'\i}a}, \citenamefont {Pinczuk}, \citenamefont
  {Goldoni}, \citenamefont {Molinari}, \citenamefont {Pfeiffer},\ and\
  \citenamefont {West}}]{Kalliakos.08}%
  \BibitemOpen
  \bibfield  {author} {\bibinfo {author} {\bibfnamefont {S.}~\bibnamefont
  {Kalliakos}}, \bibinfo {author} {\bibfnamefont {M.}~\bibnamefont {Rontani}},
  \bibinfo {author} {\bibfnamefont {V.}~\bibnamefont {Pellegrini}}, \bibinfo
  {author} {\bibfnamefont {C.~P.}\ \bibnamefont {Garc{\'\i}a}}, \bibinfo
  {author} {\bibfnamefont {A.}~\bibnamefont {Pinczuk}}, \bibinfo {author}
  {\bibfnamefont {G.}~\bibnamefont {Goldoni}}, \bibinfo {author} {\bibfnamefont
  {E.}~\bibnamefont {Molinari}}, \bibinfo {author} {\bibfnamefont {L.~N.}\
  \bibnamefont {Pfeiffer}}, \ and\ \bibinfo {author} {\bibfnamefont {K.~W.}\
  \bibnamefont {West}},\ }\href {https://doi.org/10.1038/nphys944} {\bibfield
  {journal} {\bibinfo  {journal} {Nat. Phys.}\ }\textbf {\bibinfo {volume}
  {4}},\ \bibinfo {pages} {467} (\bibinfo {year} {2008})}\BibitemShut {NoStop}%
\bibitem [{\citenamefont {Kouwenhoven}\ \emph {et~al.}(1997)\citenamefont
  {Kouwenhoven}, \citenamefont {Oosterkamp}, \citenamefont {Danoesastro},
  \citenamefont {Eto}, \citenamefont {Austing}, \citenamefont {Honda},\ and\
  \citenamefont {Tarucha}}]{Kouwenhoven.97}%
  \BibitemOpen
  \bibfield  {author} {\bibinfo {author} {\bibfnamefont {L.~P.}\ \bibnamefont
  {Kouwenhoven}}, \bibinfo {author} {\bibfnamefont {T.~H.}\ \bibnamefont
  {Oosterkamp}}, \bibinfo {author} {\bibfnamefont {M.~W.~S.}\ \bibnamefont
  {Danoesastro}}, \bibinfo {author} {\bibfnamefont {M.}~\bibnamefont {Eto}},
  \bibinfo {author} {\bibfnamefont {D.~G.}\ \bibnamefont {Austing}}, \bibinfo
  {author} {\bibfnamefont {T.}~\bibnamefont {Honda}}, \ and\ \bibinfo {author}
  {\bibfnamefont {S.}~\bibnamefont {Tarucha}},\ }\href
  {https://www.science.org/doi/abs/10.1126/science.278.5344.1788} {\bibfield
  {journal} {\bibinfo  {journal} {Science}\ }\textbf {\bibinfo {volume}
  {278}},\ \bibinfo {pages} {1788} (\bibinfo {year} {1997})}\BibitemShut
  {NoStop}%
\end{thebibliography}

\end{document}